\documentclass{amsart}
\usepackage{amsaddr}
\usepackage{amsmath}
\usepackage{amssymb}
\usepackage{gensymb}
\usepackage{dsfont}
\usepackage{breqn}
\usepackage{bm}
\usepackage{graphicx}
\usepackage{geometry}
\usepackage{cite}
\usepackage{comment}

\title{Second- and third-order properties of multidimensional Langevin equations}
\author{Yeeren I. Low}
\address{Department of Physics, McGill University, Montr\'{e}al, Qu\'{e}bec, Canada}
\curraddr{Department of Physics, University of Vermont, Burlington, Vermont, USA}
\email{yeeren.low@mail.mcgill.ca}

\begin{document}
\begin{abstract}
Recent work has addressed the problem of inferring Langevin dynamics from data. In this work, we address the problem of relating terms in the Langevin equation to statistical properties, such as moments of the probability density function and of the probability current density, as well as covariance functions. We first review the case of linear Gaussian dynamics, and then consider extensions beyond this simple case. We address the question of quantitative significance of effects. We also analyze underdamped (second-order) processes, specifically in the limit where dynamics in state space is almost Markovian. Finally, we address detection of non-Markovianity.
\end{abstract}
\maketitle

\section{Introduction}
Langevin equations, or equivalently Fokker\textendash Planck equations, play an important role in stochastic modeling of biological systems, from animal locomotion \cite{StephensDimensionalityDynamicsCElegans2008} to cell migration \cite{BrucknerLearning2024}. These stochastic differential equations describe Markov processes in continuous time having values in a continuous state space with continuous paths \cite{GardinerStochasticMethods}. The simplest process admitting a stationary probability distribution is a linear Gaussian process where drift is a linear function of the state variables, and diffusion matrix is constant. Such processes have been discussed in \cite{Weiss}, where their properties are analyzed in an coordinate-invariant manner. In this paper, we elaborate on the linear case and extend this analysis to processes having nonlinear drift and inhomogeneous diffusion.

Probability currents are related to entropy production in the case where the state variables are even under time reversal \cite{Seifert2012}. It has been proposed to use state-space coarse-graining to analyze probability currents \cite{Dieball2022}. However, this approach is not ideal for purposes of statistical testing, as it requires arbitrarily choosing a (closed) path around which to calculate the probability flux (see \cite{Battle2016}). In addition, it is data-hungry in high dimension. Instead, a quantity known as angular momentum, also known as twice the stochastic area, has been proposed to quantify circulation in linear models \cite{Gonzalez2019}. In this paper, we extend this approach to nonlinear circulations.

Recent work has addressed the problem of inferring Langevin dynamics from data \cite{Frishman, Bruckner}. The amount of information that can be statistically resolved depends on the amount of data collected \cite{SelmecziCellMotility2008, LiDictyDynamics2011}. However, just because effects can be statistically resolved does not mean that they are quantitatively significant. Effects of interest include non-Gaussian probability distributions, non-vanishing probability currents, and inhomogeneous diffusion. We introduce a framework in which to evaluate the quantitative significance of such effects, invariant under linear transformations of the state coordinates.

We discuss the case where some variables do not admit stationary probability distributions. We also discuss the case where some variables are odd under time reversal. We discuss experimentally measurable covariance functions, and prove that they are time-reversible under certain assumptions and suitable approximations. We also discuss quantitative comparison of theoretical and experimental covariance functions.

Finally, we consider underdamped (second-order) processes, specifically in the limit where dynamics in state space is almost Markovian, as well as detection of non-Markovianity.

\section{Linear Gaussian stationary systems}

\subsection{Preliminaries: Definition, covariance functions, and detailed balance}
The general time-homogeneous It\^o\textendash Langevin equation is given by:
\begin{equation} \label{eq:langevin-general}
	\dot{\mathbf x}(t) = \bm{\mathcal A}(\mathbf x(t)) + \boldsymbol \xi(t), \quad \langle \boldsymbol \xi(t) \mid \mathbf x(t) \rangle = 0, \quad \langle \boldsymbol \xi(t) \boldsymbol \xi(t')^{\mathsf T} \mid \mathbf x(t) \rangle = \bm{\mathcal B}(\mathbf x(t)) \delta(t-t'),
\end{equation}
where $\boldsymbol \xi(t)$ conditioned on $\mathbf x(t)$ (the vertical lines denote conditional expectation) is Gaussian white noise and $\delta(\cdot)$ denotes Dirac delta function. The functions $\bm{\mathcal A}(\mathbf x)$ and $\bm{\mathcal B}(\mathbf x)/2$ are called drift and diffusion, respectively. The simplest multivariate stochastic process having a stationary distribution is the multivariate Ornstein\textendash Uhlenbeck process, where:
\begin{equation} \label{eq:langevin-linear-stat}
	\bm{\mathcal A}(\mathbf x) = \mathbf A \mathbf x, \quad \bm{\mathcal B}(\mathbf x) = 2 \mathbf D,
\end{equation}
\footnote{We have omitted a possible additive constant in $\dot{\mathbf x}$. In the stationary case, a system with an additive constant may be transformed into Eq.\ \eqref{eq:langevin-linear-stat} by considering instead the variable $\mathbf x - \mathbf c$, where $\mathbf c$ is a constant (in this case, $\mathbf c = \langle \mathbf x \rangle$). This simplification is used throughout.} where $\mathbf A$ and $\mathbf D$ are constants. The covariance matrix $\mathbf C := \langle \mathbf x \mathbf x^{\mathsf T} \rangle$ is determined from the Lyapunov equation \cite{Gnesotto}:
\begin{equation}
\mathbf A \mathbf C + \mathbf C \mathbf A^{\mathsf T} + 2 \mathbf D = \mathbf 0,
\end{equation}
which can be solved using tensor notation and Einstein summation notation as:
\begin{equation} \label{eq:covariance-matrix}
C^{ij} = -2 \left[{(\mathbf A \otimes \mathds 1 + \mathds 1 \otimes \mathbf A)^{-1}}\right]^{ij}_{kl} D^{kl}.
\end{equation}
where $\mathds 1$ denotes the identity matrix of appropriate dimension, and $\otimes$ denotes tensor or Kronecker product\footnote{For a matrix representing a linear transformation (i.e., a second-rank mixed tensor), the first index is a superscript (contravariant) and the second index is a subscript (covariant). Thus the $(i,j)$ element of such a matrix $A$ is denoted $A^i_j$. The tensor or Kronecker product $A \otimes B$ of two second-rank mixed tensors $A$ and $B$ is defined as $(A \otimes B)^{ij}_{kl} = A^i_k B^j_l$. Tensors with equal number of contravariant and covariant indices may be regarded as matrices where all the contravariant (respectively covariant) indices are regarded as a single ``super-index''. The usual matrix operations can then be defined in this way, e.g.\ multiplication as $(AB)^{ij}_{kl} := A^{ij}_{i'j'} B^{i'j'}_{kl}$ and inversion correspondingly, resulting in tensors of the same type. For inversion, we may argue as follows. Suppose $A$ and $B$ are tensors of the same type with equal number of contravariant and covariant indices, and suppose they are matrix inverses of each other in a particular coordinate system. Because the equalities $AB = \mathds 1$ and $BA = \mathds 1$ are invariant under a change of basis, we conclude that $A$ and $B$ are matrix inverses of each other in any coordinate system, hence $A^{-1} = B$ transforms according to the same law as $A$. On the other hand, the covariance and diffusion matrices are second-rank contravariant tensors and thus their matrix inverses represent covariant tensors. We can explicitly compute the transformation law for such a matrix under a change of basis matrix $\mathbf R$. A second-rank contravariant tensor $\mathbf C$ transforms into $\widetilde{\mathbf C}$ where $\widetilde{C}^{ij} = (R^{-1})^i_k (R^{-1})^j_l C^{kl} = (\mathbf R^{-1} \mathbf C (\mathbf R^{-1})^{\mathsf T})^{ij}$. Its matrix inverse $\mathbf C^{-1}$ thus transforms into $\widetilde{C}^{-1}_{ij} = (\mathbf R^{\mathsf T} \mathbf C^{-1} \mathbf R)_{ij} = R^k_i R^l_j C^{-1}_{kl}$, which is the transformation law for a covariant tensor. We may establish similar rules for tensors with an even number of indices, regardless of type. Also, symmetry or antisymmetry of tensors under exchange of indices that are either both contravariant or both covariant is invariant under a change of basis.}. When comparing to experimental data, this relation is not to be considered as a test of the linear model, as it is a simple consequence of stationarity of $\mathbf x$ and the model fit, by It\^o's lemma:
\begin{equation} \label{eq:Lyapunov}
\mathbf 0 = \frac {\mathrm d}{\mathrm d t} \langle \mathbf x(t) \mathbf x(t)^{\mathsf T} \rangle = \langle \dot{\mathbf x}(t) \mathbf x(t)^{\mathsf T} \rangle + \langle \mathbf x(t) \dot{\mathbf x}(t)^{\mathsf T} \rangle + \left\langle{\frac {\mathrm d [\mathbf x, \mathbf x^{\mathsf T}](t)}{\mathrm d t}}\right\rangle
\end{equation}
where the time-derivative is interpreted in the It\^o sense, and $[\cdot, \cdot]$ is the covariation defined by:
\begin{equation}
	[x,y](t) = \lim_{\Delta \to 0} \sum_{n=0}^{N-1} (x(t_{n+1})-x(t_n)) (y(t_{n+1})-y(t_n))
\end{equation}
where $0 = t_0 < t_1 < \cdots < t_{N-1} < t_N = t$ and $\Delta := \max \limits_{0 \le n \le N-1} (t_{n+1}-t_n)$. Its time derivative is given by:
\begin{equation}
	\frac{\mathrm d[x,y](t)}{\mathrm d t} = \lim_{\tau \to 0^+} \frac{(x(t+\tau)-x(t))(y(t+\tau)-y(t))}{\tau},
\end{equation}
whose conditional expectation is $\langle \mathrm d [\mathbf x, \mathbf x^{\mathsf T}](t)/\mathrm d t \mid \mathbf x(t) \rangle = \bm{\mathcal B}(\mathbf x(t))$. Eq.\ \eqref{eq:Lyapunov} then follows simply from a rearrangement of terms:
\begin{equation} \label{eq:Lyapunov-deriv-simple}
	\begin{aligned}
		\langle (\mathbf x(\tau) - \mathbf x(0)) (\mathbf x(\tau) - \mathbf x(0))^{\mathsf T} \rangle &= \langle \mathbf x(\tau) \mathbf x(\tau)^{\mathsf T} - \mathbf x(0) \mathbf x(\tau)^{\mathsf T} \rangle - \langle (\mathbf x(\tau) - \mathbf x(0)) \mathbf x(0)^{\mathsf T} \rangle \\
		&= \langle \mathbf x(0) (\mathbf x(0) - \mathbf x(\tau))^{\mathsf T} \rangle - \langle (\mathbf x(\tau) - \mathbf x(0)) \mathbf x(0)^{\mathsf T} \rangle
	\end{aligned}
\end{equation}
upon dividing by $\tau>0$ and taking $\tau \to 0^+$, where we used $\langle \mathbf x(\tau) \mathbf x(\tau)^{\mathsf T} \rangle = \langle \mathbf x(0) \mathbf x(0)^{\mathsf T} \rangle$ and took the derivative inside the expectation (however, see Appendix B for a counterexample). This type of manipulation will become useful later.

Of interest is the covariance function \cite{GardinerStochasticMethods}:
\begin{equation}
\langle \mathbf x(\tau) \mathbf x(0)^{\mathsf T} \rangle = e^{\mathbf A \tau} \mathbf C,
\end{equation}
where $\tau \ge 0$ is always assumed. This can be derived by solving the differential equation for the conditional expectation:
\begin{equation}
	\frac{\mathrm d}{\mathrm d \tau} \langle \mathbf x(\tau) \mid \mathbf x(0) \rangle = \mathbf A \langle \mathbf x(\tau) \mid \mathbf x(0) \rangle,
\end{equation}
\footnote{For the stationary case, we require that the probability distribution approaches the stationary distribution as time passes. This means that the eigenvalues of $\mathbf A$ must have strictly negative real parts. Additionally, non-vanishing real parts of eigenvalues of $\mathbf A$ are required for $\mathbf A \otimes \mathds 1 + \mathds 1 \otimes \mathbf A$ to be invertible. We may demonstrate this as follows. Let $\mathbf v$ be an eigenvector of $\mathbf A$ with eigenvalue $\lambda + i \omega$ ($\lambda, \omega$ real). Then its complex conjugate $\mathbf v^*$ is an eigenvector of $\mathbf A$ with eigenvalue $\lambda - i \omega$. We see that $\mathbf v \otimes \mathbf v^*$ is an eigenvector of $\mathbf A \otimes \mathds 1 + \mathds 1 \otimes \mathbf A$ with eigenvalue $2 \lambda$. Thus if $\mathbf A \otimes \mathds 1 + \mathds 1 \otimes \mathbf A$ is invertible, we must have $\lambda \ne 0$.} and multiplying on the right by $\mathbf x(0)^{\mathsf T}$ and using the law of iterated expectations, again assuming that the derivative can be taken inside the expectation. Note that this result does not depend on homogeneity of diffusion.

Of interest also is the condition for detailed balance. A requirement for detailed balance is that the covariance function is symmetric, $\langle \mathbf x(\tau) \mathbf x(0)^{\mathsf T} \rangle = \langle \mathbf x(0) \mathbf x(\tau)^{\mathsf T} \rangle$. By differentiating with respect to $\tau$, we get $\mathbf A \mathbf C = \mathbf C \mathbf A^{\mathsf T}$\footnote{By expanding $e^{\mathbf A \tau}$ in a Taylor series, it is seen that this condition implies symmetry of the covariance function.}. To gain insight into this relation, we switch to ``covariance-identity'' coordinates in which $\mathbf C = \mathds 1$ \cite{Gnesotto}. (We assume that $\mathbf C$ is finite and positive definite, so that such a transformation can be made. We make this assumption throughout, except for the case of non-stationary variables to which it is not applicable.) In these coordinates, detailed balance implies that $\mathbf A$ is symmetric, which means it is diagonalizable by an orthogonal basis. In ``covariance-identity'' coordinates, an orthogonal basis means uncorrelated components (because $\langle (\mathbf v^{\mathsf T} \mathbf x) (\mathbf x^{\mathsf T} \mathbf w) \rangle = \mathbf v^{\mathsf T} \mathbf w$ for constant vectors $\mathbf v$, $\mathbf w$). By the Lyapunov equation, $\mathbf D$ must also be diagonal in this basis. Hence the system decomposes into a number of noninteracting one-dimensional linear systems \cite{Weiss}. Ref.\ \cite{Weiss} also contains good discussions about general properties of linear Gaussian systems.

The Lyapunov equation implies a relationship between the eigenvalues of $\mathbf A$ and of $\mathbf D \mathbf C^{-1}$. (These are second-rank mixed tensors, whose eigenvalues are independent of the basis in which they are represented.) We again work in a coordinate system in which $\mathbf C = \mathds 1$. Let $\lambda$ be an eigenvalue of $\mathbf A$ and $\mathbf v$ be a corresponding eigenvector. Then:
\begin{equation}
	\mathbf v^{\mathsf H} \mathbf D \mathbf v = -\mathbf v^{\mathsf H} \left({\frac{\mathbf A + \mathbf A^{\mathsf T}}{2}}\right) \mathbf v = -(\Re \lambda) (\mathbf v^{\mathsf H} \mathbf v)
\textbf{}\end{equation}
($\mathsf{H}$ denoting Hermitian conjugate and $\Re$ denoting the real part), therefore $-\Re \lambda$ is bounded by the smallest and largest eigenvalues of $\mathbf D$ \cite{52588}. In addition, the trace of $\mathbf D \mathbf C^{-1}$ equals negative the trace of $\mathbf A$.

\subsection{Comparison of experimental data with theoretical predictions}
Now, we turn to the question of comparing experimental data and theoretical predictions under a linear model. As stated in the introduction, we may not necessarily care about statistically resolvable deviations if they are quantitatively small. We thus desire a way to evaluate the size of deviations. We may consider comparing the deviation between experiment and theory of $\langle \mathbf x(\tau) \mathbf x(0)^{\mathsf T} \rangle$ to $\mathbf C$. That is, if we consider a whitened dataset for which $\mathbf C = \mathds 1$, the entries of the ``deviation matrix'' (i.e., experiment minus theory of $\langle \mathbf x(\tau) \mathbf x(0)^{\mathsf T} \rangle$) in these coordinates should be compared to unity. In the following, we develop a coordinate-independent description of this procedure.

First, we consider the covariance matrix. We denote the deviation $\mathbf M := \langle \mathbf x \mathbf x^{\mathsf T} \rangle_{\text{expt}} - \mathbf C_{\text{theo}}$, where ``expt'' and ``theo'' subscripts denote experimental and theoretical values, respectively. From now on we suppress the ``theo'' subscript and treat $\mathbf C$ as fixed. We may imagine an ensemble of stochastic systems, each having its own value of $\langle \mathbf x \mathbf x^{\mathsf T} \rangle$ which may differ from $\mathbf C$, and again denote the difference by $\mathbf M$. We assume that the ensemble mean of $\mathbf M$ is zero, and we now ask the question of what a suitable ensemble covariance of $\mathbf M$ would be that reflects significant deviations. We may then compare the elements of the experimental value of $\mathbf M$ to the square root of the ensemble variance. For biological applications, a factor of 0.3 or more may be considered significant.

In the presence of symmetries, some components of $\mathbf M$ may be constrained to be 0. Absent such restrictions, we may consider the putative relation:
\begin{equation} \label{eq:Msq}
	M^{ij} M^{i'j'} \sim C^{ii'} C^{jj'}.
\end{equation}
where $\sim$ denotes ensemble expectation. We may treat $(i,j)$ as a single ``super-index'' and similarly for $(i',j')$, and multiply the left-hand side (l.h.s.)\ by the matrix inverse of the right-hand side (r.h.s.). We denote by $\boldsymbol{\Gamma}$ the resulting matrix:
\begin{equation}
	\Gamma^{ij}_{i'j'} := M^{ij} M^{kl} C^{-1}_{i'k} C^{-1}_{j'l}.
\end{equation}
In coordinates where $\mathbf C = \mathds 1$, the matrix $\boldsymbol{\Gamma}$ takes the form $\mathbf v \mathbf v^{\mathsf T}$, and therefore has rank at most 1. Thus, $\boldsymbol{\Gamma}$ has at most one non-zero eigenvalue, which is equal to its trace. We have:
\begin{equation}
	\operatorname{tr}(\boldsymbol{\Gamma}) = \Gamma^{ij}_{ij} = M^{ij} M^{i'j'} C^{-1}_{ii'} C^{-1}_{jj'} = \operatorname{tr}((\mathbf M \mathbf C^{-1})^2).
\end{equation}
The second-rank mixed tensor $\mathbf M \mathbf C^{-1}$ has real eigenvalues because it is symmetric in ``covariance-identity'' coordinates. We see that the r.h.s.\ of the above is the sum of the squared eigenvalues of $\mathbf M \mathbf C^{-1}$. We may use this quantity to evaluate quantitative significance\footnote{It may be noticed that in ``covariance-identity'' coordinates, the ``cross-covariances'' ($M^{ij}$ where $i \ne j$) are counted with double the contribution to $\operatorname{tr}(\boldsymbol{\Gamma})$ as compared to ``self-covariances'' ($M^{ij}$ where $i=j$). This is necessary for a coordinate-independent evaluation. We may illustrate this on a two-dimensional example, with variables $x$ and $y$. We fix $\langle x^2 \rangle = \langle y^2 \rangle = 1$, while $\langle xy \rangle$ is allowed to deviate from 0. Changing coordinates to $x' := (x+y)/\sqrt{2}$ and $y' = (y-x)/\sqrt{2}$, we obtain $\langle {x'}^2 \rangle = 1 + \langle xy \rangle$, $\langle {y'}^2 \rangle = 1 - \langle xy \rangle$, and $\langle x'y' \rangle = 0$. We thus see that if squared deviations of $\langle {x'}^2 \rangle$ and $\langle {y'}^2 \rangle$ to 1 are to be added, we must add the contribution due to $\langle xy \rangle$ twice. Furthermore, for a two-dimensional Ornstein\textendash Uhlenbeck process with $-\mathbf A = \mathbf D = \mathbf C = \mathds 1$, the time-averages for a trajectory of finite length, denoted by overlines, satisfy $\langle (\overline{x^2}-1)^2 \rangle = 2 \langle (\overline{xy})^2 \rangle$.}.

First, however, we must address a problem with Eq.\ \eqref{eq:Msq}, which is that it does not obey symmetry in the indices. We may remedy this by symmetrizing its r.h.s. To justify this choice, consider a complete set of left eigenvectors of $\boldsymbol{\Gamma}$. Such a set exists because a rank-one matrix can always be diagonalized. In dimension $d$, there are $d(d-1)/2$ linearly independent left eigenvectors such that $v_{ji} = -v_{ij}$; we call these ``antisymmetric'' left eigenvectors and are associated with a zero eigenvalue. We may assume that they are included our complete set of left eigenvectors. Let $v_{ij}^{(k)}$ be the remaining left eigenvectors of $\boldsymbol{\Gamma}$ with eigenvalues $\mu^{(k)}$ ($k$ taking $d(d+1)/2$ values), so that:
\begin{equation}
	v_{ij}^{(k)} M^{ij} M^{i'j'} = \mu^{(k)} v_{ij}^{(k)} C^{ii'} C^{jj'}.
\end{equation}
Since this is symmetric in $(i',j')$, we can write this as:
\begin{equation}
	v_{ij}^{(k)} M^{ij} M^{i'j'} = \mu^{(k)} v_{ij}^{(k)} \frac{C^{ii'} C^{jj'} + C^{ij'} C^{ji'}}{2}.
\end{equation}
Now if we neglect permutations in $(i,j)$ and $(i',j')$\footnote{I.e., only one of $(i,j)$ and $(j,i)$ is included in the list of ``super-indices'' (and similarly for $(i',j')$). This is necessary to invert $(C^{ii'} C^{jj'} + C^{ij'} C^{ji'})/2$, since this quantity is invariant upon swapping $i \leftrightarrow j$ (or $i' \leftrightarrow j'$).}, we can see that the matrix product of $M^{ij} M^{i'j'}$ with the matrix inverse of $(C^{ii'} C^{jj'} + C^{ij'} C^{ji'})/2$, which we will denote $\boldsymbol{\Gamma}'$, has left eigenvectors with elements $v_{ij}^{(k)} + v_{ji}^{(k)}$ for $i \ne j$ and $v_{ij}^{(k)}$ for $i=j$ with the same eigenvalues $\mu^{(k)}$. The resulting left eigenvectors of $\boldsymbol{\Gamma}'$ are linearly independent, since if $\sum_k c^{(k)} (v_{ij}^{(k)} + v_{ji}^{(k)}) = 0$ for some scalars $c^{(k)}$, then $\sum_k c^{(k)} v_{ij}^{(k)}$ is antisymmetric and therefore $c^{(k)} = 0$ by assumption of linear independence of the original set of left eigenvectors. Thus, aside from the zero eigenvalues of $\boldsymbol{\Gamma}$ associated with antisymmetric left eigenvectors, the eigenvalues of $\boldsymbol{\Gamma}$ and $\boldsymbol{\Gamma'}$ are identical. We thus arrive at the relation:
\begin{equation}
	M^{ij} M^{i'j'} \sim \frac{C^{ii'} C^{jj'} + C^{ij'} C^{ji'}}{2}.
\end{equation}
We may assess the quantitative significance of a single element $M^{ij}$ by setting $i'=i$, $j'=j$ in the above, taking square roots, and comparing. For an assessment of all $M^{ij}$ collectively, we may calculate the trace of $\boldsymbol \Gamma$ according to this prescription:
\begin{equation}
	\operatorname{tr}(\boldsymbol{\Gamma}) \sim \frac{d(d+1)}{2}.
\end{equation}
The r.h.s.\ is the number of degrees of freedom in $\mathbf M$, accounting for symmetry. To evaluate quantitative significance, we may compare the square roots of the above. We see that it is not suitable to compare the eigenvalues of $\mathbf M \mathbf C^{-1}$ to unity, as there are only $d$ of these while the sum of their squares is compared to $d(d+1)/2$.

Now, we turn to covariance functions with a time-lag. We may again denote the deviation $\mathbf M := \langle \mathbf x(\tau) \mathbf x(0)^{\mathsf T} \rangle_{\text{expt}} - \langle \mathbf x(\tau) \mathbf x(0)^{\mathsf T} \rangle_{\text{theo}}$. Now, there is no symmetry requirement for the components of $\mathbf M$, and hence we can take them as obeying Eq.\ \eqref{eq:Msq}. Collectively, if $\sqrt{M^{ij} M^{i'j'} C^{-1}_{ii'} C^{-1}_{jj'}}$ is comparable to $d$, then $\mathbf M$ may be considered (collectively) quantitatively significant. In a sense, we are done. We may however consider what happens if we separate $\mathbf M$ into symmetric and antisymmetric components, denoted by $\mathbf S$ and $\mathbf T$ respectively:
\begin{align}
	\mathbf S &:= \frac{\mathbf M + \mathbf M^{\mathsf T}}{2}, \\
	\mathbf T &:= \frac{\mathbf M - \mathbf M^{\mathsf T}}{2}.
\end{align}
Then $\mathbf S \mathbf C^{-1}$ and $\mathbf T \mathbf C^{-1}$ have purely real and purely imaginary eigenvalues, respectively (again by arguing using coordinates in which $\mathbf C = \mathds 1$). We have:
\begin{align}
	S^{ij} S^{i'j'} C^{-1}_{ii'} C^{-1}_{jj'} &= \operatorname{tr}((\mathbf S \mathbf C^{-1})^2), \\
	T^{ij} T^{i'j'} C^{-1}_{ii'} C^{-1}_{jj'} &= -\operatorname{tr}((\mathbf T \mathbf C^{-1})^2).
\end{align}
The r.h.s.'s are the sums of the squared moduli of the eigenvalues of $\mathbf S \mathbf C^{-1}$ and $\mathbf T \mathbf C^{-1}$. Also, straightforward calculation reveals that:
\begin{equation}
	(S^{ij} S^{i'j'} + T^{ij} T^{i'j'}) C^{-1}_{ii'} C^{-1}_{jj'} = M^{ij} M^{i'j'} C^{-1}_{ii'} C^{-1}_{jj'}.
\end{equation}
The r.h.s.\ is $\operatorname{tr}(\boldsymbol{\Gamma})$, as before (although not $\operatorname{tr}((\mathbf M \mathbf C^{-1})^2)$, since $\mathbf M$ is in general not symmetric). We have thus decomposed this into contributions from the symmetric and antisymmetric components of $\mathbf M$. Using Eq.\ \eqref{eq:Msq}, we then have:
\begin{align}
	S^{ij} S^{i'j'} &\sim \frac{C^{ii'} C^{jj'} + C^{ij'} C^{ji'}}{2}, \\
	T^{ij} T^{i'j'} &\sim \frac{C^{ii'} C^{jj'} - C^{ij'} C^{ji'}}{2}, \\
	S^{ij} T^{i'j'} &\sim 0.
\end{align}
The ensemble variance of $\mathbf S$ is therefore the same as what was previously argued for $\mathbf M$ in the symmetric case. The ensemble variance of $\mathbf T$ is similar, except that it obeys antisymmetry rather than symmetry. The ensemble covariance between $\mathbf S$ and $\mathbf T$ vanishes, which makes sense because $\mathbf S$ and $\mathbf T$ acquire opposite signs upon time reversal. For ``collective'' quantitative significance, the above relations imply:
\begin{align}
	S^{ij} S^{i'j'} C^{-1}_{ii'} C^{-1}_{jj'} &\sim \frac{d(d+1)}{2}, \\
	T^{ij} T^{i'j'} C^{-1}_{ii'} C^{-1}_{jj'} &\sim \frac{d(d-1)}{2}.
\end{align}
The r.h.s.'s correspond to the number of degrees of freedom in $\mathbf S$ and $\mathbf T$, respectively. In this way, $\mathbf S$ and $\mathbf T$ may be individually evaluated for quantitative significance, if so desired. Similarly to before, it is not suitable to compare the eigenvalues of $\mathbf S \mathbf C^{-1}$ and $\mathbf T \mathbf C^{-1}$ to unity.

For the case of complex variables, see Appendix A.

\subsection{Angular momentum and probability current density}
The dynamics of the multivariate Ornstein\textendash Uhlenbeck process is determined by the matrices $\mathbf A$ and $\mathbf D$. However, to alternatively characterize the system by quantities even and odd under time reversal, we may consider the angular momentum matrix:
\begin{equation}
\mathbf L := \lim_{\tau \to 0^+} \frac{\langle \mathbf x(0) \mathbf x(\tau)^{\mathsf T} - \mathbf x(\tau) \mathbf x(0)^{\mathsf T} \rangle}{\tau} = \langle \mathbf x \dot{\mathbf x}^{\mathsf T} - \dot{\mathbf x} \mathbf x^{\mathsf T} \rangle = \mathbf C \mathbf A^{\mathsf T} - \mathbf A \mathbf C,
\end{equation}
where the time-derivatives are interpreted in It\^o sense. We also have $\mathbf L = 2 \langle \mathbf x \circ \dot{\mathbf x}^{\mathsf T} \rangle$ where the time-derivative is interpreted in Stratonovich sense (denoted by the open circle). The system may now be characterized in terms of $\mathbf C$, $\mathbf L$, and $\mathbf D$.

\subsubsection{Stochastic rotation frequencies} At stationarity (which is assumed throughout to the extent applicable), the probability density is $p(\mathbf x) \propto \exp \left({-\frac 1 2 \mathbf x^{\mathsf T} \mathbf C^{-1} \mathbf x}\right)$ and the probability current density is $\mathbf J(\mathbf x) = \boldsymbol \Omega \mathbf x p(\mathbf x)$, where $\boldsymbol \Omega = \mathbf A + \mathbf D \mathbf C^{-1}$. The matrix $\boldsymbol \Omega$ is traceless and has purely imaginary eigenvalues \cite{Gnesotto} which define stochastic rotation frequencies, and the corresponding complex eigenvectors define planes in which breaking of detailed balance occurs.

Now, we specialize to the two-dimensional case, where $\mathbf x = (x, y)^{\mathsf T}$. Here, $\boldsymbol \Omega$ has a single pair of purely imaginary eigenvalues $\pm i \omega_{\mathrm{stoch}}$. We will see that the possible values of $\omega_{\mathrm{stoch}}$ are restricted by the eigenvalues of $\mathbf A$. The angular momentum is $L := \langle x \dot y - \dot x y \rangle$. It is related to the matrix $\boldsymbol \Omega$ by the relation:
\begin{equation}
\begin{pmatrix}
0 & -L \\
L & 0
\end{pmatrix} = \mathbf A \mathbf C - \mathbf C \mathbf A^{\mathsf T} = 2 \boldsymbol \Omega \mathbf C,
\end{equation}
where we used the Lyapunov relation. Upon taking determinants, we get:
\begin{equation}
\omega_{\mathrm{stoch}} = \frac L {2\sqrt{\det(\mathbf C)}}.
\end{equation}
(In the three-dimensional case, $\boldsymbol \Omega$ has at most a single pair of non-zero purely imaginary eigenvalues, and the stochastic rotation frequency is half the magnitude of the angular momentum vector evaluated in ``covariance-identity'' coordinates.)

The stochastic rotation frequency when $\mathbf A$ has two real eigenvalues has been computed in \cite{Gladrow} and is easily derived from the above formula in coordinates where:
\begin{equation} \label{eq:A-real-eigvals}
\mathbf A =
\begin{pmatrix}
-\lambda_x & 0 \\
0 & -\lambda_y
\end{pmatrix},
\end{equation}
where $\lambda_x, \lambda_y > 0$. For completeness, the result is:
\begin{equation}
\omega_{\mathrm{stoch}} = (\lambda_x - \lambda_y) D^{xy} \left[{\frac{(\lambda_x+\lambda_y)^2}{\lambda_x \lambda_y} D^{xx} D^{yy} - 4(D^{xy})^2}\right]^{-1/2}.
\end{equation}
The stochastic rotation frequency is bounded above by the geometric mean of the relaxation rates, $\left |\omega_{\mathrm{stoch}} \right | \le \sqrt{\lambda_x \lambda_y}$, with equality attained if and only if $\mathbf D$ is singular ($D^{xx} D^{yy} = (D^{xy})^2$). Now, we turn to the case where $\mathbf A$ has a pair of complex conjugate eigenvalues. Without loss of generality, after a change of coordinates, we may write:
\begin{equation} \label{eq:A-cmplx-conj}
\mathbf A =
\begin{pmatrix}
-\lambda & -\omega \\
\omega & -\lambda
\end{pmatrix}
, \quad D^{xy} = 0,
\end{equation}
where $\lambda > 0$, $\omega > 0$. Here, $L = \omega \operatorname{tr}(\mathbf C)$, which gives
\begin{equation}
\omega_{\mathrm{stoch}} = \frac{\omega \operatorname{tr}(\mathbf C)}{2\sqrt{\det(\mathbf C)}}.
\end{equation}
Since $\mathbf C$ has real eigenvalues, $(\operatorname{tr}(\mathbf C))^2 - 4 \det(\mathbf C) \ge 0$ and therefore $\omega_{\mathrm{stoch}} \ge \omega$. Equality holds if and only if $\mathbf C$ is a multiple of the identity.

To continue the calculation, we solve the Lyapunov equation $\mathbf A \mathbf C + \mathbf C \mathbf A^{\mathsf T} = -2 \mathbf D$ for $\mathbf C$ and obtain:
\begin{equation}
\mathbf C = \frac 1 \lambda \begin{pmatrix}
D^{xx} - \delta/2 & \lambda \delta/{2 \omega} \\
\lambda \delta/{2 \omega} & D^{yy} + \delta/2
\end{pmatrix}, \quad \delta := \frac{D^{xx} - D^{yy}}{1 + \lambda^2/\omega^2}.
\end{equation}
This gives:
\begin{equation}
\omega_{\mathrm{stoch}} = \omega(D^{xx} + D^{yy}) \left[{4 D^{xx} D^{yy} + \frac{(D^{xx} - D^{yy})^2}{1 + \lambda^2/\omega^2}}\right]^{-1/2}.
\end{equation}
It can be seen from elementary algebra that $\omega_{\mathrm{stoch}} \le \sqrt{\lambda^2 + \omega^2}$, with equality holding if and only if $\mathbf D$ is singular ($D^{xx} D^{yy} = 0$).

\begin{figure}
	\begin{center}
		\includegraphics[scale=0.5]{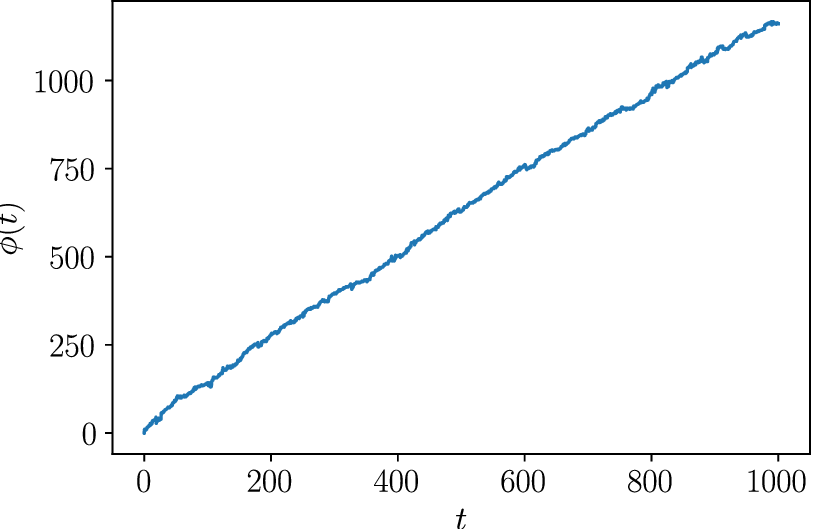}
	\end{center}
	\caption{Measured stochastic rotation frequency.}
	\label{fig:stoch-freq}
\end{figure}

As noted in \cite{Gladrow}, the stochastic rotation frequencies can be experimentally measured by averaging over angular motions. We can formally prove this by working in coordinates where $\mathbf C = \mathds 1$ (the eigenvalues of $\boldsymbol \Omega$ and the measured stochastic rotation frequencies are independent of the coordinate system chosen, up to a sign change) and use Stratonovich calculus to transform to polar coordinates $(r, \phi)$\footnote{If we use It\^o calculus, then $\dot \phi$ has a term $\propto r^{-2}$, whose expectation diverges because a 2D Gaussian has $p(r) \propto r$ as $r \to 0$, and it would appear that $\langle \dot \phi \rangle$ does not exist. However, using Stratonovich calculus, no such problem arises.}. We have:
\begin{equation}
	{} \circ \mathrm d \phi = \frac{\cos \phi \circ \mathrm d y - \sin \phi \circ \mathrm d x}{r} = \frac{x \circ \mathrm d y - y \circ \mathrm d x}{r^2}
\end{equation}
Because the phase-space velocity $\mathbf J(\mathbf x)/p(\mathbf x)$ is linear, $\langle {} \circ \dot \phi \mid r, \phi \rangle$ is independent of $r$ (see section 4 for a derivation). Also, by our choice of coordinates, the distribution of $\mathbf x$ is isotropic, which means that $r$ is independent of $\phi$. Thus $L = \langle r^2 \circ \dot \phi \rangle = \langle r^2 \rangle \langle \dot \phi \rangle = 2 \langle \dot \phi \rangle$, as desired. We have numerically verified this for a system obeying Eq.\ \eqref{eq:A-cmplx-conj} where $\lambda = \omega = 1$, $D^{xx} = 1$, $D^{yy} = 10$. We simulated this system using Euler\textendash Maruyama discretization with $\Delta t = 0.005$ for 1000 time-units. We see that the measured stochastic rotation frequency is in agreement with the theoretical value of 1.23 (Fig.\ \ref{fig:stoch-freq}).

\subsubsection{Quantitative significance of broken detailed balance} At this point, we still do not have a good dimensionless measure of when broken detailed balance is quantitatively significant. To address this, we turn to a modified definition of gain matrix \cite{Gnesotto, Weiss}:
\begin{equation}
\mathbf H := \mathds 1 + \mathbf A \mathbf C \mathbf D^{-1} = (\mathbf A \mathbf C - \mathbf C \mathbf A^{\mathsf T}) (2 \mathbf D)^{-1}.
\end{equation}
where the last equality follows from the Lyapunov equation. (We may understand the factor of 2 in the definition of $\mathbf H$ by considering that the symmetric part of $\mathbf A \mathbf C$ is $-\mathbf D$ by the Lyapunov equation, so the antisymmetric part should be compared to $\mathbf D$.) The entropy production rate $\dot S$ is related to $\mathbf H$ as $\dot S = -\operatorname{tr}(\mathbf A \mathbf H)$ \cite{Gnesotto}. By a similar argument as in \cite{Gnesotto}, in coordinates in which $\mathbf D = \mathds 1$, we see that $\mathbf H$ is antisymmetric and therefore, like $\boldsymbol \Omega$, has purely imaginary complex conjugate eigenvalues. As an alternative derivation, following \cite{GardinerStochasticMethods}, we may consider that $\mathbf D^{-1/2} \mathbf H \mathbf D^{1/2}$ is antisymmetric and therefore $\mathbf H$ is similar to an antisymmetric matrix. In two dimensions the (dimensionless) eigenvalues $\pm i h$ are given by a simple formula similar to that of $\omega_{\mathrm{stoch}}$:
\begin{equation}
h = \frac{L}{2 \sqrt{\det(\mathbf D)}}.
\end{equation}
If $h$ is at least comparable to unity (more precisely, $1/\sqrt{2}$; see end of subsubsection), then we consider that broken detailed balance is significant. We may illustrate this criterion in the cases of real or complex eigenvalues of $\mathbf A$. For two real eigenvalues $-\lambda_x$ and $-\lambda_y$ of $\mathbf A$ (Eq.\ \eqref{eq:A-real-eigvals}),
\begin{equation}
h = \frac{\lambda_x-\lambda_y}{\lambda_x+\lambda_y} \cdot \frac{D^{xy}}{\sqrt{D^{xx}D^{yy}-(D^{xy})^2}}.
\end{equation}
For broken detailed balance to be significant, one possibility is that $\lambda_x$ is far away from $\lambda_y$ and $D^{xy}/\sqrt{D^{xx}D^{yy}}$ is far from zero. The other possibility is that $D^{xy} \approx \sqrt{D^{xx}D^{yy}}$. The latter possibility allows for $\lambda_x \approx \lambda_y$, in which case $\mathbf C$ as well as $\mathbf D$ is almost singular in the chosen coordinates. In coordinates where $\mathbf C$ is well-conditioned, this corresponds to the case where $\mathbf A$ almost has a generalized eigenvector.

For the case of complex conjugate eigenvalues $-\lambda \pm i \omega$ of $\mathbf A$ with $D^{xy} = 0$ (Eq.\ \eqref{eq:A-cmplx-conj}),
\begin{equation}
h = \frac{\omega}{\lambda} \cdot \frac{D^{xx}+D^{yy}}{2\sqrt{D^{xx}D^{yy}}}.
\end{equation}
We see that $h \ge \omega/\lambda$. Again, there are two possibilities for significant broken detailed balance. One possibility is that $\omega/\lambda$ is significant compared to unity. The other possibility is $\sqrt{D^{xx}} \ll \sqrt{D^{yy}}$ (or vice versa), which allows for the possibility that $\omega \ll \lambda$. Similarly to before, this corresponds to the case where $\mathbf A$ almost has a generalized eigenvector.

For angular momenta of $d$-dimensional systems, we may follow a similar line of reasoning as in section 2.2 but with symmetry replaced by antisymmetry, and with $\mathbf C$ replaced by $2 \mathbf D$, which leads to the ``ensemble covariance'':
\begin{equation} \label{eq:L-mag}
	L^{ij} L^{i'j'} \sim 2 (D^{ii'} D^{jj'} - D^{ij'} D^{ji'}).
\end{equation}
For a collective comparison, we have:
\begin{equation}
	L^{ij} L^{i'j'} D^{-1}_{ii'} D^{-1}_{jj'} \sim 2d(d-1).
\end{equation}
Similarly to before, for high $d$ it is not suitable to compare the eigenvalues of $\mathbf H$ to unity. Furthermore, if the time-step is not small and angular momenta are measured using discrete time, then for the above comparisons, it may be suitable to use instead discrete-time estimators of the diffusion.

\subsection{Estimation of $\mathbf A$ from trajectories}
\begin{figure}
	\begin{center}
		\includegraphics[scale=0.5]{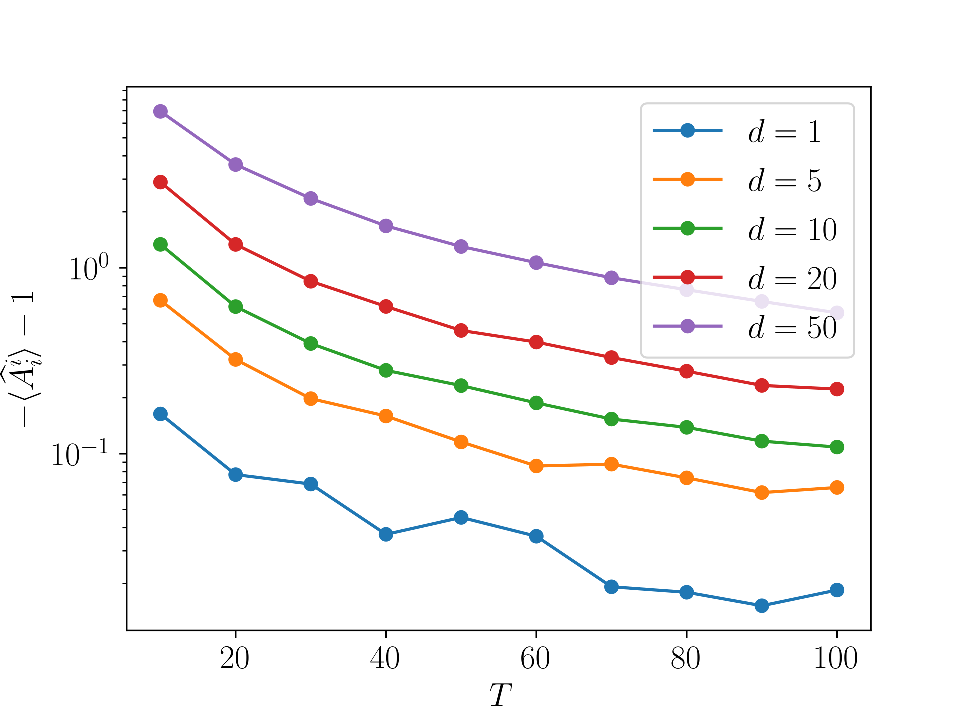}
	\end{center}
	\caption{Effect of dimensionality on mean of (inferred) $\widehat{A}^i_i$ ($i$ not summed) when $-\mathbf A = \mathbf D = \mathbf C = \mathds 1$.}
	\label{fig:mean-dimensionality}
\end{figure}

We now consider the estimation of $\mathbf A$ in multiple dimensions. It has been shown that for estimation of the force field from stochastic trajectories, there is a finite rate at which information can be extracted \cite{Frishman}. This work has shown that the mean coefficient of the inferred force field $\widehat{\mathbf A} \mathbf x$ is $\langle \widehat{\mathbf A} \rangle = \mathbf A$ with correction $\mathcal O(T^{-1})$. However, they do not discuss the dependence of $\langle \widehat{\mathbf A} \rangle$ on dimension $d$. Eq.\ (C11) in \cite{Frishman} is summed over basis function index $\alpha$, and we may guess Eq.\ (C13) in \cite{Frishman} to be of order $\mathcal O(d^3)$, leaving a $\mathcal O(d)$ bias for the inferred force field coefficient. To explore the impact of dimension, we explicitly calculate the bias and variance of $\widehat{\mathbf A}$ for a $d$-dimensional linear system in which $-\mathbf A = \mathbf D = \mathbf C = \mathds 1$. According to the information-theoretic criterion in \cite{Frishman}, the force field should start to be resolved for a trajectory length of 2 time-units. The value of $\widehat{\mathbf A}$ inferred from a trajectory of duration $T$ is\footnote{The results in this section hold for inference based on either It\^o or Stratonovich calculus in \cite{Frishman}. Also, we have neglected the inference of the true mean $\langle \mathbf x \rangle$ and have assumed this as known to be $\mathbf 0$. This simplifies the calculations while retaining interesting effects.}:
\begin{equation}
	\widehat{\mathbf A}(T) = \left({\int_0^T \mathrm d \tau \, \dot{\mathbf x}(\tau) \mathbf x(\tau)^{\mathsf T}}\right) \left({\int_0^T \mathrm d \tau' \, \mathbf x(\tau') \mathbf x(\tau')^{\mathsf T}}\right)^{-1}.
\end{equation}
We now define:
\begin{equation}
	\widehat{\mathbf C}(T) := \frac 1 T \int_0^T \mathrm d \tau \, \mathbf x(\tau) \mathbf x(\tau)^{\mathsf T}.
\end{equation}
Denoting $\Delta \widehat{\mathbf C}(T) := \widehat{\mathbf C}(T) - \mathds 1$, we expand:
\begin{equation}
	\widehat{\mathbf C}(T)^{-1} = (\mathds 1 + \Delta \widehat{\mathbf C}(T))^{-1} = \sum_{n=0}^{\infty} (-1)^n (\Delta \widehat{\mathbf C}(T))^n.
\end{equation}
The $n=0$ term contributes $-\mathds 1$ to $\langle \widehat{\mathbf A}(T) \rangle$. For the contribution of the $n=1$ term, we have from Gaussianity and Isserlis's theorem (also known as Wick's theorem \cite{Goldenfeld}):
\begin{equation}
	\begin{aligned}
		&\frac{1}{T^2} \int_0^T \mathrm d \tau \int_0^T \mathrm d \tau' \, \langle \dot x^i(\tau) x^j (\tau) (x^k(\tau') x^l(\tau') - \delta^{kl}) \rangle \\
		&\quad{} = \frac{1}{T^2} \int_0^T \mathrm d \tau \int_0^T \mathrm d \tau' \, ( \langle \dot x^i(\tau) x^k(\tau') \rangle \langle x^j(\tau) x^l(\tau') \rangle + \langle \dot x^i(\tau) x^l(\tau') \rangle \langle x^j(\tau) x^k(\tau') \rangle ).
	\end{aligned}
\end{equation}
We have:
\begin{align}
	\langle x^i(\tau) x^j(\tau') \rangle &= \delta^{ij} e^{-|\tau-\tau'|}, \\
	\langle \dot x^i(\tau) x^j(\tau') \rangle &= \delta^{ij} e^{-|\tau-\tau'|} \begin{cases}
		-1, \quad \tau' \le \tau, \\
		+1, \quad \tau' > \tau,
	\end{cases}
\end{align}
where $\delta$ is Kronecker delta. Thus the contribution of the $n=1$ term vanishes. For the contribution of the $n=2$ term, we have:
\begin{equation}
	\begin{aligned}
		&\left \langle {\frac 1 T \int_0^T \mathrm d \tau \, \dot x^i(\tau) x^j(\tau) (\Delta \widehat C(T))^{kl} (\Delta \widehat C(T))^{k'l'} } \right \rangle = \frac{1}{T^3} \int_0^T \mathrm d \tau \int_0^T \mathrm d \tau' \int_0^T \mathrm d \tau'' \\
		&\quad \Big[{\langle \dot x^i(\tau) x^j(\tau) \rangle (\langle x^k(\tau') x^{k'}(\tau'') \rangle \langle x^l(\tau') x^{l'}(\tau'') \rangle + \langle x^k(\tau') x^{l'}(\tau'') \rangle \langle x^l(\tau') x^{k'}(\tau'') \rangle)}\Big. \\
		&\qquad{} + \langle \dot x^i(\tau) x^k(\tau') \rangle (\langle x^j(\tau) x^{k'}(\tau'') \rangle \langle x^l(\tau') x^{l'}(\tau'') \rangle + \langle x^j(\tau) x^{l'}(\tau'') \rangle \langle x^l(\tau') x^{k'}(\tau'') \rangle) \\
		&\qquad{} + \langle \dot x^i(\tau) x^l(\tau') \rangle (\langle x^j(\tau) x^{k'}(\tau'') \rangle \langle x^k(\tau') x^{l'}(\tau'') \rangle + \langle x^j(\tau) x^{l'}(\tau'') \rangle \langle x^k(\tau') x^{k'}(\tau'') \rangle) \\
		&\qquad{} + \langle \dot x^i(\tau) x^{k'}(\tau'') \rangle (\langle x^j(\tau) x^k(\tau') \rangle \langle x^l(\tau') x^{l'}(\tau'') \rangle + \langle x^j(\tau) x^l(\tau') \rangle \langle x^{l''}(\tau'') x^k(\tau') \rangle) \\
		&\Big.{\qquad{} + \langle \dot x^i(\tau) x^{l'}(\tau'') \rangle (\langle x^j(\tau) x^k(\tau') \rangle \langle x^{k'}(\tau'') x^l(\tau') \rangle + \langle x^j(\tau) x^l(\tau') \rangle \langle x^k(\tau') x^{k'}(\tau'') \rangle)}\Big].
	\end{aligned}
\end{equation}
Only the integral of the first line does not vanish, as the rest of the lines are antisymmetric upon swapping $\tau$ and $\tau'$, or $\tau$ and $\tau''$, after (Kronecker) delta functions are factored out. It evaluates to:
\begin{equation}
	-\delta^{ij} (\delta^{kk'} \delta^{ll'} + \delta^{kl'} \delta^{lk'}) \frac{2T + e^{-2T} - 1}{2T^2}
\end{equation}
and therefore combining the terms accounted for thus far (the remaining terms give higher powers of $T^{-1}$):
\begin{equation}
	\langle \widehat A(T)^i_j \rangle \approx -\delta^i_j \left({1 + \frac{d+1}{T}}\right), \quad T \gg \frac 1 2.
\end{equation}
Thus in addition to $T \gg 2$, we also need $T \gg d+1$ in order to correctly resolve the force field. This is shown in Fig.\ \ref{fig:mean-dimensionality}. Statistics were obtained from 1000 trajectories with $\Delta t = 0.05$ for each point on the plots. We expect similar effects to occur when trying to infer a large number of coefficients for nonlinear drift. However, under the assumption of a linear Gaussian model, by symmetry $\mathbf x \to -\mathbf x$, the expectation of inferred coefficients of quadratic terms in $\langle \dot{\mathbf x} \mid \mathbf x \rangle$ vanishes.

\begin{figure}
	\begin{center}
		\includegraphics[scale=0.5]{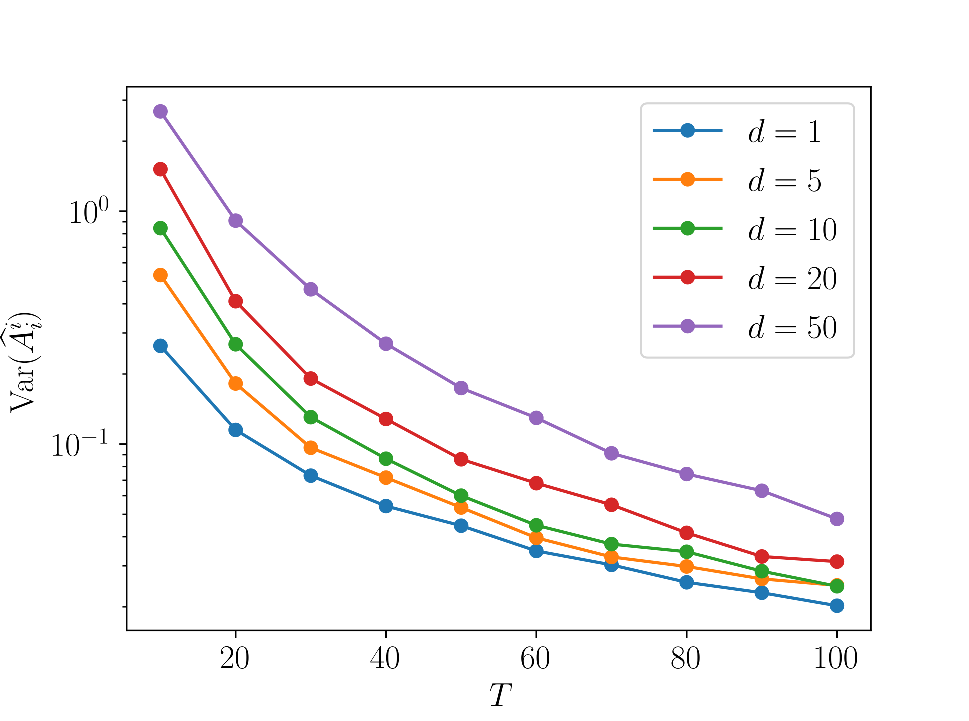}
	\end{center}
	\caption{Effect of dimensionality on variance of (inferred) $\widehat{A}^i_i$ ($i$ not summed) when $-\mathbf A = \mathbf D = \mathbf C = \mathds 1$.}
	\label{fig:var-dimensionality}
\end{figure}

For the second moment of $\widehat{\mathbf A}$, we need to evaluate the expectation:
\begin{equation}
	\left \langle{ \left({\frac 1 T \int_0^T \mathrm d \tau \, \dot x^i(\tau) x^j(\tau)}\right) \sum_{m=0}^{\infty} (-1)^m [(\Delta \widehat{\mathbf C}(T))^m]^{kl} \left({\frac 1 T \int_0^T \mathrm d \tau' \, \dot x^{i'}(\tau') x^{j'}(\tau')}\right) \sum_{n=0}^{\infty} (-1)^n [(\Delta \widehat{\mathbf C}(T))^n]^{k'l'} }\right \rangle.
\end{equation}
The $m = n = 0$ term is simply $\delta^{ij} \delta^{i'j'}$. For the $m + n = 1$ terms, we need to evaluate:
\begin{equation}
	\begin{aligned}
		&\frac{1}{T^3} \int_0^T \mathrm d \tau \int_0^T \mathrm d \tau' \int_0^T \mathrm d \upsilon \, \langle \dot x^i(\tau) x^j(\tau)  \dot x^{i'} (\tau') x^{j'}(\tau') (x^k(\upsilon) x^l(\upsilon) - \delta^{kl})\rangle = \frac{1}{T^3} \int_0^T \mathrm d \tau \int_0^T \mathrm d \tau' \int_0^T \mathrm d \upsilon \\
		&\quad \Big[{\langle \dot x^i(\tau) x^j(\tau) \rangle (\langle \dot x^{i'} (\tau') x^k(\upsilon) \rangle \langle x^{j'} (\tau') x^l(\upsilon) \rangle + \langle \dot x^{i'} (\tau') x^l(\upsilon) \rangle \langle x^{j'} (\tau') x^k(\upsilon) \rangle)}\Big. \\
		&\qquad{} + \langle \dot x^{i'}(\tau') x^{j'}(\tau') \rangle (\langle \dot x^i (\tau) x^k(\upsilon) \rangle \langle x^j (\tau) x^l(\upsilon) \rangle + \langle \dot x^i (\tau) x^l(\upsilon) \rangle \langle x^j (\tau) x^k(\upsilon) \rangle) \\
		&\qquad{} + \langle \dot x^i(\tau) \dot x^{i'}(\tau') \rangle (\langle x^j(\tau) x^k(\upsilon) \rangle \langle x^{j'}(\tau') x^l(\upsilon) \rangle + \langle x^j(\tau) x^l(\upsilon) \rangle \langle x^{j'}(\tau') x^k(\upsilon) \rangle) \\
		&\qquad{} + \langle \dot x^i(\tau) x^{j'}(\tau') \rangle (\langle x^j(\tau) x^k(\upsilon) \rangle \langle \dot x^{i'}(\tau') x^l(\upsilon) \rangle + \langle x^j(\tau) x^l(\upsilon) \rangle \langle \dot x^{i'}(\tau') x^k(\upsilon) \rangle) \\
		&\qquad{} + \langle \dot x^i(\tau) x^k(\upsilon) \rangle (\langle x^j(\tau) \dot x^{i'} (\tau') \rangle \langle x^{j'}(\tau') x^l(\upsilon) \rangle + \langle x^j(\tau) x^{j'} (\tau') \rangle \langle \dot x^{i'}(\tau') x^l(\upsilon) \rangle) \\
		&\Big.{\qquad{} + \langle \dot x^i(\tau) x^l(\upsilon) \rangle (\langle x^j(\tau) \dot x^{i'} (\tau') \rangle \langle x^{j'}(\tau') x^k(\upsilon) \rangle + \langle x^j(\tau) x^{j'} (\tau') \rangle \langle \dot x^{i'}(\tau') x^k(\upsilon) \rangle)}\Big].
	\end{aligned}
\end{equation}
The integrals of the first two lines vanish, and the remaining lines have integral $\mathcal O(T)$ (before dividing by $T^3$). For the $m + n = 2$ terms, we have:
\begin{equation}
	\begin{aligned}
		&\frac{1}{T^4} \int_0^T \mathrm d \tau \int_0^T \mathrm d \tau' \int_0^T \mathrm d \upsilon \int_0^T \mathrm d \upsilon' \, \langle \dot x^i(\tau) x^j(\tau) \dot x^{i'}(\tau') x^{j'}(\tau') (x^k(\upsilon) x^l(\upsilon) - \delta^{kl}) (x^{k'}(\upsilon') x^{l'}(\upsilon') - \delta^{k'l'}) \rangle \\
		&\quad{} = \frac{1}{T^4} \int_0^T \mathrm d \tau \int_0^T \mathrm d \tau' \int_0^T \mathrm d \upsilon \int_0^T \mathrm d \upsilon' \, \langle \dot x^i(\tau) x^j(\tau) \rangle \langle \dot x^{i'}(\tau') x^{j'}(\tau') \rangle \\
		&\qquad{} \times (\langle x^k(\upsilon) x^{k'}(\upsilon') \rangle \langle x^l(\upsilon) x^{l'}(\upsilon') \rangle + \langle x^k(\upsilon) x^{l'}(\upsilon') \rangle \langle x^l(\upsilon) x^{k'}(\upsilon') \rangle) + \mathcal O\left({\frac{1}{T^2}}\right) \\
		&\quad{} \approx \frac{\delta^{ij} \delta^{i'j'} (\delta^{kk'} \delta^{ll'} + \delta^{kl'} \delta^{lk'})}{T}, \quad T \gg \frac 1 2.
	\end{aligned}
\end{equation}
The remaining terms give higher powers of $T^{-1}$. This gives:
\begin{equation}
	\langle \widehat A(T)^i_j \widehat A(T)^{i'}_{j'} \rangle - \langle \widehat A(T)^i_j \rangle \langle \widehat A(T)^{i'}_{j'} \rangle \approx \frac{\delta^{ii'} \delta^{jj'} + \delta^{ij'} \delta^{ji'}}{T}.
\end{equation}
Perhaps surprisingly, there is no $d$-dependence to leading order in $T$ (Fig.\ \ref{fig:var-dimensionality}). This means that it is possible for the bias to dominate the variance in high dimensions. The $d$-dependence in the variance presumably occurs at higher order in $T^{-1}$, but we do not bother to calculate it.

\begin{figure}
	\begin{center}
		\includegraphics[scale=0.5]{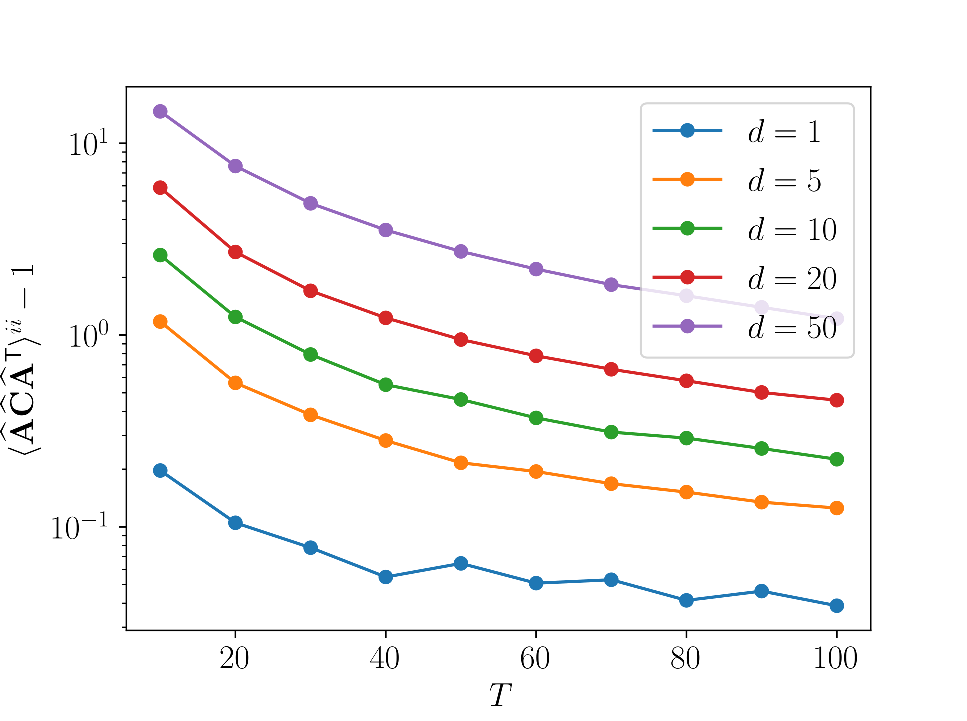}
	\end{center}
	\caption{Effect of dimensionality on variance of inferred drift ($\widehat{A}^i_k \widehat{A}^i_l \widehat{C}^{kl}$, $i$ not summed) when $-\mathbf A = \mathbf D = \mathbf C = \mathds 1$.}
	\label{fig:drift-sq-dimensionality}
\end{figure}

Finally, we can calculate the covariance of the estimated drift:
\begin{equation}
	\langle \widehat A^i_k \widehat A^j_l \widehat C^{kl} \rangle = \frac{1}{T^2} \left \langle \left({\int_0^T \mathrm d \tau \, \dot x^i(\tau) x^k(\tau) }\right) \left({\int_0^T \mathrm d \tau' \, \dot x^j(\tau') x^l(\tau')}\right) \sum_{n=0}^{\infty} [(\Delta \widehat {\mathbf C}(T))^n]^{kl} \right \rangle \approx \left({1 + \frac{2d}{T}}\right) \delta^{ij},
\end{equation}
which is an overestimate (Fig.\ \ref{fig:drift-sq-dimensionality}).

\section{Integrated variables}

\subsection{Preliminaries: Modeling, covariance functions, and detailed balance}
In the statistical literature, the term ``integrated variable'' in a stochastic process refers to a variable with no stationary distribution, but whose increments, or increments of increments, etc., do possess a stationary distribution. Here, we focus on variables integrated of order one, i.e., variables whose increments possess a stationary distribution. Such variables can be used to model e.g.\ the position and orientation of a particle in a homogeneous, isotropic medium. The dynamics of these variables may depend on degrees of freedom represented by variables possessing a stationary distribution\footnote{Our analysis does not address ratchet models where $\langle \dot x \mid x \rangle$ is periodic in $x$. In the systems we consider, any dependence on a variable not possessing a stationary distribution is disallowed.}. For linear dynamics, this type of system is modeled as:
\begin{align}
\dot{\mathbf x} &= \mathbf A \mathbf x + \boldsymbol \xi^{\mathbf x}, \\
\dot y &= \boldsymbol \alpha^{\mathsf T} \mathbf x + \xi^y, \label{eq:langevin-linear-int} \\
\langle \boldsymbol \xi^{\mathbf x}(t) \boldsymbol \xi^{\mathbf x}(t')^{\mathsf T} \rangle &= 2 \mathbf D^{\mathbf x \mathbf x} \delta(t-t'), \\
\langle \boldsymbol \xi^{\mathbf x}(t) \xi^y(t') \rangle &= 2 \mathbf D^{\mathbf x y} \delta(t-t'), \\
\langle \xi^y(t) \xi^y(t') \rangle &= 2 D^{yy} \delta(t-t').
\end{align}
\footnote{As with $\mathbf x$, we have omitted a possible additive constant in $\dot y$. A system with an additive constant may be transformed into Eq.\ \eqref{eq:langevin-linear-int} by considering instead the quantity $y(t) - ct$, where $c$ is a constant (in this case, $c = \langle \dot y \rangle$). This simplification is again used throughout. Note that such a constant must vanish when detailed balance is satisfied, since we must have in that case $\langle \dot y \rangle = 0$.} where $\boldsymbol \xi^{\mathbf x}$ and $\xi^y$ are zero-mean Gaussian white noise and all coefficients are constant. In this case, the angular momentum between $\mathbf x$ and $y$ is defined as:
\begin{equation} \label{eq:det-bal-int}
\mathbf L(\mathbf x, y) := \lim_{\tau \to 0^+} \frac {\langle (\mathbf x(0)+\mathbf x(\tau))(y(\tau)-y(0)) \rangle} {\tau}
\end{equation}
which, in this model, evaluates to:
\begin{equation}
\mathbf L(\mathbf x, y) = 2 \langle \mathbf x \dot y \rangle + \left\langle{\frac{\mathrm d [\mathbf x, y](t)}{\mathrm d t}}\right\rangle = 2 \mathbf C \boldsymbol \alpha + 2 \mathbf D^{\mathbf x y}.
\end{equation}
The condition for detailed balance is $\mathbf L(\mathbf x, y) = \mathbf 0$ and $L(x^i, x^j) = 0$ for all $i, j$, where we used the notation $L(x^i, x^j) := \langle x^i \dot x^j - \dot x^i x^j \rangle$.

To calculate covariance functions, it is useful to switch to a new variable:
\begin{equation} \label{eq:y-to-z}
z := y - \boldsymbol \alpha^{\mathsf T} \mathbf A^{-1} \mathbf x,
\end{equation}
so that $z$ obeys
\begin{equation}
\dot z = \xi^z, \quad \langle \boldsymbol \xi^{\mathbf x}(t) \xi^z(t') \rangle = 2 \mathbf D^{\mathbf x z} \delta(t-t'), \quad \langle \xi^z(t) \xi^z(t') \rangle = 2 D^{zz} \delta(t-t'),
\end{equation}
where $\xi^z$ is zero-mean Gaussian white noise, and the transformed diffusion coefficients are:
\begin{align}
\mathbf D^{\mathbf x z} &= \mathbf D^{\mathbf x y} - \mathbf D^{\mathbf x \mathbf x} (\mathbf A^{-1})^{\mathsf T} \boldsymbol \alpha, \\
D^{zz} &= D^{yy} - 2 \boldsymbol \alpha^{\mathsf T} \mathbf A^{-1} \mathbf D^{\mathbf x y} + \boldsymbol \alpha^{\mathsf T} \mathbf A^{-1} \mathbf D^{\mathbf x \mathbf x} (\mathbf A^{-1})^{\mathsf T} \boldsymbol \alpha.
\end{align}
If detailed balance is satisfied for $\mathbf x$ (i.e., $L(x^i, x^j) = 0$), then $\mathbf D^{\mathbf x \mathbf x} (\mathbf A^{-1})^{\mathsf T} = -\mathbf C$ and $\mathbf L(\mathbf x, z) = \mathbf L(\mathbf x, y)$. Moreover, the detailed balance condition for $z$ simply becomes $\mathbf D^{\mathbf x z} = \mathbf 0$.

There are two kinds of covariance functions that can be evaluated. One is the forward-difference covariance function, which for $z$ evaluates to zero:
\begin{equation}
\langle \mathbf x(0) (z(\tau)-z(0)) \rangle = \mathbf 0,
\end{equation}
where again $\tau \ge 0$ is (always) assumed. The backward-difference covariance function satisfies the equation:
\begin{equation}
\frac {\mathrm d} {\mathrm d \tau} \langle \mathbf x(\tau) (z(\tau)-z(0)) \rangle = \mathbf A \langle \mathbf x(\tau) (z(\tau)-z(0)) \rangle + 2 \mathbf D^{\mathbf x z},
\end{equation}
which has solution
\begin{equation}
\langle \mathbf x(\tau) (z(\tau)-z(0)) \rangle = 2 \mathbf A^{-1} \left({e^{\mathbf A \tau} - \mathds 1}\right) \mathbf D^{\mathbf x z}.
\end{equation}
We also have:
\begin{equation}
	\langle (z(\tau)-z(0))^2 \rangle = 2 D^{zz} \tau.
\end{equation}
Note that in the case where $\mathbf x$ is one-dimensional, the stochastic rotation frequency always vanishes because of a zero relaxation rate, but the entropy production rate does not vanish if detailed balance is broken.

To characterize the coupling between $\mathbf x$ and $y$, instead of $\boldsymbol \alpha$ and $\mathbf D^{\mathbf x y}$, we may prefer to use $\mathbf L(\mathbf x, y)$ and $\mathbf D^{\mathbf x y}$. Here, we consider $\mathbf L(y, \mathbf x) := -\mathbf L(\mathbf x, y)$ and use the same criterion as in the previous section to judge significance of angular momenta.

Between two integrated variables, there is no additional quantity governing break of detailed balance. Thus the angular momentum between them is considered to be 0, assuming that dynamics do not depend in any way on the values of those variables.

\subsection{Quantitative significance of the deterministic contribution}
The next question we wish to consider is that of quantitative significance. How do we judge whether or not the deterministic contribution $\boldsymbol \alpha^{\mathsf T} \mathbf x$ to $y$ is significant? And how do we compare whether a fitted model is a good match to experimental data, at least to linear order? This question has been addressed in the case of the usual stationary variables: compare the values of the experimentally measured covariance functions to the theoretical model prediction, and if the difference is small compared to $\mathbf C$, it is declared a good fit. However, we cannot proceed in an analogous way because the measured quantities are differences $y(\tau)-y(0)$ whose expectations scale as $\mathcal O(\tau^{1})$ whereas root-mean-square values $\langle (y(\tau)-y(0))^2 \rangle^{1/2}$ scale as $\mathcal O(\tau^{1/2})$. Hence, they are not comparable.

We specialize to the case where $\mathbf x$ is one-dimensional, i.e.:
\begin{align}
\dot x &= -\lambda x + \xi^x, \\
\dot y &= \alpha x + \xi^y,
\end{align}
where $\lambda>0$. The dimensionless combination involving $\alpha$ is $(\alpha/\lambda) \sqrt{D^{xx}/D^{yy}}$. We will explicitly arrive at this quantity by studying the function $\langle (y(\tau)-y(0))^2 \rangle$. We introduce according to Eq.\ \eqref{eq:y-to-z} the variable $z$, with transformed diffusion coefficients $D^{xz}$ and $D^{zz}$:
\begin{align}
z &:= y + \frac{\alpha}{\lambda} x, \\
D^{xz} &= D^{xy} + \frac{\alpha}{\lambda} D^{xx}, \\
D^{zz} &= D^{yy} + 2 \frac{\alpha}{\lambda} D^{xy} + \left({\frac{\alpha}{\lambda}}\right)^2 D^{xx}.
\end{align}
From the covariance function in the previous subsection, we obtain:
\begin{equation}
\langle (y(\tau)-y(0))^2 \rangle = 2D^{yy} \tau + 2\frac{\alpha}{\lambda} \left({ \frac{\alpha}{\lambda} D^{xx} + 2D^{xy} }\right) \left({\tau - \frac{1 - e^{-\lambda \tau}}{\lambda}}\right).
\end{equation}
We see that depending on $D^{xy}$, the value can be either increased or decreased relative to that due to the deterministic ($\alpha$) or stochastic ($D^{yy}$) contributions alone. Nevertheless, some bounds can be established by using $\left|{D^{xy}}\right| \le \sqrt{D^{xx}D^{yy}}$. We treat $\lambda$ and $D^{xx}$ as fixed. By minimizing with respect to $\alpha$ while keeping $D^{yy}$ fixed, we get:
\begin{equation}
\langle (y(\tau)-y(0))^2 \rangle \ge 2 D^{yy} \frac{1 - e^{-\lambda\tau}}{\lambda}.
\end{equation}
On the other hand, by minimizing with respect to $D^{yy}$ while keeping $\alpha$ fixed, we have
\begin{equation}
\langle (y(\tau)-y(0))^2 \rangle \ge 2 \left({\frac{\alpha}{\lambda}}\right)^2 D^{xx} \left({1 - \frac{1 - e^{-\lambda \tau}}{\lambda \tau}}\right) \frac{1 - e^{-\lambda \tau}}{\lambda}.
\end{equation}
Notice that both of these bounds stay finite as $\tau \to \infty$. The reason for this is that the $\tau \to \infty$ behavior of $\langle (y(\tau)-y(0))^2 \rangle$ is given by $2 D^{zz} \tau$ if $D^{zz} > 0$, and $D^{zz}$ can be made to be zero for any particular choice of either $\alpha$ and $D^{yy}$ (but not both simultaneously, because $D^{zz} \ge \left({\sqrt{D^{yy}} - |\alpha|\sqrt{D^{xx}}/\lambda}\right)^2$). Comparing these two bounds, we see that the relevant comparison is between $(\alpha/\lambda) \sqrt{D^{xx}}$ and $\sqrt{D^{yy}}$. Indeed, if $(|\alpha|/\lambda) \sqrt{D^{xx}} \ll \sqrt{D^{yy}}$, then the contribution of $D^{yy}$ to $\langle (y(\tau)-y(0))^2 \rangle$ is dominant. On the other hand, if $(|\alpha|/\lambda) \sqrt{D^{xx}} \gg \sqrt{D^{yy}}$, then for long times $\tau \gg 1/\lambda$, the contribution of $\alpha$ to $\langle (y(\tau)-y(0))^2 \rangle$ is dominant.

For multidimensional $\mathbf x$, we may compare the square roots of the quadratic variation of the two terms on the r.h.s.\ of Eq.\ \eqref{eq:y-to-z}, i.e., $\sqrt{D^{yy}}$ with $\sqrt{\boldsymbol{\alpha}^{\mathsf T} \mathbf A^{-1} \mathbf D^{\mathbf{xx}} (\mathbf A^{-1})^{\mathsf T} \boldsymbol{\alpha}}$.

\subsection{Comparison of experimental data with theoretical predictions}
To compare experimentally measured covariance functions to theoretical predictions, we can again normalize by the diffusion matrix. We may consider coordinates in which $\mathbf D = \mathds 1$ and compare the experimental and theoretical values of $\langle \dot{\mathbf x}(\tau)\mathbf x(0)^{\mathsf T} \rangle$. In terms of the ``ensemble covariance'', this amounts to:
\begin{equation}
	\Delta_{\textrm{exp\textendash theo}} \langle \dot x^i(\tau) x^j(0) \rangle \Delta_{\textrm{exp\textendash theo}} \langle \dot x^{i'}(\tau) x^{j'}(0) \rangle \sim D^{ii'} D^{jj'},
\end{equation}
where $\Delta_{\textrm{exp\textendash theo}}$ denotes the deviation of experimental to theoretical values. As mentioned for the angular momentum, if the time-step is not small, then on the r.h.s.\ of the above, it may be suitable to use discrete-time estimators of the diffusion. In case the angular momenta exceed the ``reference'' values (Eq.\ \eqref{eq:L-mag}) by some factor, we may need to relax the above comparison by the same factor, not just for the antisymmetric part but also for the symmetric part. We may illustrate this by way of two-dimensional linear Gaussian dynamics with complex conjugate eigenvalues $-\lambda \pm i \omega$ of $\mathbf A$ (Eq.\ \eqref{eq:A-cmplx-conj}) and $D^{xx} = D^{yy}$, in which case:
\begin{equation}
	(\langle \dot{\mathbf x}(\tau) \mathbf x(0)^{\mathsf T} \rangle + \langle \mathbf x(0) \dot{\mathbf x}(\tau)^{\mathsf T} \rangle)(2 \mathbf D)^{-1} = e^{-\lambda \tau} \left({-\cos(\omega \tau) - \frac{\omega}{\lambda} \sin(\omega \tau)}\right) \mathds 1.
\end{equation}
Thus if $\omega/\lambda \gtrsim 1$, deviation of experimental to theoretical values could be amplified by this factor without necessarily being considered quantitatively significant. Similarly to the comparison of experimental values to theoretical predictions of $\langle \mathbf x(\tau) \mathbf x(0)^{\mathsf T} \rangle$, we may compare separately the symmetric and antisymmetric components (with respect to indices), which correspond to time-symmetric and -antisymmetric quantities, respectively.

Now, we need to extend the notion of $\langle \dot{\mathbf x}(\tau)\mathbf x(0)^{\mathsf T} \rangle$ to integrated variables. We easily make the identification $\text{``$\langle \dot{\mathbf x}(\tau) y(0) \rangle$''} = -\langle \mathbf x(\tau) \dot y(0) \rangle$ for $\tau > 0$. (For $\tau = 0$, the time-derivative $\dot y(0)$ is interpreted in the anti-It\^o sense.) For a pair of integrated variables $y$ and $z$ (not the same $z$ as in Eq.\ \eqref{eq:y-to-z}), we manipulate the terms in the following equation, similarly to Eq.\ \eqref{eq:Lyapunov-deriv-simple}:
\begin{equation}
2D^{yz} = \lim_{\tau \to 0^+} \frac{\langle (y(\tau)-y(0))(z(\tau)-z(0)) \rangle}{\tau} = \text{``$\langle -\dot y z - y \dot z \rangle$''}.
\end{equation}
Since there is no angular momentum between $y$ and $z$, we assign \text{``$\langle \dot y z \rangle$''} and \text{``$\langle y \dot z \rangle$''} the same (fictional) value, $-D^{yz}$. For non-zero time lags, we integrate:
\begin{equation} \label{eq:dotintint}
\text{``$\langle \dot y(\tau) z(0) \rangle$''} = \text{``$\langle \dot y(0) z(0) \rangle$''} - \int_{0^+}^{\tau} \mathrm d t \, \langle \dot y(t) \dot z(0) \rangle.
\end{equation}
The above is equal to $-\langle (z(\tau)-z(0)) \circ \dot y(\tau) \rangle$. From the velocity covariance function $\langle \dot y(t) \dot z(0) \rangle$ (where $t$ is allowed to run negative), we can compute the expected value of any second-order quantity, such as the two-variable analogue of the mean squared displacement:
\begin{equation} \label{eq:msd}
	\langle (y(\tau)-y(0)) (z(\tau)-z(0)) \rangle = \int_{-\tau}^{+\tau} \mathrm d t \, (\tau - |t|) \langle \dot y(t) \dot z(0) \rangle.
\end{equation}
Note that $\langle \dot y(t) \dot z(0) \rangle$ has a contribution $2 D^{yz} \delta(t)$ at $t=0$, where $\delta(\cdot)$ is the Dirac delta function. The derivative of Eq.\ \eqref{eq:msd} is:
\begin{equation} \label{eq:msd-deriv}
	\frac{\mathrm d}{\mathrm d \tau} \langle (y(\tau)-y(0)) (z(\tau)-z(0)) \rangle = \int_{-\tau}^{+\tau} \mathrm d t \, \langle \dot y(t) \dot z(0) \rangle.
\end{equation}
The $\tau \to \infty$ limit of Eq.\ \eqref{eq:msd-deriv} is known as one-half the long-time diffusivity. Thus, for a single (integrated) variable, Eq.\ \eqref{eq:dotintint} gives as $\tau \to \infty$ the negative of the long-time diffusivity. In the proposed scheme, deviations between theory and experiment of the long-time diffusivity would be compared to the short-time diffusivity.

It should be noted, however, that the $\tau \to \infty$ limit of Eq.\ \eqref{eq:msd-deriv} cannot be taken for a trajectory of finite length. To illustrate this, suppose we have an integrated variable $x$ sampled with time-step $\Delta t$, and define $\Delta x_n := x((n+1)\Delta t)-x(n\Delta t)$ for integer $n$, and $\Delta x_n' := \Delta x_n - \langle \Delta x \rangle$ (the subscript $n$ in the expectation value is suppressed). We then define the modified mean squared displacement:
\begin{equation}
	\text{MSD}'(N \Delta t) := \langle (x(N \Delta t) - x(0) - \langle x(N \Delta t) - x(0) \rangle)^2 \rangle = (\Delta t)^2 \sum_{-N<k<N} (N-|k|) C(k)
\end{equation}
where the prime denotes that we are subtracting the mean $\langle x(N \Delta t) - x(0) \rangle$ before squaring, and $C(k) := \langle \Delta x_{n+k}' \Delta x_n' \rangle/(\Delta t)^2$ is the (discrete) velocity autocovariance function. We then have:
\begin{equation} \label{eq:msd-diff}
	\frac{\text{MSD}'(N\Delta t) - \text{MSD}'((N-1)\Delta t)}{\Delta t} = (\Delta t) \sum_{-N < k < N} C(k).
\end{equation}
The long-time diffusivity is one-half the $N \to \infty$ limit of Eq.\ \eqref{eq:msd-diff}. We now want to estimate this from a trajectory of length $N+1$. The velocity autocovariance function may be estimated as \cite{PriestleyTimeSeries}:
\begin{equation} \label{eq:vacf}
	\widehat{C}(k) := \frac{1}{N (\Delta t)^2} \sum_{\substack{0 \le n < N \\ 0 \le n+k < N }} (\Delta x_{n+k} - \overline{\Delta x}) (\Delta x_n - \overline{\Delta x}),
\end{equation}
where
\begin{equation}
	\overline{\Delta x} := \frac{1}{N} \sum_{0 \le n < N} \Delta x_n.
\end{equation}
However, substituting the estimate Eq.\ \eqref{eq:vacf} into Eq.\ \eqref{eq:msd-diff} gives:
\begin{equation} \label{eq:zero-diffusion}
	(\Delta t) \sum_{-N<k<N} \widehat{C}(k) = \frac{1}{N \Delta t} \sum_{0 \le m < N} \sum_{0 \le n < N} (\Delta x_m - \overline{\Delta x}) (\Delta x_n - \overline{\Delta x}) = 0,
\end{equation}
where we re-indexed $m = n+k$. Thus, although the estimate Eq.\ \eqref{eq:vacf} for the velocity autocovariance function is self-averaging for any given finite time lag, its sum over all time lags is not\footnote{In the case where $\langle \Delta x \rangle$ is treated as known and can be used in place of $\overline{\Delta x}$ in Eq.\ \eqref{eq:zero-diffusion}, the resulting estimate of long-time diffusivity is $(N/2\Delta t) ( \overline{\Delta x} - \langle \Delta x \rangle )^2$, which has the correct expectation value, but is not self-averaging.}. If we denote the limits of summation by $\pm M$, then the limits $M \to \infty$, $N \to \infty$ do not commute. The same considerations apply when integrating the covariance functions of stationary variables.

\section{Third-order properties}

So far, we have dealt with linear Gaussian systems. These are characterized by the quantities $\mathbf A$ and $\mathbf D$ which give rise to the second-order moments $\langle \mathbf x \mathbf x^{\mathsf T} \rangle$ and $\langle \dot{\mathbf x} \mathbf x^{\mathsf T} \rangle$. In this section, we will establish general properties of Langevin equations beyond second order. To address the possibility of third-order effects, we may consider additional terms in the model:
\begin{align}
\langle \dot x^i(t) \mid \mathbf x(t) \rangle &= A^i_j x^j + a^i_{jk} (x^j x^k - C^{jk}) \label{eq:third-order-model-a} \\
\left \langle \frac {\mathrm d [x^i, x^j](t)} {\mathrm d t} \mid \mathbf x(t) \right \rangle &= 2 D^{ij} + 2 b^{ij}_k x^k \label{eq:third-order-model-b}
\end{align}
where all coefficients are constant, we have suppressed $t$-dependence on the r.h.s., and Einstein summation notation is used. Neither addition to the model can stand on its own: in general, $a^i_{jk}$ by itself leads to divergences of $\mathbf x \to \infty$, and $b^{ij}_k$ by itself renders $\langle \mathrm d [x^i, x^j]/\mathrm d t \mid \mathbf x(t) \rangle$ not positive semidefinite. In section 6, we will discuss the magnitude of additional terms needed to result in an admissible model. A third possible contribution,
\begin{equation} \label{eq:cubic-var}
\lim_{\tau \to 0^+} \frac{\langle (x^i(\tau)-x^i(0))(x^j(\tau)-x^j(0))(x^k(\tau)-x^k(0)) \rangle}{\tau} = 0,
\end{equation}
vanishes in the case where $\mathbf x$ has continuous paths\footnote{The $x^i$, $x^j$, and $x^k$ need not be state coordinates; they can be any functions.}. In such a situation, the probability density evolution satisfies a Fokker\textendash Planck equation \cite{GardinerStochasticMethods}, leading to vanishing of the ``cubic variation'' (Eq. \eqref{eq:cubic-var}) \cite{VanKampenStochasticProcesses}. This gives a relation between the angular momenta, as:
\begin{equation} \label{eq:LLL0}
L(x^i x^j, x^k) + L(x^j x^k, x^i) + L(x^k x^i, x^j) = 0.
\end{equation}
This may also be understood by taking the time-derivative of (the time-independent quantity) $\langle x^i x^j x^k \rangle$ and interpreting in the Stratonovich sense.

The angular momenta are related to probability currents. We start with the Fokker\textendash Planck equation for the probability density $p(\mathbf x, t)$ corresponding to the It\^o\textendash Langevin equation Eq.\ \eqref{eq:langevin-general}:
\begin{equation} \label{eq:FPeqn}
\frac{\partial}{\partial t} p(\mathbf x, t) = -\frac{\partial}{\partial x^i} [\mathcal A^i(\mathbf x) p(\mathbf x, t)] + \frac 1 2 \frac{\partial^2}{\partial x^i \partial x^j} [\mathcal B^{ij}(\mathbf x) p(\mathbf x, t)],
\end{equation}
for which the probability current density $\mathbf J(\mathbf x)$ reads \cite{GardinerStochasticMethods}:
\begin{equation} \label{eq:J}
J^i(\mathbf x) = \mathcal A^i(\mathbf x) p(\mathbf x) - \frac 1 2 \partial_j [\mathcal B^{ij}(\mathbf x) p(\mathbf x)],
\end{equation}
where $p(\mathbf x)$ is the stationary distribution and $\partial_j := \partial/\partial x^j$. Multiplying by an arbitrary function $f(\mathbf x)$ and integrating, we get:
\begin{equation} \label{eq:J-f}
	\begin{aligned}
		\int \mathrm d \mathbf x \, J^i(\mathbf x) f(\mathbf x) &= \langle \mathcal A^i(\mathbf x) f(\mathbf x) \rangle + \frac 1 2 \langle \mathcal B^{ij}(\mathbf x) \partial_j f(\mathbf x) \rangle \\
		&= \langle f(\mathbf x) \dot x^i \rangle + \frac 1 2 \left \langle \frac {\mathrm d [f(\mathbf x), x^i](t)}{\mathrm d t} \right \rangle \\
		&= \langle f(\mathbf x) \circ \dot x^i \rangle \\
		&= \frac 1 2 L(f(\mathbf x), x^i)
	\end{aligned}
\end{equation}
\footnote{This assumes that $\mathcal B^{ij}(\mathbf x) p(\mathbf x) f(\mathbf x) \to 0$ as $\mathbf x \to \infty$. It is true when $f(\cdot)$ has bounded support, e.g.\ if we take $f(\mathbf x) = \delta (\mathbf x - \mathbf x_0)$, then we get $\mathbf J(\mathbf x_0) = p(\mathbf x_0) \langle {} \circ \dot{\mathbf x} \mid \mathbf x = \mathbf x_0 \rangle$, as expected. However, in certain cases we may have $\mathcal B^{ij}(\mathbf x) p(\mathbf x) \to $ constant as $\mathbf x \to \infty$ and thus Eq.\ \eqref{eq:J-f} does not hold for polynomial $f(\mathbf x)$. See Appendix B for an example.}. Thus $L(x^i x^j, x^k)$ gives complete information about probability currents to ``third order'' (counting two orders for quadratic $f(\mathbf x)$, and one order for $x^i$).

The relation Eq.\ \eqref{eq:J-f} can also be used to prove Eq.\ \eqref{eq:LLL0} by using the vanishing of the divergence of the probability current:
\begin{equation}
	\begin{aligned}
		0 &= -\int \mathrm d \mathbf x \, x^i x^j x^k \frac{\partial J^l(\mathbf x)}{\partial x^l} \\
		&= \int \mathrm d \mathbf x \, J^l (\mathbf x) \frac{\partial}{\partial x^l} (x^i x^j x^k) \\
		&= \int \mathrm d \mathbf x \, \left({x^i x^j J^k (\mathbf x) + x^j x^k J^i (\mathbf x) + x^k x^i J^j (\mathbf x)}\right).
	\end{aligned}
\end{equation}

We can define other quantities that are odd under time reversal:
\begin{equation}
\widetilde{L}(x^i,x^j,x^k) := \lim_{\tau \to 0^+} \frac{\langle x^i(\tau/2)(x^j(0)x^k(\tau)-x^j(\tau)x^k(0))\rangle}{\tau}.
\end{equation}
We can use an It\^o\textendash Taylor expansion \cite{Bruckner} to compute this as:
\begin{equation} \label{eq:L-tilde}
\widetilde{L}(x^i,x^j,x^k) = \langle x^i (x^j \mathcal A^k(\mathbf x) - \mathcal A^j(\mathbf x) x^k) \rangle + \frac{\langle \mathcal B^{ik}(\mathbf x) x^j - \mathcal B^{ij}(\mathbf x) x^k \rangle}{2} = \frac{L(x^i x^j, x^k) - L(x^i x^k, x^j)}{2}.
\end{equation}
We can invert this relationship to give $L$ in terms of $\widetilde{L}$:
\begin{equation}
L(x^i x^j, x^k) = \frac{2}{3} \left[{\widetilde{L}(x^i,x^j,x^k) + \widetilde{L}(x^j,x^i,x^k)}\right].
\end{equation}
Thus $\widetilde{L}$ also gives complete information about time-reversal asymmetry to third order. This will become useful when we consider underdamped processes. Like $L$, it satisfies the relation:
\begin{equation} \label{eq:Ltilde-sum-0}
	\widetilde{L}(x^i, x^j, x^k) + \widetilde{L}(x^j, x^k, x^i) + \widetilde{L}(x^k, x^i, x^j) = 0.
\end{equation}

Now, we address the case of integrated variables. The angular momentum involving $y$ is again defined by Eq.\ \eqref{eq:det-bal-int} and the requirement for detailed balance is $L(x^i x^j, y) = 0$. We can compute:
\begin{equation}
L(x^i x^j, y) = 2 \langle x^i x^j \mathcal A^y(\mathbf x) \rangle + \langle x^i \mathcal B^{jy}(\mathbf x) \rangle + \langle x^j \mathcal B^{iy}(\mathbf x) \rangle.
\end{equation}
It can be shown that:
\begin{equation} \label{eq:L-alt}
\lim_{\tau \to 0^+} \frac{\langle (x^i(0)x^j(\tau)+x^i(\tau)x^j(0))(y(\tau)-y(0)) \rangle}{\tau} = \lim_{\tau \to 0^+} \frac{\langle x^i(\tau/2)(x^j(0)+x^j(\tau))(y(\tau)-y(0)) \rangle}{\tau} = L(x^i x^j, y).
\end{equation}
Thus $L(x^i x^j, y)$ are the only third-order quantities odd under time reversal involving an integrated variable. For inhomogeneous diffusion, we have from Eq.\ \eqref{eq:cubic-var}:
\begin{equation}
	\lim_{\tau \to 0^+} \frac{\langle x^i(\tau) (x^j(\tau)-x^j(0))(x^k(\tau)-x^k(0)) \rangle}{\tau} = \lim_{\tau \to 0^+} \frac{\langle x^i(0) (x^j(\tau)-x^j(0))(x^k(\tau)-x^k(0)) \rangle}{\tau}.
\end{equation}
Thus in the expression $\langle x^i \mathrm d[x^j, x^k]/\mathrm d t \rangle$, the time at which $x^i$ is evaluated has no impact.

\section{Inhomogeneous diffusion}

In this section, we shall restrict ourselves to the case where the drift $\bm{\mathcal A}(\mathbf x)$ is a linear function $\mathbf A \mathbf x$. In addition to being easier to deal with, in the case of weak nonlinearity, coordinates can be chosen via Koopman eigenfunctions \cite{MezicKoopman, WilliamsKoopman} to transform the system into this form, as will be done in the next section. Additional terms in inhomogeneous diffusion are assumed to be small. In this section, we will discuss the effect of inhomogeneous diffusion on third-order covariance functions.

\subsection{Stationary case} Our first task is the compute the third moments $\langle x^i x^j x^k \rangle$. In analogy with the derivation of the Lyapunov equation, we have:
\begin{equation}
0 = \frac{\mathrm d}{\mathrm d t} \langle x^i x^j x^k \rangle = \langle (\mathbf A \mathbf x)^i x^j x^k \rangle + \langle x^i (\mathbf A \mathbf x)^j x^k \rangle + \langle x^i x^j (\mathbf A \mathbf x)^k \rangle + 2 b^{ij}_l C^{kl} + 2 b^{ik}_l C^{jl} + 2 b^{jk}_l C^{il},
\end{equation}
which has solution
\begin{equation}
\langle x^i x^j x^k \rangle = -\left[{(\mathbf A \otimes \mathds 1 \otimes \mathds 1 + \mathds 1 \otimes \mathbf A \otimes \mathds 1 + \mathds 1 \otimes \mathds 1 \otimes \mathbf A)^{-1}}\right]^{ijk}_{i'j'k'} \left({2 b^{i'j'}_l C^{k'l} + 2 b^{i'k'}_l C^{j'l} + 2 b^{j'k'}_l C^{i'l}}\right).
\end{equation}

The second task is to compute covariance functions. In the current model, $\langle \mathbf x(\tau) \mid \mathbf x(0) \rangle = e^{\mathbf A \tau} \mathbf x(0)$ as derived before, so by the law of iterated expectations:
\begin{equation}
\langle x^i(\tau) x^j(0) x^k(0) \rangle = (e^{\mathbf A \tau})^i_{i'} \langle x^{i'} x^j x^k \rangle.
\end{equation}
Now, the function $\langle x^i(\tau) x^j(\tau) \mid \mathbf x(0) \rangle$ satisfies the equation:
\begin{equation}
	\begin{aligned}
		\frac {\mathrm d}{\mathrm d \tau} \langle x^i(\tau) x^j(\tau) \mid \mathbf x(0) \rangle &= \langle (\mathbf A \mathbf x)^i(\tau) x^j(\tau) \mid \mathbf x(0) \rangle + \langle x^i(\tau) (\mathbf A \mathbf x)^j(\tau) \mid \mathbf x(0) \rangle \\
		&\quad{} + 2 D^{ij} + 2 b^{ij}_k (e^{\mathbf A \tau})^k_{k'} x^{k'}(0)
	\end{aligned}
\end{equation}
which has solution
\begin{equation} \label{eq:third-order-b-solution}
\begin{aligned}
&\langle x^i(\tau) x^j(\tau) \mid \mathbf x(0) \rangle = C^{ij} + (e^{\mathbf A \tau})^i_{i'} (e^{\mathbf A \tau})^j_{j'} \left({x^{i'}(0) x^{j'}(0) - C^{i'j'}}\right) \\
&\qquad{} + 2 \left[{(\mathbf A \otimes \mathds 1 \otimes \mathds 1 + \mathds 1 \otimes \mathbf A \otimes \mathds 1 - \mathds 1 \otimes \mathds 1 \otimes \mathbf A)^{-1} (e^{\mathbf A \tau} \otimes e^{\mathbf A \tau} \otimes \mathds 1 - \mathds 1 \otimes \mathds 1 \otimes e^{\mathbf A \tau})}\right]^{ijk}_{i'j'k'} b^{i'j'}_k x^{k'}(0).
\end{aligned}
\end{equation}
The covariance function $\langle x^i(\tau) x^j(\tau) x^k(0) \rangle$ is then easily obtained.

The last task is to understand time-reversal asymmetry. We can easily compute the angular momenta:
\begin{equation}
L(x^i x^j, x^k) = \langle x^i x^j (\mathbf A \mathbf x)^k \rangle - \langle (\mathbf A \mathbf x)^i x^j x^k \rangle - \langle x^i (\mathbf A \mathbf x)^j x^k \rangle - 2 b^{ij}_l C^{kl}.
\end{equation}
The question is, if this quantity is 0 for all $i,j,k$, are the covariance functions time-reversible? We will see that we need the lower-order time-reversibility to hold, i.e., $\mathbf A \mathbf C = \mathbf C \mathbf A^{\mathsf T}$. Calling $\widetilde{\mathbf A}(\tau)$ the quantity in square brackets in Eq.\ \eqref{eq:third-order-b-solution}, we have the covariance function:
\begin{equation}
\langle x^i(\tau) x^j(\tau) x^k(0) \rangle = (e^{\mathbf A \tau})^i_{i'} (e^{\mathbf A \tau})^j_{j'} \langle x^{i'} x^{j'} x^k \rangle + 2 \widetilde A(\tau)^{ijl}_{i'j'k'} b^{i'j'}_{l} C^{kk'}.
\end{equation}
From $\mathbf A \mathbf C = \mathbf C \mathbf A^{\mathsf T}$, we have:
\begin{equation}
\widetilde A(\tau)^{ijl}_{i'j'k'} C^{kk'} = \widetilde A(\tau)^{ijk}_{i'j'k'} C^{k'l}.
\end{equation}
Now in $\langle x^i(0) x^j(0) x^k(\tau) \rangle - \langle x^i(\tau) x^j(\tau) x^k(0) \rangle$ we can pull out a factor of $e^{\mathbf A \tau} \otimes e^{\mathbf A \tau} \otimes \mathds 1 - \mathds 1 \otimes \mathds 1 \otimes e^{\mathbf A \tau}$ and thus the condition for this quantity to be 0 for all $\tau$ and all $i,j,k$ reduces to $L(x^i x^j, x^k) = 0$.

So far, we have understood how the quantities $b$ give rise to the quantities $\langle x^i x^j x^k \rangle$ and $L(x^i x^j, x^k)$. The remaining degree of freedom, $\langle x^i x^j \dot x^k \rangle$ is accounted for by nonlinearities in the drift, considered in the next section.

\subsection{Two-dimensional example} Now we turn to an example in two dimensions, with variables $x$ and $y$. Assume that there is a symmetry $y \to -y$, so that $\mathbf A = \operatorname{diag}(-\lambda_x, -\lambda_y)$ with $D^{xy} = 0$ and $b^{xx}_y = b^{xy}_x = b^{yy}_y = 0$. Then $\langle x^2 \rangle = D^{xx}/\lambda_x$, $\langle y^2 \rangle = D^{yy}/\lambda_y$, and
\begin{align}
\langle x^3 \rangle &= \frac {2 b^{xx}_x \langle x^2 \rangle} {\lambda_x}, \\
L(x^2, x) &= 0, \\
\langle xy^2 \rangle &= \frac{2 b^{yy}_x \langle x^2 \rangle + 4 b^{xy}_y \langle y^2 \rangle}{\lambda_x+2\lambda_y}, \\
L(x, y^2) &= \frac{4\lambda_x b^{yy}_x \langle x^2 \rangle + 4(\lambda_x-2\lambda_y) b^{xy}_y \langle y^2 \rangle}{\lambda_x+2\lambda_y}, \\
L(xy,y) &= \frac{L(x,y^2)}{2}.
\end{align}
We see that there is only one ``mode of time-reversal asymmetry'', corresponding to $L(x,y^2)$. An angular momentum $L(x,y^2)>0$ represents probability current which is anticlockwise for $y>0$ and clockwise for $y<0$. We see that $b^{yy}_x$ and $b^{xy}_y$ contribute positively to $\langle x y^2 \rangle$, and that $b^{yy}_x$ contributes positively to $L(x,y^2)$. However, the sign of the contribution of $b^{xy}_y$ to $L(x,y^2)$ depends on the ratio $\lambda_x/\lambda_y$.

The question now arises: if $L(x,y^2)=0$, is time-reversal symmetry truly satisfied? To answer this question, we investigate Eq.\ \eqref{eq:J-f} where $i = x$ and $f(x,y) = y^4$. First, we write for the moment:
\begin{equation}
0 = \frac {\mathrm d} {\mathrm d t} \langle x(t)y(t)^4 \rangle = (-\lambda_x - 4 \lambda_y) \langle x y^4 \rangle + 8 b^{xy}_y \langle y^4 \rangle + 12 b^{yy}_x \langle x^2 y^2 \rangle.
\end{equation}
Now treating the added term in the model as a perturbation, the fourth moments obey:
\begin{align}
\langle y^4 \rangle &= 3 \langle y^2 \rangle + \mathcal O(b^2) \\
\langle x^2 y^2 \rangle &= \langle x^2 \rangle \langle y^2 \rangle + \mathcal O(b^2).
\end{align}
Now we write the condition for detailed balance for this choice of $f$:
\begin{equation}
-\lambda_x \langle xy^4 \rangle = -4 b^{xy}_y \langle y^4 \rangle.
\end{equation}
We see that the above is not generally satisfied, even to order $\mathcal O(b)$, when $L(x, y^2) = 0$. Thus $L(x, y^2) = 0$ is not a sufficient condition for detailed balance in the model; higher-order terms in $\mathbf x$ need to be included in the dynamics in order to truly satisfy detailed balance. However, this effect is absent from the third-order covariance functions.

\subsection{Integrated variables} Now, we address the case of integrated variables. We again assume a linear drift function. Following the previous section, we can make a transformation so that $y$ obeys the law $\langle \dot y \mid \mathbf x \rangle = 0$. We now use the model:
\begin{align}
\left \langle \frac{\mathrm d [x^i, y](t)}{\mathrm d t} \mid \mathbf x(t) \right \rangle &= 2 D^{iy} + 2 b^{iy}_j x^j, \\
\left \langle \frac{\mathrm d [y, y](t)}{\mathrm d t} \mid \mathbf x(t) \right \rangle &= 2 D^{yy} + 2 b^{yy}_i x^i,
\end{align}
where all coefficients are constant. Then:
\begin{equation}
L(x^i x^j, y) = 2 b^{iy}_k C^{jk} + 2 b^{jy}_k C^{ik}.
\end{equation}
When stationary variables are expressed in ``covariance-identity'' coordinates, $b^{iy}_j$ and $b^{jy}_i$ both contribute positively to $L(x^i x^j, y)$.

We can solve for the covariance functions in the same manner as before. Therefore, we focus on the condition for detailed balance. Similarly to before, we have:
\begin{equation}
\langle x^i(0) x^j(0) (y(\tau)-y(0)) \rangle = 0.
\end{equation}
The following covariance function satisfies the equation:
\begin{equation}
\begin{aligned}
\frac{\mathrm d}{\mathrm d \tau} \langle x^i(\tau) x^j(\tau) (y(\tau)-y(0)) \rangle &= \langle (\mathbf A \mathbf x)^i(\tau) x^j(\tau) (y(\tau)-y(0)) \rangle + \langle x^i(\tau) (\mathbf A \mathbf x)^j(\tau) (y(\tau)-y(0)) \rangle \\
&\qquad{} + 2 b^{ij}_k \langle x^k(\tau) (y(\tau)-y(0)) \rangle + 2 b^{iy}_k C^{jk} + 2 b^{jy}_k C^{ik}.
\end{aligned}
\end{equation}
We note that if $D^{ky} = 0$ for $k$ stationary, which is a condition for detailed balance, then $\langle x^k(\tau) (y(\tau)-y(0)) \rangle = 0$. If in addition $2 b^{iy}_k C^{jk} + 2 b^{jy}_k C^{ik} = 0$, then $\langle x^i(\tau) x^j(\tau) (y(\tau)-y(0)) \rangle = 0$.

The other covariance function is of the form:
\begin{equation}
\langle x^i(\tau) x^j(0) (y(\tau)-y(0)) \rangle = 2 \left[{\lim_{\nu \to 1} (\nu \mathbf A \otimes \mathds 1 - \mathds 1 \otimes \mathbf A)^{-1} (e^{\nu \mathbf A \tau} \otimes \mathds 1 - \mathds 1 \otimes e^{\mathbf A \tau})}\right]^{ik}_{i'l} b^{i'y}_k C^{jl}.
\end{equation}
\footnote{The limit is needed because $\mathbf A \otimes \mathds 1 - \mathds 1 \otimes \mathbf A$ is not invertible. To see this, let $\mathbf v$ be an eigenvector of $\mathbf A$. Then $(\mathbf A \otimes \mathds 1 - \mathds 1 \otimes \mathbf A)(\mathbf v \otimes \mathbf v) = \mathbf 0$.} If $\mathbf A \mathbf C = \mathbf C \mathbf A^{\mathsf T}$, then calling $\widetilde{\mathbf A}(\tau)$ the quantity in square brackets above:
\begin{equation}
\widetilde{A}(\tau)^{ik}_{i'l} C^{jl} = \widetilde{A}(\tau)^{ij}_{i'l} C^{kl}.
\end{equation}
If in addition $b^{iy}_k C^{jk} = -b^{jy}_k C^{ik}$, then we may write:
\begin{equation}
\langle x^i(\tau) x^j(0) (y(\tau)-y(0)) \rangle = -2 \widetilde{A}(\tau)^{ij}_{i'l} b^{ly}_k C^{i'k} = -2 \widetilde{A}(\tau)^{kj}_{i'l} b^{ly}_k C^{ii'} = -\langle x^i(0) x^j(\tau) (y(\tau)-y(0)) \rangle
\end{equation}
where we used $\widetilde{A}(\tau)^{ij}_{kl} = \widetilde{A}(\tau)^{ji}_{lk}$. Thus we see that under assumption of detailed balance to second and third order, the investigated third-order covariance functions are time-reversible.

At this point, we introduce a second integrated variable $z$ obeying $\langle \dot z \mid \mathbf x \rangle = 0$ as well as analogous laws as for $y$ with corresponding notation. The term $b^{yz}_i$ does not give rise to breaking detailed balance. However, detailed balance can still be broken to lower order in $\mathbf x, y, z$ and as before, this will be reflected in the three-point covariance functions. Specifically, we can compute:
\begin{equation}
\langle x^i(0) (y(\tau)-y(0))(z(\tau)-z(0)) \rangle = 2 \left[{\mathbf A^{-1} (e^{\mathbf A \tau} - \mathds 1)}\right]^j_k b^{yz}_j C^{ik}.
\end{equation}
We can write down the differential equation for the covariance function in the other direction:
\begin{equation}
\begin{aligned}
\frac{\mathrm d}{\mathrm d \tau} \langle x^i(\tau) (y(\tau)-y(0))(z(\tau)-z(0)) \rangle &= \langle (\mathbf A \mathbf x)^i (y(\tau)-y(0))(z(\tau)-z(0)) \rangle + 2 b^{iy}_j \langle x^j(\tau)(z(\tau)-z(0)) \rangle \\
&\qquad{} + 2 b^{iz}_j \langle x^j(\tau) (y(\tau)-y(0)) \rangle + 2 b^{yz}_j C^{ij}.
\end{aligned}
\end{equation}
We will again not solve this equation in full, but will simply assume $D^{iy} = D^{iz} = 0$ for all stationary $i$, corresponding to detailed balance. Under this condition, the covariance function becomes:
\begin{equation}
\langle x^i(\tau)(y(\tau)-y(0))(z(\tau)-z(0)) \rangle = 2 \left[{\mathbf A^{-1} (e^{\mathbf A \tau} - \mathds 1)}\right]^i_k b^{yz}_j C^{jk}.
\end{equation}
If also $\mathbf A \mathbf C = \mathbf C \mathbf A^{\mathsf T}$, it is seen that the two covariance functions are equal.

\section{Nonlinear drift}

\subsection{Preliminaries}
Processes with nonlinear drift suffer from the fact that the determination of lower-order moments depends on higher-order moments. We first address the case in one dimension. We set the relaxation time and the diffusion coefficient to 1. Then to a first approximation, $\langle x^2 \rangle = 1$. As mentioned before, a quadratic function for $\langle \dot x \mid x \rangle$ gives rise to diverging $x$. Thus a cubic contribution is needed for stability. However, if this cubic contribution is too small, $x$ will no longer be localized around 0. We consider an ``extreme'' case where $\langle \dot x \mid x \rangle$ has a double zero. Such a case is attained by the function
\begin{equation} \label{eq:a-extreme}
\langle \dot x \mid x \rangle = -x \left({\frac{ax}{2} - 1}\right)^2 = -x + a x^2 - \frac{a^2}{4} x^3,
\end{equation}
showing that we only need the coefficient of $x^3$ to be $\mathcal O(a^2)$ for the model to make sense, a result which could have been anticipated by symmetry. Thus, for small $a$, we can interpret the quadratic model for $\langle \dot x \mid x \rangle$ irrespective of any higher-order contributions. We also assume the linear model for inhomogeneous diffusion, noting similarly that in order for the diffusion function to be non-negative we can add a term $(b^2/4) x^2$, which is $\mathcal O(b^2)$. Thus, to order $\mathcal O((a,b)^2)$\footnote{We write $\mathcal O((a,b)^2)$ to stand for $\mathcal O(a^2, ab, b^2)$.}, our model is:
\begin{align}
\langle \dot x \mid x \rangle &= -x + a (x^2-1), \\
\left \langle \frac{\mathrm d [x, x](t)}{\mathrm d t} \mid x \right \rangle &= 2(1 + bx).
\end{align}
The multidimensional version is given by Eqs.\ \eqref{eq:third-order-model-a}\textendash \eqref{eq:third-order-model-b}. In sections 6.2\textendash 6.6, we will present the solution of this model, and in sections 6.7\textendash 6.10, we will discuss inference, quantitative significance, and comparison between theory and experiment.

\subsection{One-dimensional case}
Our first task is to compute the moments. We do this by introducing the probabilist's Hermite polynomials ${He}_n(x)$ and calculating their time-derivatives using It\^o's lemma:
\begin{equation} \label{eq:Hen-dynamics}
\begin{aligned}
\left \langle \frac{\mathrm d}{\mathrm d t} {He}_n(x) \mid x \right \rangle &= -n {He}_n(x) + na \left[{{He}_{n+1}(x) + 2(n-1) {He}_{n-1}(x) + (n-1)(n-2) {He}_{n-3}(x)}\right] \\
&\qquad{} + n(n-1)b \left[{{He}_{n-1}(x) + (n-2) {He}_{n-3}(x)}\right].
\end{aligned}
\end{equation}
This gives rise to a system of equations for $\langle {He}_n(x) \rangle$, as:
\begin{equation}
\begin{pmatrix}
1 & & & & & \\
& -1 & a & & & \\
& 2a+b & -1 & a & & \\
2(a+b) & & 2(2a+b) & -1 & a & \\
& 6(a+b) & & 3(2a+b) & -1 & \ddots \\
& & \ddots & & \ddots & \ddots
\end{pmatrix} \begin{pmatrix}
\langle {He}_0(x) \rangle \\
\langle {He}_1(x) \rangle \\
\langle {He}_2(x) \rangle \\
\langle {He}_3(x) \rangle \\
\langle {He}_4(x) \rangle \\
\vdots
\end{pmatrix} = \begin{pmatrix}
1 \\
0 \\
0 \\
0 \\
0 \\
\vdots
\end{pmatrix}.
\end{equation}
We solve this by truncation, which gives
\begin{align}
\langle x \rangle &= a^2 \mathcal O(a,b), \\
\langle {He}_2(x) \rangle &= a \mathcal O(a,b), \\
\langle {He}_3(x) \rangle &= 2(a+b) + a \mathcal O((a,b)^2), \\
|\langle {He}_n(x) \rangle| &\le \mathcal O((a,b)^2), \quad n \ge 4.
\end{align}

This is similar to an asymptotic expansion in that the above behavior holds as $a, b \to 0$ for any finite maximum degree $N$, but the solutions do not converge for fixed $a \ne 0$ as $N \to \infty$, as expected since trajectories obeying the Langevin equation diverge. Notice that the actual values of $\langle x \rangle$ and $\langle x^2 \rangle$ differ from the ``nominal'' values of 0 and 1, respectively, used to construct the equation for the dynamics. We may understand this by rescaling $x$ and $t$ and attempting to set the mean and variance of $x$ to $\mu$ and $\nu$ respectively, obtaining:
\begin{align}
	\langle \dot x \mid x \rangle &= A \left\{{x - \mu + \frac{a}{\sqrt{D/A}} \left[{(x - \mu)^2 - \nu}\right] }\right\}, \\
	\left \langle \frac{\mathrm d[x,x]}{\mathrm d t} \mid x \right \rangle &= 2D \left[{1 + b(x-\mu)}\right].
\end{align}
Apparently, there are 6 parameters ($A$, $D$, $a$, $b$, $\mu$, $\nu$); however, there are actually only 5 (quadratic drift has only 3 coefficients). Thus, if $\mu$ and $\nu>0$ are the actual mean and variance of $x$, then given these together with $D>0$ and $b$, the possible values of the pair $(A, a)$ must lie along a one-dimensional manifold. Thus in Eq.\ \eqref{eq:third-order-model-a}, $\mathbf C$ should be interpreted as a ``nominal'' value not necessarily corresponding to the actual covariance matrix. In particular, we will take it to be the value obeying the Lyapunov equation for the linear part of the dynamics (Eq.\ \eqref{eq:covariance-matrix}).

Now we introduce the (stochastic) Koopman operator $\mathcal K$ \cite{MezicKoopman, WilliamsKoopman}. It is defined as an operator on functions of state space $f(\mathbf x)$, as follows:
\begin{equation}
(\mathcal K f) (\mathbf x) = \left \langle \frac{\mathrm d f(\mathbf x(t))}{\mathrm d t} \mid \mathbf x(t) = \mathbf x \right \rangle.
\end{equation}
Although the dynamics of $\mathbf x$ may have arbitrary nonlinearities, the Koopman operator is a linear operator. The significance of the eigenfunctions of the Koopman operator (hereafter called ``Koopman eigenfunctions'') is that their dynamics follows a linear law in expectation. (Techniques and results for estimation of the Koopman operator are presented in \cite{LearningDynamicalSystemsKoopman, SharpSpectralRatesKoopman}.) The Koopman operator $\mathcal K$ has the same eigenvalues as the Perron\textendash Frobenius operator $\mathcal P$ appearing in the Fokker\textendash Planck equation ($\partial p(\mathbf x, t)/\partial t = \mathcal P p(\mathbf x, t)$, where $\mathcal P$ operates on $\mathbf x$), because $\mathcal K$ is the adjoint of $\mathcal P$ \cite{WilliamsKoopman} with respect to the inner product defined by:
\begin{equation}
	\langle f, g \rangle := \int \mathrm d \mathbf x \, f(\mathbf x)^* g(\mathbf x)
\end{equation}
(asterisk denoting the complex conjugate), which can be easily shown using It\^o's lemma (with suitable assumptions on $f, g$). When $a = b = 0$, the Koopman eigenfunctions are the probabilist's Hermite polynomials ${He}_n(x)$ with eigenvalues $-n$. In the present case, the Koopman eigenfunctions $f_n(x)$ and the corresponding eigenvalues $\lambda_n$ are:
\begin{align}
f_1(x) &= a^2 \mathcal O(a,b) {He}_0(x) + {He}_1(x) + (a + a^2 \mathcal O(a,b)) {He}_2(x) \\ \nonumber
&\qquad{} + \mathcal O(a^2) {He}_3(x) + \mathcal O(a^3) {He}_4(x) + \cdots, \\
\lambda_1 &= -1 + a \mathcal O(a,b); \\
f_2(x) &= a \mathcal O(a,b) {He}_0(x) + (-2(2a+b) + a^2 \mathcal O(a,b)) {He}_1(x) + {He}_2(x) \\ \nonumber
&\qquad{} + (2a + a^2 \mathcal O(a,b)) {He}_3(x) + \mathcal O(a^2) {He}_4(x) + \cdots, \\
\lambda_2 &= -2 + a \mathcal O(a,b).
\end{align}

Now to calculate covariance functions, we use the linearity property of eigenfunctions:
\begin{equation}
\langle f_n(x(\tau)) g(x(0)) \rangle = e^{\lambda_n \tau} \langle f_n(x(0)) g(x(0)) \rangle.
\end{equation}
First, we apply this to $n = 1$ and $g(x) = x$, leading to:
\begin{equation}
\langle (x(\tau) + a{He}_2(x(\tau))) x(0) \rangle = e^{-\tau} \langle (x(0) + a{He}_2(x(0))) x(0) \rangle + a \mathcal O(a,b).
\end{equation}
Now, just like the moments, $\langle {He}_2(x(\tau))x(0) \rangle = \mathcal O(a,b)$\footnote{This is easily seen by taking $a, b \to 0$ in which case $\langle {He}_2(x(\tau))x(0) \rangle \to e^{-2 \tau} \langle {He}_2(x) x \rangle = 0$. This quantity is odd in $x$, leading to the estimate $\mathcal O(a,b)$. An even quantity such as $\langle {He}_3(x(\tau))x(0) \rangle \to e^{-3\tau} \langle {He}_3(x) x \rangle = 0$ would be $\mathcal O((a,b)^2)$, by orthogonality of the Hermite polynomials.} and so
\begin{equation}
\langle x(\tau) x(0) \rangle = e^{-\tau} + a \mathcal O(a,b),
\end{equation}
where the quadratic correction could have been anticipated by symmetry. Thus the second-order covariance functions can be expected to be a poor proxy for nonlinearity.

Next, we would like the covariance function with $g(x) = x^2$. However, we need to use Isserlis's theorem to calculate $\langle (x(\tau)^2-1)x(0)^2 \rangle$ to zeroth-order in $a,b$. In fact, by symmetry, this has corrections of order only $\mathcal O((a,b)^2)$, and we could show this explicitly by calculating $\langle f_2(x(\tau)) f_2(x(0)) \rangle$ (which we will not do). The result is:
\begin{equation} \label{eq:tau-0-0}
\langle x(\tau) x(0)^2 \rangle = 2(a+b)e^{-\tau} + 2a(e^{-\tau} - e^{-2\tau}) + a \mathcal O((a,b)^2).
\end{equation}
Addressing the other covariance function $\langle x(\tau)^2 x(0) \rangle$, in one dimension this is necessarily identical because the probability current vanishes \cite{VanKampenStochasticProcesses} implying time-reversal invariance \cite{GardinerStochasticMethods}, but we explicitly calculate it here for illustration. The eigenfunction $f_2(x(\tau))$ involves ${He}_3(x(\tau))$, but when multiplying by $x(0)$ and taking the expectation, it vanishes to zeroth order. Thus Isserlis's theorem holds again to zeroth order. The result is:
\begin{equation} \label{eq:tau-tau-0}
\langle x(\tau)^2 x(0) \rangle = 2(a+b)e^{-2\tau} - 2(2a+b) (e^{-2\tau} - e^{-\tau}) + a \mathcal O((a,b)^2),
\end{equation}
which is indeed identical to Eq.\ \eqref{eq:tau-0-0}.

\begin{figure}
	\begin{center}
		\includegraphics[scale=0.5]{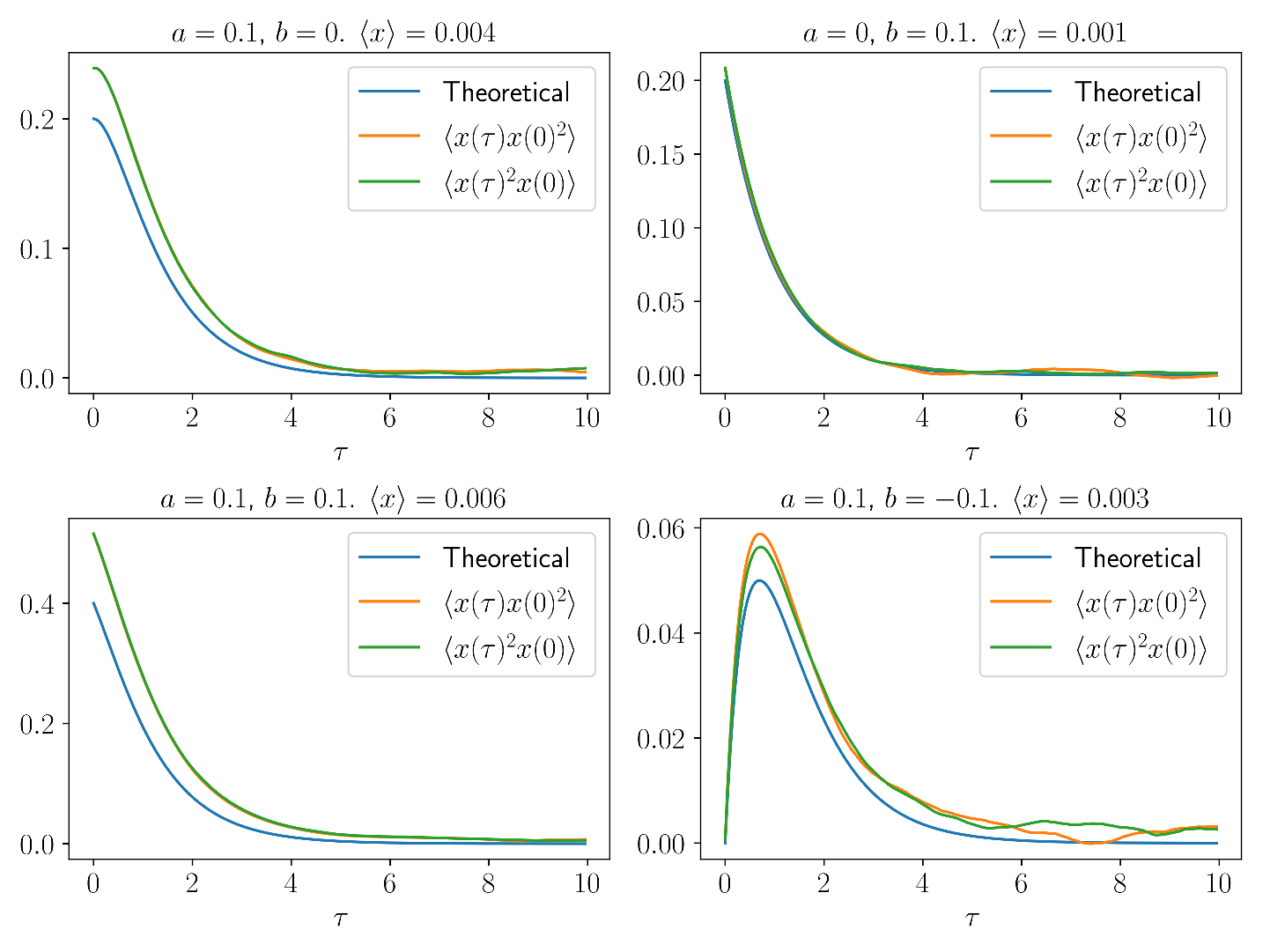}
	\end{center}
	\caption{Third-order covariance functions.}
	\label{fig:cov3}
\end{figure}

\begin{figure}
	\begin{center}
		\includegraphics[scale=0.5]{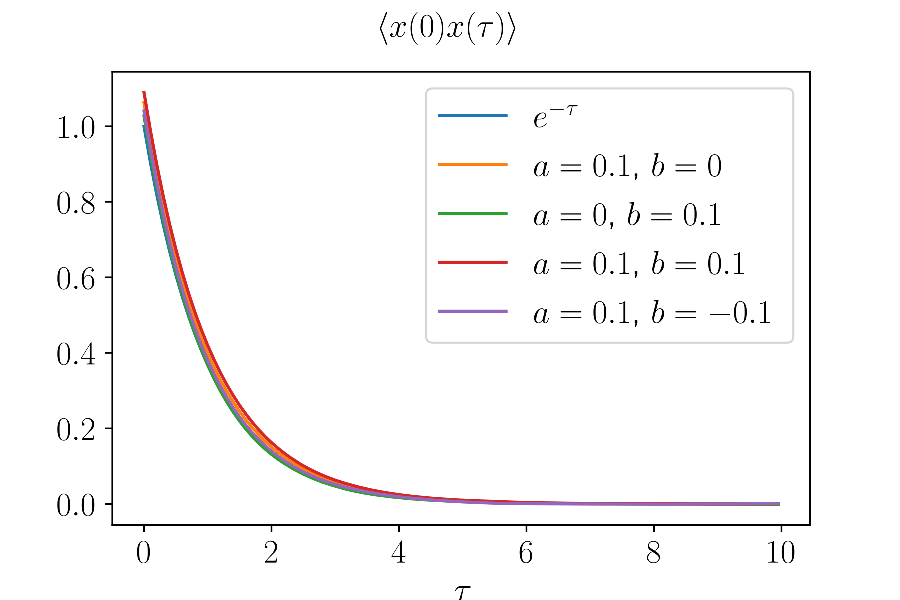}
	\end{center}
	\caption{Second-order covariance functions.}
	\label{fig:cov}
\end{figure}

We now present some numerical simulations to confirm the above results. We simulate a one-dimensional stochastic process where $\langle \dot x \mid x \rangle$ follows Eq.\ \eqref{eq:a-extreme} with $x \to x+a$ and inhomogeneous diffusion $(1+bx/2)^2$. We use an Euler\textendash Maruyama discretization with time-step $\Delta t = 0.05$ and trajectory length $10^6$ time-units. Although corrections to the third-order covariance functions are only cubic order in $a,b$, they turn out to be quite large, as shown in Fig.\ \ref{fig:cov3} (the exception being when $a=0$). The second-order covariance function is also affected, although less so (Fig.\ \ref{fig:cov}), but the deviation is still larger than what one might expect for a quadratic correction. However, the deviations are limited to approximate multiplication by a constant, and thus nonlinearity cannot be effectively discerned from the second-order covariance functions, as expected. For larger values of $|a|$ ($\sim$ 0.3), the covariance functions are nowhere near the theoretically predicted values (not shown).

\subsection{Multidimensional case}
Now we consider the multidimensional case, given by Eqs.\ \eqref{eq:third-order-model-a}\textendash \eqref{eq:third-order-model-b}. Without loss of generality, we assume $a^i_{jk} = a^i_{kj}$. We could expand functions in terms of multidimensional Hermite polynomials, defined as:
\begin{equation}
x^i, \quad x^i x^j - C^{ij}, \quad x^i x^j x^k - C^{ij} x^k - C^{ik} x^j - C^{jk} x^i, \quad \ldots.
\end{equation}
We could use these to calculate multidimensional generalizations of the Koopman eigenfunctions satisfying $\langle \mathrm d \mathbf f(\mathbf x(t))/\mathrm d t \mid \mathbf x(t) \rangle = \mathbf A \mathbf f(\mathbf x(t))$ to transform into the case of the previous section ($a = 0$). However, we opt for a more direct approach here and calculate covariance functions by means of differential equations, as in the previous sections. We must first calculate the third moments to order $\mathcal O(a,b)$:
\begin{equation}
\begin{aligned}
\langle x^i x^j x^k \rangle &= -2 \left[{(\mathbf A \otimes \mathds 1 \otimes \mathds 1 + \mathds 1 \otimes \mathbf A \otimes \mathds 1 + \mathds 1 \otimes \mathds 1 \otimes \mathbf A)^{-1}}\right]^{ijk}_{i'j'k'} \\
&\qquad{} \times \left({a^{i'}_{lm} C^{j'l} C^{k'm} + a^{j'}_{lm} C^{i'l} C^{k'm} + a^{k'}_{lm} C^{i'l} C^{j'm} + b^{i'j'}_l C^{k'l} + b^{i'k'}_l C^{j'l} + b^{j'k'}_l C^{i'l}}\right) \\
&\quad{} + a \mathcal O((a,b)^2).
\end{aligned}
\end{equation}
For the covariance functions, we solve a differential equation, replacing fourth-order moments using Isserlis's theorem, with $\mathcal O((a,b)^2)$ corrections. Without going through the details of the calculations, we use again the quantity in square brackets in Eq.\ \eqref{eq:third-order-b-solution}, which we call $\widetilde{\mathbf A}(\tau)$, and the result is:
\begin{align}
\langle x^i(0) x^j(0) x^k(\tau) \rangle &= (e^{\mathbf A \tau})^k_{k'} \langle x^i x^j x^{k'} \rangle + 2 \widetilde{A}(\tau)^{i'j'k}_{lmk'} a^{k'}_{i'j'} C^{il} C^{jm} + a \mathcal O((a,b)^2), \label{eq:third-order-1} \\
\langle x^i(\tau) x^j(\tau) x^k(0) \rangle &= (e^{\mathbf A \tau})^i_{i'} (e^{\mathbf A \tau})^j_{j'} \langle x^{i'}x^{j'}x^k \rangle + 2 \widetilde{A}(\tau)^{ijk'}_{i'j'l} \left({b^{i'j'}_{k'} + a^{i'}_{k'm} C^{j'm} + a^{j'}_{k'm} C^{i'm}}\right) C^{kl} \nonumber \\
&\quad{} + a \mathcal O((a,b)^2). \label{eq:third-order-2}
\end{align}
As expected, we see that as in the case where $a = 0$, the third-order covariance functions are symmetric (with $a \mathcal O((a,b)^2)$ corrections) if $L(x^i, x^j) = 0$ and $L(x^i x^j, x^k) = 0$ for all $i,j,k$.

At this point, we note that by a similar calculation, if there are cubic terms in the drift or quadratic terms in the diffusivity, then the corrections to the second- and third-order covariance functions are quadratic in these coefficients (multiplying multidimensional Hermite polynomials). Beyond this, the calculations may not work as expected (see Appendix B).

We can expand the stationary probability distribution into a sum of terms, each of which is the product of a Hermite polynomial with a Gaussian function \cite{RiskenFokkerPlanck}:
\begin{equation}
	\begin{aligned}
		p(\mathbf x) &= \frac{1}{\sqrt{\det (2 \pi \mathbf C)}} \exp \left({-\frac{1}{2} \mathbf x^{\mathsf T} \mathbf C^{-1} \mathbf x}\right) \Big[{1 + p^{(1)}_i x^i + p^{(2)}_{ij} (x^i x^j - C^{ij})}\Big. \\
		&\Big.{\qquad{} + p^{(3)}_{ijk} (x^i x^j x^k - C^{ij} x^k - C^{ik} x^j - C^{jk} x^i) + \cdots}\Big],
	\end{aligned}
\end{equation}
where $p^{(n)}$ is symmetric. From the moments, we have:
\begin{align}
	p^{(1)} &= a^2 \mathcal O(a,b), \\
	p^{(2)} &= a \mathcal O(a,b), \\
	p^{(3)}_{ijk} &= \frac 1 6 C^{-1}_{ii'} C^{-1}_{jj'} C^{-1}_{kk'} \langle x^{i'} x^{j'} x^{k'} \rangle, \\
	|p^{(n)}| &\le \mathcal O((a,b)^2), \quad n \ge 4.
\end{align}
We can calculate the probability current density $\mathbf J(\mathbf x)$ from Eq.\ \eqref{eq:J}.

One may characterize the third-order dynamics by the $a$ and $b$ coefficients, or alternatively by the combination of the third-order moments $\langle x^i x^j x^k \rangle$, the angular momenta $L(x^i x^j, x^k)$, and the $b^{ij}_k$ coefficients.

\subsection{Generalized Koopman eigenfunctions}
In case the time-step is not small compared to dynamics, and we want to convert between continuous-time and discrete-time quantities, it is useful to calculate the multidimensional generalization of Koopman eigenfunctions. To order $\mathcal O(a,b)$, they are:
\begin{align}
	f_1^i(\mathbf x) &= x^i + \alpha^i_{jk} (x^j x^k - C^{jk}), \\
	\left \langle \frac{\mathrm d}{\mathrm d t} f_1^i(\mathbf x) \mid \mathbf x \right \rangle &= A^i_j f_1^j(\mathbf x) + a \mathcal O(a,b),
\end{align}
where
\begin{equation}
	\alpha^i_{jk} := \left[{(\mathbf A \otimes \mathds 1 \otimes \mathds 1 - \mathds 1 \otimes \mathbf A \otimes \mathds 1 - \mathds 1 \otimes \mathds 1 \otimes \mathbf A)^{-1}}\right]^{ij'k'}_{i'jk} a^{i'}_{j'k'},
\end{equation}
and
\begin{align}
	f_2^{ij}(\mathbf x) &= \beta^{ij}_k x^k + (x^i x^j - C^{ij}) + \gamma^{ij}_{klm} (x^k x^l x^m - C^{kl} x^m - C^{km} x^l - C^{lm} x^k), \\
	\left \langle \frac{\mathrm d}{\mathrm d t} f_2^{ij}(\mathbf x) \mid \mathbf x \right \rangle &= (A^i_k \delta^j_l + \delta^i_k A^j_l) f_2^{kl}(\mathbf x) + a \mathcal O(a,b),
\end{align}
where
\begin{align}
	\beta^{ij}_k &:= 2 \left[{(\mathbf A \otimes \mathds 1 \otimes \mathds 1 + \mathds 1 \otimes \mathbf A \otimes \mathds 1 - \mathds 1 \otimes \mathds 1 \otimes \mathbf A)^{-1}}\right]^{ijk'}_{i'j'k} \left({b^{i'j'}_{k'} + a^{i'}_{k'l} C^{j'l} + a^{j'}_{k'l} C^{i'l}}\right), \\
	\gamma^{ij}_{klm} &:= \left\{{\left[{(\mathbf A \otimes \mathds 1 + \mathds 1 \otimes \mathbf A) \otimes \mathds 1 \otimes \mathds 1 \otimes \mathds 1 - \mathds 1 \otimes \mathds 1 \otimes (\mathbf A \otimes \mathds 1 \otimes \mathds 1 + \mathds 1 \otimes \mathbf A \otimes \mathds 1 + \mathds 1 \otimes \mathds 1 \otimes \mathbf A)}\right]^{-1}}\right\}^{ijk'l'm'}_{i'j'klm} \nonumber \\
	&\quad{} \times \left({a^{i'}_{k'l'} \delta^{j'}_{m'} + a^{j'}_{k'l'} \delta^{i'}_{m'}}\right).
\end{align}
These equalities yield:
\begin{equation} \label{eq:expec-x}
	\begin{aligned}
		\langle x^i(\tau) \mid \mathbf x(0) \rangle &= (e^{\mathbf A \tau})^i_j x^j(0) + \left[{(e^{\mathbf A \tau})^i_{i'} \delta^j_{j'} \delta^k_{k'} - \delta^i_{i'} (e^{\mathbf A \tau})^j_{j'} (e^{\mathbf A \tau})^k_{k'}}\right] \alpha^{i'}_{jk} \left({x^{j'}(0) x^{k'}(0) - C^{j'k'}}\right) \\
		&\quad{} + a \mathcal O(a,b),
	\end{aligned}
\end{equation}
\begin{equation}
	\begin{aligned} \label{eq:expec-x-x}
		\langle x^i(\tau) x^j(\tau) \mid \mathbf x(0) \rangle &= C^{ij} + (e^{\mathbf A \tau})^i_{i'} (e^{\mathbf A \tau})^j_{j'} \left({x^{i'}(0) x^{j'}(0) - C^{i'j'}}\right) \\
		&\quad{} + \left[{(e^{\mathbf A \tau})^i_{i'} (e^{\mathbf A \tau})^j_{j'} \delta^k_{k'} - \delta^i_{i'} \delta^j_{j'} (e^{\mathbf A \tau})^k_{k'}}\right] \beta^{i'j'}_k x^{k'}(0) \\
		&\quad{} + \left[{(e^{\mathbf A \tau})^i_{i'} (e^{\mathbf A \tau})^j_{j'} \delta^k_{k'} \delta^l_{l'} \delta^m_{m'} - \delta^i_{i'} \delta^j_{j'} (e^{\mathbf A \tau})^k_{k'} (e^{\mathbf A \tau})^l_{l'} (e^{\mathbf A \tau})^m_{m'}}\right] \gamma^{i'j'}_{klm} \\
		&\qquad{} \times \left({x^{k'}(0) x^{l'}(0) x^{m'}(0) - C^{k'l'} x^{m'}(0) - C^{k'm'} x^{l'}(0) - C^{l'm'} x^{k'}(0)}\right) \\
		&\quad{} + a \mathcal O(a,b).
	\end{aligned}
\end{equation}
It is not desirable to try to estimate $b$ from the quantity:
\begin{equation}
	\langle (x^i(\tau) - \langle x^i(\tau) \mid \mathbf x(0) \rangle) (x^j(\tau) - \langle x^j(\tau) \mid \mathbf x(0) \rangle) x^k(0) \rangle
\end{equation}
with $\langle \mathbf x(\tau) \mid \mathbf x(0) \rangle$ a quadratic function, since this would involve fifth moments. Also, note that if $\mathbf A$ has eigenvalues $\lambda_1, \lambda_2, \lambda_3$ such that $\lambda_1 = \lambda_2 + \lambda_3$, then $\mathbf A \otimes \mathds 1 \otimes \mathds 1 - \mathds 1 \otimes \mathbf A \otimes \mathds 1 - \mathds 1 \otimes \mathds 1 \otimes \mathbf A$ is not invertible\footnote{If $\mathbf v_i$ ($i=1,2,3$) are eigenvectors of $\mathbf A$ with eigenvalues $\lambda_i$, respectively, then $(\mathbf A \otimes \mathds 1 \otimes \mathds 1 - \mathds 1 \otimes \mathbf A \otimes \mathds 1 - \mathds 1 \otimes \mathds 1 \otimes \mathbf A) (\mathbf v_1 \otimes \mathbf v_2 \otimes \mathbf v_3) = \mathbf 0$.} and Koopman eigenfunctions cannot in general be found.

\subsection{Integrated variables}
We now address the case of an integrated variable $y$ obeying:
\begin{equation} \label{eq:integrated-nonlinear}
\langle \dot y \mid \mathbf x \rangle = A^y_i x^i + a^y_{ij} (x^i x^j - C^{ij}),
\end{equation}
where all coefficients are constant. We can define:
\begin{align}
y^{(0)} &:= y - A^y_i (\mathbf A^{-1})^i_j x^j, \\
D^{iy^{(0)}} &:= \frac 1 2 \left \langle \frac{\mathrm d [x^i, y^{(0)}](t)}{\mathrm d t} \right \rangle = D^{iy} - A^y_j (\mathbf A^{-1})^j_k D^{ik}, \\
y' &:= y^{(0)} - \left({a^y_{ij} - A^y_{i'} (\mathbf A^{-1})^{i'}_{j'} a^{j'}_{ij}}\right) \left[{(\mathbf A \otimes \mathds 1 + \mathds 1 \otimes \mathbf A)^{-1}}\right]^{ij}_{kl} (x^k x^l - C^{kl}), \label{eq:y-prime}
\end{align}
and similarly for $b^{iy^{(0)}}$, where the notation of $\mathbf A$ by itself excludes the dynamics of $y$, so that $y'$ satisfies the law:
\begin{equation}
\langle \dot y' \mid \mathbf x \rangle = a \mathcal O(a,b).
\end{equation}
We change coordinates to ${x'}^i := f_1^i(\mathbf x)$, where
\begin{equation}
\langle {x'}^i {x'}^j \rangle = C^{ij} + a \mathcal O(a,b),
\end{equation}
\begin{equation}
\begin{aligned}
\left \langle \frac{\mathrm d [{x'}^i, y'](t)}{\mathrm d t} \mid \mathbf x(t) \right \rangle &= 2 D^{iy^{(0)}} + 2 b^{iy^{(0)}}_j x^j \\
&\qquad{} - 4 \left({a^y_{jk} - A^y_{j'} (\mathbf A^{-1})^{j'}_{k'} a^{k'}_{jk}}\right) \left[{(\mathbf A \otimes \mathds 1 + \mathds 1 \otimes \mathbf A)^{-1}}\right]^{jk}_{lm} D^{il} x^m \\
&\qquad{} + 4 \left[{(\mathbf A \otimes \mathds 1 \otimes \mathds 1 - \mathds 1 \otimes \mathbf A \otimes \mathds 1 - \mathds 1 \otimes \mathds 1 \otimes \mathbf A)^{-1}}\right]^{ij'k'}_{i'jk} a^{i'}_{j'k'} D^{jy^{(0)}} x^k \\
&\qquad{} + a \mathcal O(a,b).
\end{aligned}
\end{equation}
The above shows the values of $b^{iy'}$. On the r.h.s., $x$ can be replaced by $x'$ with an error $a \mathcal O(a,b)$. We can now calculate third-order covariance functions with $a \mathcal O((a,b)^2)$ correction by evaluating fourth-order quantities by applying Isserlis's theorem. The condition for detailed balance (with $a \mathcal O((a,b)^2)$ correction) is also now ready to be formulated. When writing the time-reversibility for third-order covariance functions of the primed variables, one obtains those for the unprimed variables by adding $\mathcal O(a)$ terms involving fourth-order quantities. The fourth-order quantities satisfy time-reversibility with correction $\mathcal O((a,b)^2)$ by applying Isserlis's theorem and the time-reversibility of second-order moments. The condition for detailed balance then reduces to the vanishing of the angular momenta for primed variables, which for the same reason reduces to the vanishing of the angular momenta for unprimed variables.

\subsection{Second-order covariance functions to quadratic order}

We may calculate second-order covariance functions to order $a \mathcal O(a,b)$, with $a^2 \mathcal O((a,b)^2)$ corrections. We may consider what happens if second-order angular momenta are $a^2 \mathcal O((a,b)^2)$ and third-order angular momenta are $a \mathcal O((a,b)^2)$. We again change coordinates to ${x'}^i := f_1^i(\mathbf x)$. We have:
\begin{align}
	L({x'}^i, {x'}^j) &= a^2 \mathcal O((a,b)^2), \\
	L({x'}^i {x'}^j, {x'}^k) &= a \mathcal O((a,b)^2),
\end{align}
and therefore:
\begin{align}
	\langle {x'}^i(0) {x'}^j(\tau) \rangle - \langle {x'}^i(\tau) {x'}^j(0) \rangle &= a^2 \mathcal O((a,b)^2), \\
	\langle {x'}^i(0) {x'}^j(0) {x'}^k(\tau) \rangle - \langle {x'}^i(\tau) {x'}^j(\tau) {x'}^k(0) \rangle &= a \mathcal O((a,b)^2).
\end{align}
Switching back to the original coordinates:
\begin{equation}
	x^i = \mathcal N^i_j \left[{ {x'}^j - \alpha^j_{kl} \mathcal O(1) \left({{x'}^k {x'}^l - {C'}^{kl}}\right) + a^2 \mathcal O(1) {He}_3'(\mathbf x') + a^3 \mathcal O(1) {He}_4'(\mathbf x') + \cdots }\right],
\end{equation}
where $\mathcal N^i_j = \delta^i_j + a^2 \mathcal O(1)$ is a normalization factor, $\mathbf C' := \langle \mathbf x' {\mathbf x'}^{\mathsf T} \rangle$, and ${He}_n'$ denotes the multidimensional Hermite polynomial defined with respect to covariance matrix $\mathbf C'$. We then have:
\begin{equation}
	\langle x^i(0) x^j(\tau) \rangle - \langle x^i(\tau) x^j(0) \rangle = a^2 \mathcal O((a,b)^2),
\end{equation}
i.e., the second-order covariance functions are symmetric to quadratic order. A similar result holds for integrated variables.

\subsection{Finite time interval inhomogeneous diffusion}

The coefficients describing inhomogeneous diffusion can be estimated by way of the quantities:
\begin{equation} \label{eq:tau-inhomo-diff}
	\langle (x^i(0) + x^i(\tau)) (x^j(\tau) - x^j(0)) (x^k(\tau) - x^k(0)) \rangle.
\end{equation}
However, for finite $\tau$, the above quantity can be non-zero for homogeneous diffusion because of non-vanishing third moments. This can be illustrated by the example of Eq.\ \eqref{eq:third-order-model-a} in one dimension with $-A = D = 1$ and $b = 0$. From Eqs.\ \eqref{eq:tau-0-0}\textendash \eqref{eq:tau-tau-0}, we have:
\begin{equation}
	\langle (x(0) + x(\tau)) (x(\tau) - x(0))^2 \rangle = 4a (1 - 2 e^{-\tau} + e^{-2\tau}) = \mathcal O(\tau^2).
\end{equation}
It can be easily seen that if the $b$ coefficients vanish, then Eq.\ \eqref{eq:tau-inhomo-diff} is $\mathcal O(\tau^2)$.

\subsection{Quantitative significance of third-order effects}
Now, we address the problem of determining if the third-order quantities are quantitatively significant. Following the discussion for the second moments, we may consider a putative relation for the ``ensemble covariance'' of the third moments as:
\begin{equation}
	\langle x^i x^j x^k \rangle \langle x^{i'} x^{j'} x^{k'} \rangle \sim C^{ii'} C^{jj'} C^{kk'}.
\end{equation}
Following the discussion of deviations of experiment from theory for the second moments, we may treat $(i,j,k)$ as a single ``super-index'' and similarly for $(i',j',k')$, then multiply the l.h.s.\ by the matrix inverse of the r.h.s. For the same reason as before, the resulting matrix has at most one non-zero eigenvalue equal to its trace. To satisfy symmetry, using the same argument as before, we may consider an ensemble of stochastic systems whose ``ensemble covariance'' of third moments (l.h.s.\ of the above) is:
\begin{equation}
	\frac{C^{ii'} C^{jj'} C^{kk'} + C^{ii'} C^{jk'} C^{kj'} + C^{ik'} C^{jj'} C^{ki'} + C^{ij'} C^{ji'} C^{kk'} + C^{ij'} C^{jk'} C^{ki'} + C^{ik'} C^{ji'} C^{kj'}}{6}.
\end{equation}
In the absence of symmetry constraints, we may calculate the ``expected value'':
\begin{equation}
	\langle x^i x^j x^k \rangle \langle x^{i'} x^{j'} x^{k'} \rangle C^{-1}_{ii'} C^{-1}_{jj'} C^{-1}_{kk'} \sim \frac{d^3 + 3 d^2 + 2d}{6},
\end{equation}
where $d$ is the dimension. The r.h.s.\ of the above is the degrees of freedom of the third moments.

Now, we turn to the coefficients for inhomogeneous diffusion. A putative relation for the ``ensemble covariance'' of the $b$ coefficients can be written as:
\begin{equation}
	b^{ij}_k b^{i'j'}_{k'} \sim D^{ii'} D^{jj'} C^{-1}_{kk'}.
\end{equation}
For fixed $k$ and $k'$, we may consider $(i,j)$ as a single ``super-index'' and similarly for $(i',j')$ and take the l.h.s.\ multiplied by the matrix inverse of the r.h.s. Similarly to before, in coordinates where $\mathbf D = \mathds 1$ (without transforming $k$ and $k'$), the resulting matrix has the form $\mathbf v \mathbf w^{\mathsf T}$ and therefore has rank at most 1. Thus, we may use the same argument to deduce that the eigenvalues not associated with antisymmetric left eigenvectors remain the same when symmetrizing the r.h.s., which is inverted after neglecting permutations. In the absence of symmetry constraints, we may then consider the ``ensemble covariance'' for an ensemble of stochastic systems to be:
\begin{equation}
	b^{ij}_k b^{i'j'}_{k'} \sim \frac{D^{ii'} D^{jj'} + D^{ij'} D^{ji'}}{2} C^{-1}_{kk'}.
\end{equation}
An alternative choice would be to divide the above by the dimension $d$. However, this choice is inconsistent with the scaling with the dimension of the ``ensemble covariance'' of the third moments. We now calculate the ``ensemble covariance'' for $b$ collectively:
\begin{equation}
	b^{ij}_k b^{i'j'}_{k'} D^{-1}_{ii'} D^{-1}_{jj'} C^{kk'} \sim \frac{d_0 d(d+1)}{2}.
\end{equation}
where $d_0$ is the dimension of the stationary variables only and $d$ is the total dimension including integrated variables. In the above, $k$ and $k'$ are summed over stationary variables only, whereas $i$, $j$, $i'$, and $j'$ are summed over all variables including integrated variables\footnote{We may however be concerned about the possibility for a non-negative diffusivity function to have a mean far less than the standard deviation of its ``linear component'' (scaling as $\sqrt{d}$). To this end, we may consider the function $e^{cx}$, where $c$ is a constant, for $x \sim \mathcal{N}(0,1)$. Its mean value is $\langle e^{cx} \rangle = e^{c^2/2}$. When performing linear regression of $e^{cx}$ with regressor $x$, the coefficient is $\langle x e^{cx} \rangle = c e^{c^2/2}$. We see that the ratio between the two can be made arbitrarily large, and thus there is no issue.}. The r.h.s.\ is the number of degrees of freedom of $b^{ij}_k$. As for second-order quantities, if the time-step is not small, it may be suitable to use discrete-time estimators of the diffusion.

Next, we address the nonlinear drift. For the coefficients $a^i_{jk}$, we would like to compare $a^i_{jk} (x^j x^k - C^{jk})$ with $A^i_j x^j$. We are thus led to consider the matrix:
\begin{equation}
	M^{jkj'k'} := \langle (x^j x^k - C^{jk}) (x^{j'} x^{k'} - C^{j'k'}) \rangle = C^{jj'} C^{kk'} + C^{jk'} C^{kj'},
\end{equation}
evaluated in the linear Gaussian model, where $(j,k)$ is considered as a single ``super-index'', as is $(j',k')$. We need to invert the above matrix neglecting permutations of $(j,k)$ and $(j',k')$. This is done by introducing a matrix $N_{jkj'k'}$ symmetric under the swappings $j \leftrightarrow k$, $j' \leftrightarrow k'$, and $(j,k) \leftrightarrow (j',k')$ such that:
\begin{equation}
	M^{jkj'k'} N_{j'k'lm} = \frac{\delta^j_l \delta^k_m + \delta^j_m \delta^k_l}{2},
\end{equation}
and specifying:
\begin{equation}
	a^i_{jk} a^{i'}_{j'k'} \sim A^i_l A^{i'}_{l'} C^{ll'} N_{jkj'k'}.
\end{equation}
We then have the ``ensemble covariance'' for the $a$ coefficients collectively:
\begin{equation}
	a^i_{jk} a^{i'}_{j'k'} M^{jkj'k'} (\mathbf A^{-1})^l_i (\mathbf A^{-1})^{l'}_{i'} C^{-1}_{ll'} \sim \frac{d_0^2(d_0+1)}{2}.
\end{equation}
The r.h.s.\ is the number of degrees of freedom of $a^i_{jk}$. The above applies only to stationary variables. Alternatively, we may compare the quadratic variations of the linear and quadratic terms in the drift function. We thus introduce the quantity:
\begin{equation}
	{\widetilde{M}}^{jkj'k'} := \left \langle{\frac{\mathrm d [x^j x^k, x^{j'} x^{k'}]}{\mathrm d t}}\right \rangle = 2 \left({C^{jj'} D^{kk'} + C^{jk'} D^{kj'} + C^{kj'} D^{jk'} + C^{kk'} D^{jj'}}\right),
\end{equation}
evaluated in the linear Gaussian model. Similarly to before, we need the ``symmetric inverse'' $\widetilde{N}$ of the above, neglecting permutations:
\begin{equation}
	{\widetilde{M}}^{jkj'k'} {\widetilde{N}}_{j'k'lm} = \frac{\delta^j_l \delta^k_m + \delta^j_m \delta^k_l}{2}.
\end{equation}
We then specify:
\begin{equation}
	a^i_{jk} a^{i'}_{j'k'} \sim 2 A^i_l A^{i'}_{l'} D^{ll'} {\widetilde{N}}_{jkj'k'}.
\end{equation}
The advantage of this formulation is that it can be extended to integrated variables. For a description independent of linear transformations of coordinates, we compare the linear and quadratic terms in Eq.\ \eqref{eq:y-prime}. For two integrated variables $y$ and $z$, we have the ``ensemble covariances'':
\begin{equation}
	a^i_{jk} \left({a^y_{i'j'} - A^y_{k'} (\mathbf A^{-1})^{k'}_{l'} a^{l'}_{i'j'}}\right) \sim 2 {\widetilde{N}}_{jkk'l'} A^i_l (A^{k'}_{i'} \delta^{l'}_{j'} + \delta^{k'}_{i'} A^{l'}_{j'}) \left({D^{ly} - A^y_p (\mathbf A^{-1})^p_q D^{lq}}\right),
\end{equation}
\begin{equation}
	\begin{aligned}
		&\left({a^y_{ij} - A^y_k (\mathbf A^{-1})^k_l a^l_{ij}}\right) \left({a^z_{i'j'} - A^z_{k'} (\mathbf A^{-1})^{k'}_{l'} a^{l'}_{i'j'}}\right) \sim 2 {\widetilde{N}}_{klk'l'} (A^k_i \delta^l_j + \delta^k_i A^l_j) (A^{k'}_{i'} \delta^{l'}_{j'} + \delta^{k'}_{i'} A^{l'}_{j'}) \\
		&\qquad \qquad \qquad {} \times \left({D^{yz} - A^y_p (\mathbf A^{-1})^p_q D^{qz} - A^z_r (\mathbf A^{-1})^r_s D^{sy} + A^y_p A^z_r (\mathbf A^{-1})^p_q (\mathbf A^{-1})^r_s D^{qs}}\right).
	\end{aligned}
\end{equation}

Lastly, we may consider the angular momenta $L(x^i x^j, x^k)$. Motivated by the case of angular momenta between two variables, we may suppose an ``ensemble covariance'' of the form:
\begin{equation} \label{eq:L-third-order-eval}
	\begin{aligned}
		L(x^i x^j, x^k) L(x^{i'} x^{j'}, x^{k'}) &\sim \frac 1 2 \left \langle {\frac{\mathrm d [x^i x^j, x^{i'} x^{j'}]}{\mathrm d t} \frac{\mathrm d [x^k, x^{k'}]}{\mathrm d t} - \frac{\mathrm d [x^i x^j, x^{k'}]}{\mathrm d t} \frac{\mathrm d [x^{i'} x^{j'}, x^k]}{\mathrm d t}} \right \rangle \\
		&= 2 \Big[{C^{ii'} (D^{jj'} D^{kk'} - D^{jk'} D^{kj'}) + C^{ij'} (D^{ji'} D^{kk'} - D^{jk'} D^{ki'})}\Big. \\
		&\qquad \Big.{{}+ C^{ji'} (D^{ij'} D^{kk'} - D^{ik'} D^{kj'}) + C^{jj'} (D^{ii'} D^{kk'} - D^{ik'} D^{ki'})}\Big].
	\end{aligned}
\end{equation}
where the expectation is evaluated in the linear Gaussian model. It is seen that the above satisfies Eq.\ \eqref{eq:LLL0}. We then have the ``ensemble covariance'' of these quantities collectively:
\begin{equation}
	L(x^i x^j, x^k) L(x^{i'} x^{j'}, x^{k'}) C^{-1}_{ii'} C^{-1}_{jj'} D^{-1}_{kk'} \sim 4 (d_0+1) (d-1) D^{jj'} C^{-1}_{jj'},
\end{equation}
where $d_0$ is the dimension of the stationary variables only and $d$ is the total dimension including integrated variables. In the above, $i$, $i'$, $j$, and $j'$ are summed over the stationary variables only, whereas $k$ and $k'$ are summed over all the variables including integrated variables.

\subsection{Finite time interval ``cubic variations''}
We now briefly address the issue of measuring ``cubic variations'' at finite time intervals $\tau$. As mentioned before, this vanishes faster than $\mathcal O(\tau)$. We can solve a differential equation to obtain:
\begin{equation}
\begin{aligned}
&\frac{\mathrm d}{\mathrm d \tau} \langle (x^i(\tau)-x^i(0))(x^j(\tau)-x^j(0))(x^k(\tau)-x^k(0)) \rangle = 2 A^i_l b^{jk}_m C^{lm} \tau + 2 A^j_l b^{ik}_m C^{lm} \tau + 2 A^k_l b^{ij}_m C^{lm} \tau \\
&\quad{} + 2 b^{jk}_l (2D^{il} + A^i_m C^{lm}) \tau + 2 b^{ik}_l (2D^{jl} + A^j_m C^{lm}) \tau + 2 b^{ij}_l (2D^{kl} + A^k_m C^{lm}) \tau + a \mathcal O((a,b)^2) \tau + \mathcal O(\tau^2).
\end{aligned}
\end{equation}
where we have organized terms by application of It\^{o}'s rule. We can swap $l$ and $m$ in the first line and combine terms using the Lyapunov equation:
\begin{equation}
2D^{il} + 2A^i_m C^{lm} = A^i_m C^{lm} - A^l_m C^{im} = -L(x^i, x^l).
\end{equation}
and similarly for $j$ and $k$. We see that if second-order angular momenta vanish, then the ``cubic variation'' at finite time intervals $\tau$ vanishes, with corrections $a \mathcal O((a,b)^2) \tau^2$ and $\mathcal O(\tau^3)$. We may evaluate quantitative significance by treating $L$ and $b$ as uncorrelated for the ``ensemble covariance'', because they acquire opposite signs under time reversal.

\subsection{Comparison of experimental data and theoretical predictions}
We now address the issue of comparing experimental and theoretical values. While a fit to a Langevin equation can be performed, it remains a question whether the measured system actually obeys such dynamics. We limit our discussion to three-point covariance functions where two of the times are equal, or, in the case of integrated variables, may differ by at most a single time-step. The quantities $\langle x^i(\tau) x^j(0) x^k(0) \rangle$ and $\langle x^i(0) x^j(\tau) x^k(\tau) \rangle$ can be evaluated like the third-order moments, but only symmetrizing with respect to $(j,k)$. For an integrated variable $y$ measured at a short frame interval $\Delta t$, we have the quantities
\begin{align}
	&\langle x^i(0) x^j(0) (y(\tau+\Delta t)-y(\tau)) \rangle, \\
	&\langle x^i(\tau+\Delta t) x^j(\tau+\Delta t) (y(\Delta t)-y(0)) \rangle.
\end{align}
We consider the sum and difference of these quantities at $\tau = 0$. The former is equal to $(\Delta t) L(x^i x^j, y)$, while the latter is equal to $-2 (\Delta t) (C^{ik} b^{jy}_k + C^{jk} b^{iy}_k)$. We can calculate the ``ensemble variances'' of these quantities using the aforementioned prescriptions. We may consider using them for comparison at arbitrary $\tau$. However, we may be interested in comparing the above quantities directly. They may be obtained by taking one-half the sum and difference of the sum and difference. We may treat the sum and difference of the above quantities as uncorrelated in the ``ensemble covariance'' as they acquire opposite signs under time reversal. Thus we conclude that the above quantities should be compared according to one-fourth the sum of the ``ensemble variances'' of $(\Delta t) L(x^i x^j, y)$ and $-2 (\Delta t) (C^{ik} b^{jy}_k + C^{jk} b^{iy}_k)$. As in the linear case, and as is analogously true for all quantities, if the theoretical values at $\tau = 0$ exceed the prescriptions of the ``ensemble variances'' (for either second- or third-order quantities), we may need to increase the tolerance for comparison accordingly.

Next, there are the quantities:
\begin{align}
	&\langle x^i(\tau+\Delta t) (x^j(0)+x^j(\Delta t)) (y(\Delta t)-y(0)) \rangle, \\
	&\langle x^i(0) (x^j(\tau)+x^j(\tau+\Delta t)) (y(\tau+\Delta t)-y(\tau)) \rangle.
\end{align}
Again we consider the sum and difference at $\tau = 0$ and generalize to arbitrary $\tau$. These are $2 (\Delta t) L(x^i x^j, y)$ and $4 (\Delta t) b^{iy}_k C^{jk}$, respectively. As in the previous case, we may use combinations of the sum and difference that result in the above quantities.

Now we replace $x^i$ with an integrated variable $z$. In this case there is no angular momentum. Thus we may compare:
\begin{align}
	&\langle (z(\tau+\Delta t)-z(\Delta t)) (x^j(0)+x^j(\Delta t)) (y(\Delta t)-y(0)) \rangle, \\
	&\langle (z(\tau)-z(0)) (x^j(\tau)+x^j(\tau+\Delta t)) (y(\tau+\Delta t)-y(\tau)) \rangle
\end{align}
to the prescription for $2 (\Delta t) b^{yz}_k C^{jk}$.

Next, we consider the quantities:
\begin{align}
	&\langle x^i(\tau+\Delta t) (x^j(\Delta t)-x^j(0)) (y(\Delta t)-y(0)), \\
	&\langle x^i(0) (x^j(\tau+\Delta t)-x^j(\tau)) (y(\tau+\Delta t)-y(\tau)) \rangle,
\end{align}
where $x^j$ may be an integrated variable. These are simply compared to the prescription for $2 (\Delta t) b^{jy}_k C^{ik}$.

Finally, we again replace $x^i$ with an integrated variable $z$. We compare:
\begin{align}
	&\langle (z(\tau+\Delta t)-z(\tau)) (x^j(\Delta t)-x^j(0)) (y(\Delta t)-y(0)) \rangle, \\
	&\langle (z(\Delta t)-z(0)) (x^j(\tau+\Delta t)-x^j(\tau)) (y(\tau+\Delta t)-y(\tau)) \rangle
\end{align}
to the prescription for:
\begin{equation}
	\langle (z(\Delta t)-z(0)) (x^j(\Delta t)-x^j(0)) (y(\Delta t)-y(0)) \rangle,
\end{equation}
discussed in section 6.9.

\section{Odd variables}
So far, we have considered variables that are even under time reversal. However, some so-called ``odd'' variables, like velocity, change sign upon time reversal. In this section, we will discuss the conditions for time-reversal symmetry in the presence of odd variables, to third order.

\subsection{Linear Gaussian systems}
We first consider linear Gaussian systems. Our state variable contains some even variables, denoted $\mathbf x^{+}$, and some odd variables, denoted $\mathbf x^{-}$. The dynamical matrix $\mathbf A$ can be partitioned into various components depending on which variables they connect, viz.:
\begin{align}
\dot{\mathbf x}^{+} &= \mathbf A^{+}_{+} \mathbf x^{+} + \mathbf A^{+}_{-} \mathbf x^{-} + \boldsymbol \xi^{+}, \\
\dot{\mathbf x}^{-} &= \mathbf A^{-}_{+} \mathbf x^{+} + \mathbf A^{-}_{-} \mathbf x^{-} + \boldsymbol \xi^{-},
\end{align}
\begin{equation}
	\quad \langle \boldsymbol \xi^{\pm}(t) {\boldsymbol \xi^{\pm}(t')}^{\mathsf T} \rangle = 2 \mathbf D^{\pm \pm} \delta(t-t'), \quad \langle \boldsymbol \xi^{+}(t) {\boldsymbol \xi^{-}(t')}^{\mathsf T} \rangle = 2 \mathbf D^{+-} \delta(t-t'),
\end{equation}
where $\boldsymbol \xi^{\pm}$ are zero-mean Gaussian white noise and all coefficients are constant. To have time-reversal symmetry, we must have $\mathbf D^{+-} = \mathbf 0$ and $\langle \mathbf x^{+} {\mathbf x^{-}}^{\mathsf T} \rangle = \mathbf 0$. These imply:
\begin{equation} \label{eq:A+-}
\mathbf 0 = \frac{\mathrm d}{\mathrm d t} \langle \mathbf x^{+}(t) \mathbf x^{-}(t)^{\mathsf T} \rangle = \mathbf A^{+}_{-} \langle \mathbf x^{-} {\mathbf x^{-}}^{\mathsf T} \rangle + \langle \mathbf x^{+} \mathbf {x^{+}}^{\mathsf T}\rangle {\mathbf A^{-}_{+}}^{\mathsf T},
\end{equation}
i.e., it automatically satisfies the relation:
\begin{equation}
\langle \dot{\mathbf x}^{+} {\mathbf x^{-}}^{\mathsf T} \rangle = -\langle \mathbf x^{+} {{}\dot{\mathbf x}^{-}}^{\mathsf T} \rangle
\end{equation}
(with time-derivatives interpreted in It\^o sense). We must also have the relations from before:
\begin{equation}
\langle \dot{\mathbf x}^{\pm} {\mathbf x^{\pm}}^{\mathsf T} \rangle = \langle \mathbf x^{\pm} {{}\dot{\mathbf x}^{\pm}}^{\mathsf T} \rangle
\end{equation}
which implies:
\begin{equation} \label{eq:A++}
\mathbf A^{\pm}_{\pm} \langle \mathbf x^{\pm} {\mathbf x^{\pm}}^{\mathsf T} \rangle = \langle \mathbf x^{\pm} {\mathbf x^{\pm}}^{\mathsf T} \rangle {\mathbf A^{\pm}_{\pm}}^{\mathsf T}.
\end{equation}
Next, we address the time-reversibility of the second moments. Similarly as before, this depends only on the linearity of the drift and not on the homogeneity of diffusion. For time-reversibility, we must have:
\begin{align}
\langle \mathbf x^{\pm}(\tau) \mathbf x^{\pm}(0)^{\mathsf T} \rangle &= \langle \mathbf x^{\pm}(0) \mathbf x^{\pm}(\tau)^{\mathsf T} \rangle \\
\langle \mathbf x^{+}(\tau) \mathbf x^{-}(0)^{\mathsf T} \rangle &= -\langle \mathbf x^{+}(0) \mathbf x^{-}(\tau)^{\mathsf T} \rangle
\end{align}
which is equivalent to:
\begin{align}
{e^{\mathbf A \tau}}^{\pm}_{\pm} \langle \mathbf x^{\pm} {\mathbf x^{\pm}}^{\mathsf T} \rangle &= \langle \mathbf x^{\pm} {\mathbf x^{\pm}}^{\mathsf T} \rangle {{e^{\mathbf A \tau}}^{\pm}_{\pm}}^{\mathsf T} \\
{e^{\mathbf A \tau}}^{+}_{-} \langle \mathbf x^{-} {\mathbf x^{-}}^{\mathsf T} \rangle &= -\langle \mathbf x^{+} {\mathbf x^{+}}^{\mathsf T} \rangle {{e^{\mathbf A \tau}}^{-}_{+}}^{\mathsf T}
\end{align}
which is in turn equivalent to, for $n \ge 1$:
\begin{align}
{\mathbf A^n}^{\pm}_{\pm} \langle \mathbf x^{\pm} {\mathbf x^{\pm}}^{\mathsf T} \rangle &= \langle \mathbf x^{\pm} {\mathbf x^{\pm}}^{\mathsf T} \rangle {{\mathbf A^n}^{\pm}_{\pm}}^{\mathsf T} \\
{\mathbf A^n}^{+}_{-} \langle \mathbf x^{-} {\mathbf x^{-}}^{\mathsf T} \rangle &= -\langle \mathbf x^{+} {\mathbf x^{+}}^{\mathsf T} \rangle {{\mathbf A^n}^{-}_{+}}^{\mathsf T}.
\end{align}
The base case $n = 1$ is guaranteed by our assumptions and the mathematical induction step is easily performed.

For the case of an integrated variable, for simplicity we consider an even variable $y$ obeying the law:
\begin{equation}
\dot y = \mathbf A^{y}_{+} \mathbf x^{+} + \mathbf A^{y}_{-} \mathbf x^{-} + \xi^y,
\end{equation}
\begin{equation}
\langle \xi^y(t) \boldsymbol \xi^{\pm}(t') \rangle = 2 \mathbf D^{y \pm} \delta(t-t'), \quad \langle \xi^y(t) \xi^y(t') \rangle = 2 D^{yy} \delta(t-t'),
\end{equation}
where $\xi^y$ is zero-mean Gaussian white noise and all coefficients are constant. For time-reversibility, we need $\mathbf D^{y-} = \mathbf 0$. We write $\Delta y(\tau) := y(\tau)-y(0)$, and:
\begin{equation} \label{eq:linear-dynamics-integrated-matrix}
\left\langle{\frac{\mathrm d}{\mathrm d \tau} \begin{pmatrix}
\mathbf x^{+} \\
\mathbf x^{-} \\
\Delta y
\end{pmatrix} \mid \mathbf x}\right\rangle = \begin{pmatrix}
\mathbf A^{+}_{+} & \mathbf A^{+}_{-} & \mathbf 0 \\
\mathbf A^{-}_{+} & \mathbf A^{-}_{-} & \mathbf 0 \\
\mathbf A^{y}_{+} & \mathbf A^{y}_{-} & 0
\end{pmatrix} \begin{pmatrix}
\mathbf x^{+} \\
\mathbf x^{-} \\
\Delta y
\end{pmatrix}.
\end{equation}
Writing $\widehat{\mathbf A}$ for the matrix appearing on the r.h.s., the conditional expectation of the vector appearing in the above differential equation given $\mathbf x(0)$ is equal to $e^{\widehat{\mathbf A} \tau}$ multiplied by its value at $\tau = 0$. Also:
\begin{equation}
\frac{\mathrm d}{\mathrm d \tau} \begin{pmatrix}
1 \\
\langle \mathbf x^{+} \Delta y \rangle \\
\langle \mathbf x^{-} \Delta y \rangle
\end{pmatrix} = \begin{pmatrix}
0 & \mathbf 0 & \mathbf 0 \\
\widehat{\mathbf A}^{y(1)+}_{(0)} & \mathbf A^{+}_{+} & \mathbf A^{+}_{-} \\
\widehat{\mathbf A}^{y(1)-}_{(0)} & \mathbf A^{-}_{+} & \mathbf A^{-}_{-}
\end{pmatrix} \begin{pmatrix}
1 \\
\langle \mathbf x^{+} \Delta y \rangle \\
\langle \mathbf x^{-} \Delta y \rangle
\end{pmatrix}
\end{equation}
where we labeled $\mathbf x^{\pm} \Delta y$ as $y^{(1)\pm}$. Again, we write $\widehat{\mathbf A}$ for the matrix on the r.h.s.\ and the vector\footnote{An ordered collection of quantities, not a true physical vector obeying the transformation laws under a change of basis.} appearing in the above differential equation is equal to $e^{\widehat{\mathbf A} \tau}$ multiplied by its value when $\tau = 0$. The elements are:
\begin{align}
	\widehat{\mathbf A}^{y(1)+}_{(0)} &= 2 \mathbf D^{y+} + \langle \mathbf x^{+} {\mathbf x^{+}}^{\mathsf T} \rangle {\mathbf A^{y}_{+}}^{\mathsf T}, \\
	\widehat{\mathbf A}^{y(1)-}_{(0)} &= \langle \mathbf x^{-} {\mathbf x^{-}}^{\mathsf T} \rangle {\mathbf A^{y}_{-}}^{\mathsf T}
\end{align}
For time-reversibility, we require:
\begin{equation}
2 \mathbf D^{y+} + 2 \langle \mathbf x^{+} {\mathbf x^{+}}^{\mathsf T} \rangle {\mathbf A^{y}_{+}}^{\mathsf T} = \mathbf 0.
\end{equation}
We need to show that, for $n \ge 1$:
\begin{align}
\langle \mathbf x^{+} {\mathbf x^{+}}^{\mathsf T} \rangle {{{}\widehat{\mathbf A}^n}^{y}_{+}}^{\mathsf T} &= -{{}\widehat{\mathbf A}^n}^{y(1)+}_{(0)} \\
\langle \mathbf x^{-} {\mathbf x^{-}}^{\mathsf T} \rangle {{{}\widehat{\mathbf A}^n}^{y}_{-}}^{\mathsf T} &= {{}\widehat{\mathbf A}^n}^{y(1)-}_{(0)}
\end{align}
which can be easily shown by induction provided that the $\mathbf x$ variables satisfy detailed balance. This establishes the time-reversibility of $\langle (y(\tau)-y(0)) \mathbf x(0) \rangle$. The case for an odd integrated variable is analogous.

\subsection{Detailed balance}
Here we discuss the conditions for detailed balance in the Fokker\textendash Planck equation Eq.\ \eqref{eq:FPeqn} in the presence of odd variables. Each variable $x^i$ has a parity $\epsilon^i = \pm 1$, where under time-reversal $x^i \to \epsilon^i x^i$ ($i$ not summed over), and the product $\boldsymbol \epsilon \mathbf x$ is defined as $(\boldsymbol \epsilon \mathbf x)^i = \epsilon^i x^i$ ($i$ not summed over), following \cite{GardinerStochasticMethods}. In terms of the stationary distribution $p(\mathbf x)$, the conditions for detailed balance are \cite{GardinerStochasticMethods}:
\begin{align}
	\epsilon^i \mathcal A^i(\boldsymbol \epsilon \mathbf x) p(\mathbf x) &= -\mathcal A^i(\mathbf x) p(\mathbf x) + \partial_j [\mathcal B^{ij}(\mathbf x) p(\mathbf x)] \label{eq:FPeqnDBOdd} \\
	\epsilon^i \epsilon^j \mathcal B^{ij}(\boldsymbol \epsilon \mathbf x) &= \mathcal B^{ij}(\mathbf x). \label{eq:diffDBOdd}
\end{align}
where $i$ or $j$ appearing twice because of $\boldsymbol \epsilon$ is not summed over. Now, as before we multiply Eq.\ \eqref{eq:FPeqnDBOdd} by a test function $f(\mathbf x)$ and integrate. We consider test functions that are either even or odd under time reversal, i.e., $f(\boldsymbol \epsilon \mathbf x) = \pm f(\mathbf x)$. We also have a necessary condition for detailed balance:
\begin{equation} \label{eq:FPeqnOddPStat}
	p(\boldsymbol \epsilon \mathbf x) = p(\mathbf x),
\end{equation}
which means that we may write $\langle g(\boldsymbol \epsilon \mathbf x) \rangle = \langle g(\mathbf x) \rangle$ for any function $g(\cdot)$. If $f(\cdot)$ and $x^i$ have the same parity, then we arrive back to the same condition as the case where there are no odd variables. On the other hand, if $f(\cdot)$ and $x^i$ have the opposite parity, then upon multiplying by $f(\mathbf x)$ and integrating, the l.h.s.\ of Eq.\ \eqref{eq:FPeqnDBOdd} cancels the first term on the r.h.s.\ of Eq.\ \eqref{eq:FPeqnDBOdd}, and the remaining term reads:
\begin{equation}
	\langle \mathcal B^{ij}(\mathbf x) \partial_j f(\mathbf x) \rangle = 0
\end{equation}
which is guaranteed by substituting $\mathbf x \to \boldsymbol \epsilon \mathbf x$ (allowed because of Eq.\ \eqref{eq:FPeqnOddPStat}) and applying Eq.\ \eqref{eq:diffDBOdd}. Thus Eq.\ \eqref{eq:FPeqnDBOdd} gives no further information besides that contained in Eq.\ \eqref{eq:diffDBOdd} and Eq.\ \eqref{eq:FPeqnOddPStat} in the case where odd quantities are considered.

\subsection{Inhomogeneous diffusion}
For inhomogeneous diffusion, we introduce the notations:
\begin{align}
\mathbf x^{(2)+} &:= \{{x^{+}}^i {x^{+}}^j - \langle {x^{+}}^i {x^{+}}^j \rangle, {x^{-}}^i {x^{-}}^j - \langle {x^{-}}^i {x^{-}}^j \rangle\}_{i,j} \\
\mathbf x^{(2)-} &:= \{{x^{+}}^i {x^{-}}^j - \langle {x^{+}}^i {x^{-}}^j \rangle\}_{i,j}
\end{align}
which are the even and odd generalizations of the Hermite polynomials. In this way we can write:
\begin{equation} \label{eq:inhomogeneous-diffusion-mat}
\left\langle{\frac{\mathrm d}{\mathrm d \tau} \begin{pmatrix}
\mathbf x^{+} \\
\mathbf x^{-} \\
\mathbf x^{(2)+} \\
\mathbf x^{(2)-}
\end{pmatrix} \mid \mathbf x}\right\rangle = \begin{pmatrix}
\mathbf A^{+}_{+} & \mathbf A^{+}_{-} & & \\
\mathbf A^{-}_{+} & \mathbf A^{-}_{-} & & \\
\widehat{\mathbf A}^{(2)+}_{+} & \widehat{\mathbf A}^{(2)+}_{-} & \widehat{\mathbf A}^{(2)+}_{(2)+} & \widehat{\mathbf A}^{(2)+}_{(2)-} \\
\widehat{\mathbf A}^{(2)-}_{+} & \widehat{\mathbf A}^{(2)-}_{-} & \widehat{\mathbf A}^{(2)-}_{(2)+} & \widehat{\mathbf A}^{(2)-}_{(2)-}
\end{pmatrix} \begin{pmatrix}
\mathbf x^{+} \\
\mathbf x^{-} \\
\mathbf x^{(2)+} \\
\mathbf x^{(2)-}
\end{pmatrix}.
\end{equation}
We again assume, as throughout this section, that $\mathbf D^{+-} = \mathbf 0$ and $\langle \mathbf x^{+} {\mathbf x^{-}}^{\mathsf T} \rangle = \mathbf 0$, and we denote the matrix on the r.h.s.\ by $\widehat{\mathbf A}$. To have time-reversibility of the covariance functions, we first need that:
\begin{equation} \label{eq:third-order-odd}
\langle \mathbf x^{(2)\pm} {\mathbf x^{\mp}}^{\mathsf T} \rangle = \mathbf 0.
\end{equation}
In this model we also must have:
\begin{equation} \label{eq:diffusion-odd}
b^{++}_{-} = b^{--}_{-} = b^{+-}_{+} = 0.
\end{equation}
where the signs denote the parities of the corresponding variables. We also need for $n \ge 1$:
\begin{align}
{{}\widehat{\mathbf A}^n}^{(2)\pm}_{\pm} \langle \mathbf x^{\pm} {\mathbf x^{\pm}}^{\mathsf T} \rangle + {{}\widehat{\mathbf A}^n}^{(2)\pm}_{(2)\pm} \langle \mathbf x^{(2)\pm} {\mathbf x^{\pm}}^{\mathsf T} \rangle &= \langle \mathbf x^{(2)\pm} {\mathbf x^{\pm}}^{\mathsf T} \rangle {{\mathbf A^n}^{\pm}_{\pm}}^{\mathsf T} \label{eq:inhomogeneous-diffusion-a} \\
{{}\widehat{\mathbf A}^n}^{(2)\pm}_{\mp} \langle \mathbf x^{\mp} {\mathbf x^{\mp}}^{\mathsf T} \rangle + {{}\widehat{\mathbf A}^n}^{(2)\pm}_{(2)\mp} \langle \mathbf x^{(2)\mp} {\mathbf x^{\mp}}^{\mathsf T} \rangle &= -\langle \mathbf x^{(2)\pm} {\mathbf x^{\pm}}^{\mathsf T} \rangle {{\mathbf A^n}^{\mp}_{\pm}}^{\mathsf T} \label{eq:inhomogeneous-diffusion-b}.
\end{align}
The base case $n = 1$ for Eq.\ \eqref{eq:inhomogeneous-diffusion-a} is an assumption needed for time-reversal symmetry. The base case $n = 1$ for Eq.\ \eqref{eq:inhomogeneous-diffusion-b} follows from differentiating Eq.\ \eqref{eq:third-order-odd} and using Eq.\ \eqref{eq:diffusion-odd}, as the latter implies:
\begin{equation}
\left \langle \frac{\mathrm d [\mathbf x^{(2)\pm}, {\mathbf x^{\mp}}^{\mathsf T}](t)}{\mathrm d t} \right \rangle = \mathbf 0.
\end{equation}
Then mathematical induction can be performed when time-reversibility for the second moments holds. This establishes the time-reversibility of $\langle x^i(\tau) x^j(0) x^k(0) \rangle$.

Now, we consider an integrated variable $y$, again assumed to be even under time reversal. The dynamics of various quantities obey the laws:
\begin{equation}
\left\langle{\frac{\mathrm d}{\mathrm d \tau} \begin{pmatrix}
1 \\
\mathbf x^{+} \\
\mathbf x^{-} \\
\mathbf x^{(2)+} \\
\mathbf x^{(2)-} \\
\mathbf x^{+} \Delta y \\
\mathbf x^{-} \Delta y
\end{pmatrix} \mid \mathbf x}\right\rangle = \begin{pmatrix}
0 & & & & & & \\
& \mathbf A^{+}_{+} & \mathbf A^{+}_{-} & & & & \\
& \mathbf A^{-}_{+} & \mathbf A^{-}_{-} & & & & \\
& \widehat{\mathbf A}^{(2)+}_{+} & \widehat{\mathbf A}^{(2)+}_{-} & \widehat{\mathbf A}^{(2)+}_{(2)+} & \widehat{\mathbf A}^{(2)+}_{(2)-} & & \\
& \widehat{\mathbf A}^{(2)-}_{+} & \widehat{\mathbf A}^{(2)-}_{-} & \widehat{\mathbf A}^{(2)-}_{(2)+} & \widehat{\mathbf A}^{(2)-}_{(2)-} & & \\
\widehat{\mathbf A}^{y(1)+}_{(0)} & \widehat{\mathbf A}^{y(1)+}_{+} & \widehat{\mathbf A}^{y(1)+}_{-} & \widehat{\mathbf A}^{y(1)+}_{(2)+} & \widehat{\mathbf A}^{y(1)+}_{(2)-} & \mathbf A^{+}_{+} & \mathbf A^{+}_{-} \\
\widehat{\mathbf A}^{y(1)-}_{(0)} & \widehat{\mathbf A}^{y(1)-}_{+} & \widehat{\mathbf A}^{y(1)-}_{-} & \widehat{\mathbf A}^{y(1)-}_{(2)+} & \widehat{\mathbf A}^{y(1)-}_{(2)-} & \mathbf A^{-}_{+} & \mathbf A^{-}_{-}
\end{pmatrix} \begin{pmatrix}
1 \\
\mathbf x^{+} \\
\mathbf x^{-} \\
\mathbf x^{(2)+} \\
\mathbf x^{(2)-} \\
\mathbf x^{+} \Delta y \\
\mathbf x^{-} \Delta y
\end{pmatrix}.
\end{equation}
To have time-reversibility we must additionally have:
\begin{equation} \label{eq:diffusion-y-odd}
b^{y+}_{-} = b^{y-}_{+} = 0.
\end{equation}
We must also have, for $n \ge 1$:
\begin{align}
{{}\widehat{\mathbf A}^n}^{y(1)\pm}_{\pm} \langle \mathbf x^{\pm} {\mathbf x^{\pm}}^{\mathsf T} \rangle + {{}\widehat{\mathbf A}^n}^{y(1)\pm}_{(2)\pm} \langle \mathbf x^{(2)\pm} {\mathbf x^{\pm}}^{\mathsf T} \rangle &= -\langle \mathbf x^{\pm} {\mathbf x^{\pm}}^{\mathsf T} \rangle {{{}\widehat{\mathbf A}^n}^{y(1)\pm}_{\pm}}^{\mathsf T} - \langle \mathbf x^{\pm} {\mathbf x^{(2)\pm}}^{\mathsf T} \rangle {{{}\widehat{\mathbf A}^n}^{y(1)\pm}_{(2)\pm}}^{\mathsf T} \label{eq:inhomogeneous-diffusion-integrated-a} \\
{{}\widehat{\mathbf A}^n}^{y(1)+}_{-} \langle \mathbf x^{-} {\mathbf x^{-}}^{\mathsf T} \rangle + {{}\widehat{\mathbf A}^n}^{y(1)+}_{(2)-} \langle \mathbf x^{(2)-} {\mathbf x^{-}}^{\mathsf T} \rangle &= \langle \mathbf x^{+} {\mathbf x^{+}}^{\mathsf T} \rangle {{{}\widehat{\mathbf A}^n}^{y(1)-}_{+}}^{\mathsf T} + \langle \mathbf x^{+} {\mathbf x^{(2)+}}^{\mathsf T} \rangle {{{}\widehat{\mathbf A}^n}^{y(1)-}_{(2)+}}^{\mathsf T} \label{eq:inhomogeneous-diffusion-integrated-b}.
\end{align}
Similarly as before, the base case $n=1$ of Eq.\ \eqref{eq:inhomogeneous-diffusion-integrated-a} is an assumption needed for time-reversal symmetry, while the base case $n = 1$ of Eq.\ \eqref{eq:inhomogeneous-diffusion-integrated-b} is a consequence of Eq.\ \eqref{eq:diffusion-y-odd}, as:
\begin{equation}
	\lim_{\tau \to 0^+} \frac{\langle \mathbf x^{-}(0) \mathbf x^{+}(\tau)^{\mathsf T} (y(\tau)-y(0)) \rangle}{\tau} = \lim_{\tau \to 0^+} \frac{\langle \mathbf x^{-}(\tau) \mathbf x^{+}(0)^{\mathsf T} (y(\tau)-y(0)) \rangle}{\tau} = \langle \mathbf x^{-} {\mathbf x^{+}}^{\mathsf T} \dot y \rangle.
\end{equation}
Eq.\ \eqref{eq:inhomogeneous-diffusion-integrated-a}\textendash \eqref{eq:inhomogeneous-diffusion-integrated-b} can then be shown by mathematical induction on $n$. This establishes the time-reversibility of $\langle \Delta y(\tau) \mathbf x(\tau) \mathbf x(0)^{\mathsf T} \rangle$.

For the final demonstration, we note that Eq.\ \eqref{eq:linear-dynamics-integrated-matrix} still holds. We also have:
\begin{equation}
\frac{\mathrm d}{\mathrm d \tau} \begin{pmatrix}
1 \\
\langle \mathbf x^{+} \Delta y \rangle \\
\langle \mathbf x^{-} \Delta y \rangle \\
\langle \mathbf x^{(2)+} \Delta y \rangle \\
\langle \mathbf x^{(2)-} \Delta y \rangle
\end{pmatrix} = \begin{pmatrix}
0 & & & & \\
{\widehat{\mathbf A}}^{y(1)+}_{(0)} & \mathbf A^{+}_{+} & \mathbf A^{+}_{-} & & \\
{\widehat{\mathbf A}}^{y(1)-}_{(0)} & \mathbf A^{-}_{+} & \mathbf A^{-}_{-} & & \\
{\widehat{\mathbf A}}^{y(2)+}_{(0)} & \widehat{\mathbf A}^{(2)+}_{+} & \widehat{\mathbf A}^{(2)+}_{-} & \widehat{\mathbf A}^{(2)+}_{(2)+} & \widehat{\mathbf A}^{(2)+}_{(2)-} \\
{\widehat{\mathbf A}}^{y(2)-}_{(0)} & \widehat{\mathbf A}^{(2)-}_{+} & \widehat{\mathbf A}^{(2)-}_{-} & \widehat{\mathbf A}^{(2)-}_{(2)+} & \widehat{\mathbf A}^{(2)-}_{(2)-}
\end{pmatrix} \begin{pmatrix}
1 \\
\langle \mathbf x^{+} \Delta y \rangle \\
\langle \mathbf x^{-} \Delta y \rangle \\
\langle \mathbf x^{(2)+} \Delta y \rangle \\
\langle \mathbf x^{(2)-} \Delta y \rangle
\end{pmatrix},
\end{equation}
where the superscript $y(2)\pm$ stands for $\mathbf x^{(2)\pm} \Delta y$. To have time-reversibility of the third-order covariance functions, we must have for $n \ge 1$:
\begin{align}
{{}\widehat{\mathbf A}^n}^{y(2)+}_{(0)} &= -\langle \mathbf x^{(2)+} {\mathbf x^{+}}^{\mathsf T} \rangle {{{}\widehat{\mathbf A}^n}^{y}_{+}}^{\mathsf T} \label{eq:inhomogeneous-diffusion-integrated-c} \\
{{}\widehat{\mathbf A}^n}^{y(2)-}_{(0)} &= \langle \mathbf x^{(2)-} {\mathbf x^{-}}^{\mathsf T} \rangle {{{}\widehat{\mathbf A}^n}^{y}_{-}}^{\mathsf T} \label{eq:inhomogeneous-diffusion-integrated-d}
\end{align}
Again similarly, the case $n=1$ for Eq.\ \eqref{eq:inhomogeneous-diffusion-integrated-c} is an assumption needed for time-reversal symmetry, while the case $n = 1$ for Eq.\ \eqref{eq:inhomogeneous-diffusion-integrated-d} is a consequence of Eqs.\ \eqref{eq:diffusion-y-odd} and \eqref{eq:third-order-odd}. Eq.\ \eqref{eq:inhomogeneous-diffusion-integrated-c}\textendash \eqref{eq:inhomogeneous-diffusion-integrated-d} can be shown by mathematical induction on $n$. This establishes the time-reversibility of $\langle \Delta y(\tau) \mathbf x^{(2)}(0) \rangle$.

We may also consider the case of two integrated variables. We will not discuss that here, but we expect similar methods to be applicable and similar results to hold.

\subsection{Nonlinear drift}
Now for dynamics obeying Eqs.\ \eqref{eq:third-order-model-a}\textendash \eqref{eq:third-order-model-b}, in place of Eq.\ \eqref{eq:inhomogeneous-diffusion-mat} we have the multivariate analog of Eq.\ \eqref{eq:Hen-dynamics}:
\begin{equation}
\left\langle{\frac{\mathrm d}{\mathrm d \tau} \begin{pmatrix}
1 \\
\mathbf x^{+} \\
\mathbf x^{-} \\
\mathbf x^{(2)+} \\
\mathbf x^{(2)-} \\
\mathbf x^{(3)+} \\
\mathbf x^{(3)-} \\
\vdots
\end{pmatrix} \mid \mathbf x}\right\rangle = \begin{pmatrix}
0 & & & & & & & \\
& \mathbf A^{+}_{+} & \mathbf A^{+}_{-} & {\widehat{\mathbf A}}^{+}_{(2)+} & {\widehat{\mathbf A}}^{+}_{(2)-} & & & \\
& \mathbf A^{-}_{+} & \mathbf A^{-}_{-} & {\widehat{\mathbf A}}^{-}_{(2)+} & {\widehat{\mathbf A}}^{-}_{(2)-} & & & \\
& {\widehat{\mathbf A}}^{(2)+}_{+} & {\widehat{\mathbf A}}^{(2)+}_{-} & {\widehat{\mathbf A}}^{(2)+}_{(2)+} & {\widehat{\mathbf A}}^{(2)+}_{(2)-} & {\widehat{\mathbf A}}^{(2)+}_{(3)+} & {\widehat{\mathbf A}}^{(2)+}_{(3)-} & \\
& {\widehat{\mathbf A}}^{(2)-}_{+} & {\widehat{\mathbf A}}^{(2)-}_{-} & {\widehat{\mathbf A}}^{(2)-}_{(2)+} & {\widehat{\mathbf A}}^{(2)-}_{(2)-} & {\widehat{\mathbf A}}^{(2)-}_{(3)+} & {\widehat{\mathbf A}}^{(2)-}_{(3)-} & \\
{\widehat{\mathbf A}}^{(3)+}_{(0)} & {\widehat{\mathbf A}}^{(3)+}_{+} & {\widehat{\mathbf A}}^{(3)+}_{-} & {\widehat{\mathbf A}}^{(3)+}_{(2)+} & {\widehat{\mathbf A}}^{(3)+}_{(2)-} & {\widehat{\mathbf A}}^{(3)+}_{(3)+} & {\widehat{\mathbf A}}^{(3)+}_{(3)-} & \ddots \\
{\widehat{\mathbf A}}^{(3)-}_{(0)} & {\widehat{\mathbf A}}^{(3)-}_{+} & {\widehat{\mathbf A}}^{(3)-}_{-} & {\widehat{\mathbf A}}^{(3)-}_{(2)+} & {\widehat{\mathbf A}}^{(3)-}_{(2)-} & {\widehat{\mathbf A}}^{(3)-}_{(3)+} & {\widehat{\mathbf A}}^{(3)-}_{(3)-} & \ddots \\
\vdots & \vdots & \vdots & \vdots & \vdots & \ddots & \ddots & \ddots
\end{pmatrix} \begin{pmatrix}
1 \\
\mathbf x^{+} \\
\mathbf x^{-} \\
\mathbf x^{(2)+} \\
\mathbf x^{(2)-} \\
\mathbf x^{(3)+} \\
\mathbf x^{(3)-} \\
\vdots
\end{pmatrix}.
\end{equation}
We have that for $n \ge 1$, $i \ge 1$, $j \ge 1$:
\begin{align}
	{{}\widehat{\mathbf A}^n}^{(i)}_{(i+j)} &= \mathcal O(a^j) [1 + a \mathcal O(a,b)], \label{eq:Aiij} \\
	{{}\widehat{\mathbf A}^n}^{(i+1)}_{(i)} &= \mathcal O(a,b), \label{eq:Ai1i}
\end{align}
which can be proven by induction on $n$\footnote{The index ``(1)'' refers to $\mathbf x$ and is sometimes omitted, and the index ``(0)'' refers to the constant 1, so that $\langle \mathrm d \mathbf x^{(i)}(\tau)/\mathrm d \tau \mid \mathbf x(\tau) \rangle = \sum_{j=0}^{\infty} {\widehat{\mathbf A}}^{(i)}_{(j)} \mathbf x^{(j)}(\tau)$ and $\langle \mathbf x^{(i)}(\tau) \mid \mathbf x(0) \rangle = \sum_{j=0}^{\infty} (e^{\widehat{\mathbf A} \tau})^{(i)}_{(j)} \mathbf x^{(j)}(0)$.}. Also:
\begin{align}
	\langle \mathbf x(\tau) \rangle &= a^2 \mathcal O(a,b) \\
	\langle \mathbf x^{(2)}(\tau) \rangle &= a \mathcal O(a,b) \\
	\langle \mathbf x^{(3)}(\tau) \mathbf x(0)^{\mathsf T} \rangle &= \mathcal O((a,b)^2) \\
	\langle \mathbf x^{(4)}(\tau) \mathbf x(0)^{\mathsf T} \rangle &= \mathcal O(a,b).
\end{align}
Now, we have:
\begin{equation}
	\langle \mathbf x(\tau) \rangle = (e^{\widehat{\mathbf A} \tau})^{(1)}_{(0)} + (e^{\widehat{\mathbf A} \tau})^{(1)}_{(1)} \langle \mathbf x(0) \rangle + (e^{\widehat{\mathbf A} \tau})^{(1)}_{(2)} \langle \mathbf x^{(2)}(0) \rangle + (e^{\widehat{\mathbf A} \tau})^{(1)}_{(3)} \langle \mathbf x^{(3)}(0) \rangle + \cdots,
\end{equation}
with:
\begin{equation}
	\langle \mathbf x(0) \rangle = a^2 \mathcal O(a,b), \quad \langle \mathbf x^{(2)}(0) \rangle = a \mathcal O(a,b), \quad \langle \mathbf x^{(3)}(0) \rangle = \mathcal O(a,b).
\end{equation}
We Taylor expand $e^{\widehat{\mathbf A} \tau}$ and use Eqs.\ \eqref{eq:Aiij}\textendash \eqref{eq:Ai1i} to conclude that for $n \ge 1$:
\begin{equation}
	{{}\widehat{\mathbf A}^n}^{(1)}_{(0)} = a^2 \mathcal O(a,b).
\end{equation}
Similarly, we have:
\begin{align}
	{{}\widehat{\mathbf A}^n}^{(2)}_{(0)} &= a \mathcal O(a,b) \\
	{{}\widehat{\mathbf A}^n}^{(3)}_{(1)} &= \mathcal O((a,b)^2) \\
	{{}\widehat{\mathbf A}^n}^{(4)}_{(1)} &= \mathcal O(a,b).
\end{align}
We need for $n \ge 1$:
\begin{align}
{{}\widehat{\mathbf A}^n}^{(2)\pm}_{\pm} \langle \mathbf x^{\pm} {\mathbf x^{\pm}}^{\mathsf T} \rangle + {{}\widehat{\mathbf A}^n}^{(2)\pm}_{(2)\pm} \langle \mathbf x^{(2)\pm} {\mathbf x^{\pm}}^{\mathsf T} \rangle &= \langle \mathbf x^{(2)\pm} {\mathbf x^{\pm}}^{\mathsf T} \rangle {{{}\widehat{\mathbf A}^n}^{\pm}_{\pm}}^{\mathsf T} + \langle \mathbf x^{(2)\pm} {\mathbf x^{(2)\pm}}^{\mathsf T} \rangle {{{}\widehat{\mathbf A}^n}^{\pm}_{(2)\pm}}^{\mathsf T} + a \mathcal O((a,b)^2) \label{eq:non-linear-a} \\
{{}\widehat{\mathbf A}^n}^{(2)\pm}_{\mp} \langle \mathbf x^{\mp} {\mathbf x^{\mp}}^{\mathsf T} \rangle + {{}\widehat{\mathbf A}^n}^{(2)\pm}_{(2)\mp} \langle \mathbf x^{(2)\mp} {\mathbf x^{\mp}}^{\mathsf T} \rangle &= -\langle \mathbf x^{(2)\pm} {\mathbf x^{\pm}}^{\mathsf T} \rangle {{{}\widehat{\mathbf A}^n}^{\mp}_{\pm}}^{\mathsf T} - \langle \mathbf x^{(2)\pm} {\mathbf x^{(2)\pm}}^{\mathsf T} \rangle {{{}\widehat{\mathbf A}^n}^{\mp}_{(2)\pm}}^{\mathsf T} + a \mathcal O((a,b)^2). \label{eq:non-linear-b}
\end{align} 
Similar as before, the base case $n=1$ for Eq.\ \eqref{eq:non-linear-a} is an assumption needed for time-reversal symmetry, while the base case $n = 1$ for Eq.\ \eqref{eq:non-linear-b} follows from Eqs.\ \eqref{eq:third-order-odd}\textendash \eqref{eq:diffusion-odd}. To carry out the proof by induction for Eq.\ \eqref{eq:non-linear-a}\textendash \eqref{eq:non-linear-b}, we need to use, for $i \ge 1$ and $n \ge 1$:
\begin{equation}
{{}\widehat{\mathbf A}^n}^{(i)}_{(i)} = {{{}\widehat{\mathbf A}}^{(i)}_{(i)}}^n + a \mathcal O(a,b)
\end{equation}
which follows from Eqs.\ \eqref{eq:Aiij}\textendash \eqref{eq:Ai1i}. We also need to use that $\widehat{\mathbf A}^{(2)}_{(2)} = \mathbf A \otimes \mathbf A$ which means that the same rules of time-reversibility that apply to $\mathbf x$ and $\mathbf A$ also apply to $\mathbf x^{(2)}$ and $\widehat{\mathbf A}^{(2)}_{(2)}$ (respectively). This establishes the time-reversibility of $\langle x^i(\tau) x^j(0) x^k(0) \rangle$ with correction $a \mathcal O((a,b)^2)$.

\subsection{Integrated variables}

For an integrated variable $y$ obeying Eq.\ \eqref{eq:integrated-nonlinear}, we have for $n \ge 1$, $i \ge 1$, $j \ge 1$:
\begin{equation}
	{{}\widehat{\mathbf A}^n}^{y(i-1)}_{(i+j)} = \mathcal O(a^j)[1 + a \mathcal O(a,b)],
\end{equation}
where the superscript ``$y(i)$'' refers to $\mathbf x^{(i)} \Delta y$ (``$y(0)$'' being synonymous with ``$y$''). Also, for $n \ge 1$, $i \ge 0$, $j \ge 0$:
\begin{equation}
	{{}\widehat{\mathbf A}^n}^{y(i)}_{y(j)} = {{}\widehat{\mathbf A}^n}^{(i)}_{(j)},
\end{equation}
and for $i \ge 1$:
\begin{equation}
	{{}\widehat{\mathbf A}^n}^{y(i)}_{(i)} = \mathcal O(a,b).
\end{equation}
We also have $\langle \dot y \rangle = a^2 \mathcal O(a,b)$, which implies that $\langle \Delta y(\tau) \rangle = a^2 \mathcal O(a,b)$, and hence:
\begin{equation}
	{{}\widehat{\mathbf A}^n}^{y}_{(0)} = a^2 \mathcal O(a,b).
\end{equation}
Additionally:
\begin{align}
	\langle \Delta y(\tau) \mathbf x^{(2)}(\tau) \rangle &= \mathcal O(a,b) \\
	\langle \Delta y(\tau) \mathbf x^{(3)}(\tau) \mathbf x(0)^{\mathsf T} \rangle &= \mathcal O(a,b),
\end{align}
which implies:
\begin{align}
	{{}\widehat{\mathbf A}^n}^{y(2)}_{(0)} &= \mathcal O(a,b), \\
	{{}\widehat{\mathbf A}^n}^{y(3)}_{(1)} &= \mathcal O(a,b).
\end{align}
We now assume for simplicity that $y$ is of even parity. We also need to use the fact that under the assumption of time-symmetric second moments, fourth-order moments are also time-symmetric with error $\mathcal O((a,b)^2)$ from Isserlis's theorem. This is expressed as:
\begin{align}
{{}\widehat{\mathbf A}^n}^{y(2)\pm}_{\pm} \langle \mathbf x^{\pm} {\mathbf x^{\pm}}^{\mathsf T} \rangle &= -\langle \mathbf x^{(2)\pm} {\mathbf x^{(2)\pm}}^{\mathsf T} \rangle {{{}\widehat{\mathbf A}^n}^{y(1)\pm}_{(2)\pm}}^{\mathsf T} + \mathcal O((a,b)^2) \\
{{}\widehat{\mathbf A}^n}^{y(2)\pm}_{\mp} \langle \mathbf x^{\mp} {\mathbf x^{\mp}}^{\mathsf T} \rangle &= \langle \mathbf x^{(2)\pm} {\mathbf x^{(2)\pm}}^{\mathsf T} \rangle {{{}\widehat{\mathbf A}^n}^{y(1)\mp}_{(2)\pm}}^{\mathsf T} + \mathcal O((a,b)^2)
\end{align}
for $n \ge 1$. The above can be used to show that, for $n \ge 1$:
\begin{align}
	{{}\widehat{\mathbf A}^n}^{y(1)\pm}_{\pm} \langle \mathbf x^{\pm} {\mathbf x^{\pm}}^{\mathsf T} \rangle + {{}\widehat{\mathbf A}^n}^{y(1)\pm}_{(2)\pm} \langle \mathbf x^{(2)\pm} {\mathbf x^{\pm}}^{\mathsf T} \rangle &= -\langle \mathbf x^{\pm} {\mathbf x^{\pm}}^{\mathsf T} \rangle {{{}\widehat{\mathbf A}^n}^{y(1)\pm}_{\pm}}^{\mathsf T} - \langle \mathbf x^{\pm} {\mathbf x^{(2)\pm}}^{\mathsf T} \rangle {{{}\widehat{\mathbf A}^n}^{y(1)\pm}_{(2)\pm}}^{\mathsf T} + a \mathcal O((a,b)^2) \label{eq:y1samesign} \\
	{{}\widehat{\mathbf A}^n}^{y(1)+}_{-} \langle \mathbf x^{-} {\mathbf x^{-}}^{\mathsf T} \rangle + {{}\widehat{\mathbf A}^n}^{y(1)+}_{(2)-} \langle \mathbf x^{(2)-} {\mathbf x^{-}}^{\mathsf T} \rangle &= \langle \mathbf x^{+} {\mathbf x^{+}}^{\mathsf T} \rangle {{{}\widehat{\mathbf A}^n}^{y(1)-}_{+}}^{\mathsf T} + \langle \mathbf x^{+} {\mathbf x^{(2)+}}^{\mathsf T} \rangle {{{}\widehat{\mathbf A}^n}^{y(1)-}_{(2)+}}^{\mathsf T} + a \mathcal O((a,b)^2), \label{eq:y1diffsign}
\end{align}
where the base case $n=1$ for Eq.\ \eqref{eq:y1samesign} follows from assuming $L(\mathbf x^{(2)+},y)=0$ and using Eq.\ \eqref{eq:L-alt}, while the base case $n=1$ for Eq.\ \eqref{eq:y1diffsign} follows from Eq.\ \eqref{eq:diffusion-y-odd}. This shows the time-reversibility of the covariance function $\langle \Delta y(\tau) \mathbf x(\tau) \mathbf x(0)^{\mathsf T} \rangle$. For $\langle \Delta y(\tau) \mathbf x^{(2)}(0) \rangle$, we need to show that for $n \ge 1$:
\begin{equation}
	{{}\widehat{\mathbf A}^n}^{y(2)\pm}_{(0)} + {{}\widehat{\mathbf A}^n}^{y(2)\pm}_{(3)+} \langle {\mathbf x}^{(3)+} \rangle = \mp \langle \mathbf x^{(2)\pm} {\mathbf x^{\pm}}^{\mathsf T} \rangle {{{}\widehat{\mathbf A}^n}^y_{\pm}}^{\mathsf T} \mp \langle \mathbf x^{(2)\pm} {\mathbf x^{(2)\pm}}^{\mathsf T} \rangle {{{}\widehat{\mathbf A}^n}^y_{(2)\pm}}^{\mathsf T} + a \mathcal O((a,b)^2).
\end{equation}
where the base case $n=1$ for the upper sign follows from assuming $L(\mathbf x^{(2)+},y)=0$ and for the lower sign follows from Eq.\ \eqref{eq:diffusion-y-odd} and \eqref{eq:third-order-odd}. Here, we need to use:
\begin{align}
	{{}\widehat{\mathbf A}^n}^{(4)}_{(0)} &= \mathcal O((a,b)^2), \\
	{{}\widehat{\mathbf A}^n}^{(5)}_{(0)} &= \mathcal O(a,b), \\
	{{}\widehat{\mathbf A}^n}^{y(3)}_{(0)} &= \mathcal O((a,b)^2), \\
	{{}\widehat{\mathbf A}^n}^{y(4)}_{(0)} &= \mathcal O(a,b).
\end{align}
We also need to use the identity from stationarity:
\begin{equation}
	0 = \frac{\mathrm d}{\mathrm d \tau} \langle \mathbf x^{(3)} \rangle = {{}\widehat{\mathbf A}}^{(3)}_{(0)} + {{}\widehat{\mathbf A}}^{(3)}_{(3)} \langle \mathbf x^{(3)} \rangle + a \mathcal O((a,b)^2).
\end{equation}
With these, together with the detailed balance conditions to linear order, the equality is readily proven.

\subsection{Second-order covariance functions to quadratic order}

We now consider the case where second-order time-antisymmetric quantities are $a^2 \mathcal O((a,b)^2)$ and third-order time-antisymmetric quantities are $a \mathcal O((a,b)^2)$. We seek to show that second-order covariance functions obey time-reversal symmetry, with $a^2 \mathcal O((a,b)^2)$ corrections. We need to show that, for $n \ge 1$:
\begin{align}
	{{}\widehat{\mathbf A}^n}^{\pm}_{\pm} \langle \mathbf x^{\pm} {\mathbf x^{\pm}}^{\mathsf T} \rangle + {{}\widehat{\mathbf A}^n}^{\pm}_{(2)\pm} \langle \mathbf x^{(2)\pm} {\mathbf x^{\pm}}^{\mathsf T} \rangle &= \langle \mathbf x^{\pm} {\mathbf x^{\pm}}^{\mathsf T} \rangle {{{}\widehat{\mathbf A}^n}^{\pm}_{\pm}}^{\mathsf T} + \langle \mathbf x^{\pm} {\mathbf x^{(2)\pm}}^{\mathsf T} \rangle {{{}\widehat{\mathbf A}^n}^{\pm}_{(2)\pm}}^{\mathsf T} + a^2 \mathcal O((a,b)^2), \label{eq:second-order-same-sign} \\
	{{}\widehat{\mathbf A}^n}^{+}_{-} \langle \mathbf x^{-} {\mathbf x^{-}}^{\mathsf T} \rangle + {{}\widehat{\mathbf A}^n}^{+}_{(2)-} \langle \mathbf x^{(2)-} {\mathbf x^{-}}^{\mathsf T} \rangle &= -\langle \mathbf x^{+} {\mathbf x^{+}}^{\mathsf T} \rangle {{{}\widehat{\mathbf A}^n}^{-}_{+}}^{\mathsf T} - \langle \mathbf x^{+} {\mathbf x^{(2)+}}^{\mathsf T} \rangle {{{}\widehat{\mathbf A}^n}^{-}_{(2)+}}^{\mathsf T} + a^2 \mathcal O((a,b)^2), \label{eq:second-order-opp-sign}
\end{align}
where the base case $n=1$ is assumed for Eq.\ \eqref{eq:second-order-same-sign}, and for Eq.\ \eqref{eq:second-order-opp-sign} follows from differentiating $\langle \mathbf x^{+} {\mathbf x^{-}}^{\mathsf T} \rangle$. Using the properties stated in section 7.4, the induction step is readily performed. For the case of an integrated variable $y$ of even parity, we additionally need the equalities in section 7.5 and:
\begin{equation}
	{{}\widehat{\mathbf A}^n}^{y(1)\pm}_{(0)} + {{}\widehat{\mathbf A}^n}^{y(1)\pm}_{(2)+} \langle \mathbf x^{(2)+} \rangle + {{}\widehat{\mathbf A}^n}^{y(1)\pm}_{(3)+} \langle \mathbf x^{(3)+} \rangle = \mp \langle \mathbf x^{\pm} {\mathbf x^{\pm}}^{\mathsf T} \rangle {{{}\widehat{\mathbf A}^n}^y_{\pm}}^{\mathsf T} \mp \langle \mathbf x^{\pm} {\mathbf x^{(2)\pm}}^{\mathsf T} \rangle {{{}\widehat{\mathbf A}^n}^y_{(2)\pm}}^{\mathsf T} + a^2 \mathcal O((a,b)^2).
\end{equation}
We also need to use, from stationarity:
\begin{equation}
	0 = \frac{\mathrm d}{\mathrm d \tau} \langle \mathbf x^{(2)} \rangle = {{}\widehat{\mathbf A}}^{(2)}_{(2)} \langle \mathbf x^{(2)} \rangle + {{}\widehat{\mathbf A}}^{(2)}_{(3)} \langle \mathbf x^{(3)} \rangle + a^2 \mathcal O((a,b)^2).
\end{equation}

\section{Underdamped processes}

\subsection{Linear Gaussian processes: Covariance functions and detailed balance}
With the discussion of odd variables, we are ready to introduce processes obeying a second-order Langevin equation where the variable with non-zero quadratic variation is not the (observable) state or ``position'' variable $\mathbf x$, but its (unobservable) time derivative (i.e., ``velocity'') $\mathbf v := \dot{\mathbf x}$. We restrict our discussion to ``position'' variables that are even under time reversal. We start with linear Gaussian processes:
\begin{equation}
\dot{\mathbf v} = \mathbf A_{\mathbf x} \mathbf x + \mathbf A_{\mathbf v} \mathbf v + \boldsymbol \xi, \quad \langle \boldsymbol \xi(t) {\boldsymbol \xi(t')}^{\mathsf T} \rangle = 2 \mathbf D \delta(t-t'),
\end{equation}
where $\boldsymbol \xi$ is zero-mean Gaussian white noise and all coefficients are constant. The first property of interest is:
\begin{equation}
\frac{\mathrm d}{\mathrm d \tau} \langle \mathbf x(\tau) \mathbf x(\tau)^{\mathsf T} \rangle = \mathbf 0 = \langle \mathbf x \mathbf v^{\mathsf T} \rangle + \langle \mathbf v \mathbf x^{\mathsf T} \rangle,
\end{equation}
so that $\langle \mathbf x \mathbf v^{\mathsf T} \rangle$ is antisymmetric. We want to understand the quantities that characterize the system in terms of the covariance function. Its derivative is the position-velocity covariance function:
\begin{equation}
\frac{\mathrm d}{\mathrm d \tau} \langle \mathbf x(\tau) \mathbf x(0)^{\mathsf T} \rangle = \langle \mathbf v(\tau) \mathbf x(0)^{\mathsf T} \rangle,
\end{equation}
or equivalently:
\begin{equation}
\frac{\mathrm d}{\mathrm d \tau} \langle \mathbf x(0) \mathbf x(-\tau)^{\mathsf T} \rangle = -\langle \mathbf x(0) \mathbf v(-\tau)^{\mathsf T} \rangle = -\langle \mathbf x(\tau) \mathbf v(0)^{\mathsf T} \rangle.
\end{equation}
By similar reasoning, the second derivative of the position-position covariance function is negative the velocity-velocity covariance function:
\begin{equation}
\frac{\mathrm d^2}{\mathrm d \tau^2} \langle \mathbf x(\tau) \mathbf x(0)^{\mathsf T} \rangle = -\langle \mathbf v(\tau) \mathbf v(0)^{\mathsf T} \rangle.
\end{equation}
The quantities that characterize this system are then the position-position covariances $\langle \mathbf x \mathbf x^{\mathsf T} \rangle$, constrained by symmetry, the position-velocity covariances $\langle \mathbf x \mathbf v^{\mathsf T} \rangle$, constrained by antisymmetry, the velocity-velocity covariances $\langle \mathbf v \mathbf v^{\mathsf T} \rangle$, constrained by symmetry, and the quantities $\langle \dot{\mathbf v} \mathbf v^{\mathsf T} \rangle$ interpreted in It\^o sense. In place of $\langle \dot{\mathbf v} \mathbf v^{\mathsf T} \rangle$, we may take the angular momenta $L(v^i, v^j)$ together with the diffusivity $\mathbf D$.

For the condition for detailed balance, Eq.\ \eqref{eq:A+-} gives:
\begin{equation} \label{eq:vvxx}
\langle \mathbf v \mathbf v^{\mathsf T} \rangle = -\langle \mathbf x \mathbf x^{\mathsf T} \rangle \mathbf A_{\mathbf x}^{\mathsf T},
\end{equation}
while Eq.\ \eqref{eq:A++} gives:
\begin{equation} \label{eq:Avvv}
\mathbf A_{\mathbf v} \langle \mathbf v {\mathbf v}^{\mathsf T} \rangle = \langle \mathbf v {\mathbf v}^{\mathsf T} \rangle \mathbf A_{\mathbf v}^{\mathsf T}.
\end{equation}
However, Eq.\ \eqref{eq:vvxx} is not properly a condition solely for detailed balance, since contained in it is also the solution for $\langle \mathbf x \mathbf x^{\mathsf T} \rangle$, which cannot be solved by Eq.\ \eqref{eq:covariance-matrix}. Rather, the proper statement of the condition for detailed balance is given by manipulating Eq.\ \eqref{eq:vvxx} to obtain:
\begin{equation} \label{eq:Axvv}
\mathbf A_{\mathbf x} \langle \mathbf v \mathbf v^{\mathsf T} \rangle = \langle \mathbf v \mathbf v^{\mathsf T} \rangle \mathbf A_{\mathbf x}^{\mathsf T}.
\end{equation}
These two conditions Eqs.\ \eqref{eq:Avvv}\textendash \eqref{eq:Axvv} then provide the correct number of equalities needed for detailed balance.

\subsection{Almost Markovian dynamics}
Now we specialize to the situation where the dynamics of $\mathbf x$ is almost Markovian. To this end, we make the substitutions $\mathbf A_{\mathbf v} \to \mu \mathbf A_{\mathbf v}$ and $\mathbf D \to \mu \mathbf D$ and put $\mu^2 \gg 1$. We now introduce multivariate generalizations of Koopman modes:
\begin{align}
\widetilde{\mathbf x} &= \mathbf x - \mu^{-1} \mathbf A_{\mathbf v}^{-1} \mathbf v + \mathcal O(\mu^{-3}), \\
\widetilde{\mathbf v} &= \mathbf v + \mu^{-1} \mathbf A_{\mathbf v}^{-1} \mathbf A_{\mathbf x} \mathbf x + \mathcal O(\mu^{-3}),
\end{align}
which satisfy:
\begin{align}
\dot{\widetilde{\mathbf x}} &= (-\mu^{-1} \mathbf A_{\mathbf v}^{-1} \mathbf A_{\mathbf x} + \mathcal O(\mu^{-3})) \widetilde{\mathbf x}, \\
\dot{\widetilde{\mathbf v}} &= (\mu \mathbf A_{\mathbf v} + \mu^{-1} \mathbf A_{\mathbf v}^{-1} \mathbf A_{\mathbf x} + \mathcal O(\mu^{-3})) \widetilde{\mathbf v}.
\end{align}
We see that the condition $\mu^2 \gg 1$ amounts to separation of time-scales. More specifically, the squares of the real parts of the eigenvalues of $\mu \mathbf A_{\mathbf v}$ should be large compared to the complex moduli of the eigenvalues of $\mu^{-1} \mathbf A_{\mathbf v}^{-1} \mathbf A_{\mathbf x}$. We shall assume that the real parts of the eigenvalues of $\mathbf A_{\mathbf v}$ and of $\mathbf A_{\mathbf v}^{-1} \mathbf A_{\mathbf x}$ are at least of order 1 in magnitude. The inverse transformation is given by:
\begin{align}
\mathbf x &= \widetilde{\mathbf x} + \mu^{-1} \mathbf A_{\mathbf v}^{-1} \widetilde{\mathbf v} - \mu^{-2} \mathbf A_{\mathbf v}^{-2} \mathbf A_{\mathbf x} \widetilde{\mathbf x} + \mathcal O(\mu^{-3}), \\
\mathbf v &= \widetilde{\mathbf v} - \mu^{-1} \mathbf A_{\mathbf v}^{-1} \mathbf A_{\mathbf x} \widetilde{\mathbf x} - \mu^{-2} \mathbf A_{\mathbf v}^{-1} \mathbf A_{\mathbf x} \mathbf A_{\mathbf v}^{-1} \widetilde{\mathbf v} + \mathcal O(\mu^{-3}).
\end{align}
We can then solve for the modified covariances:
\begin{align}
\langle \widetilde x^i \widetilde x^j \rangle &= 2 \left[{(\mathbf A_{\mathbf v}^{-1} \mathbf A_{\mathbf x} \otimes \mathds 1 + \mathds 1 \otimes \mathbf A_{\mathbf v}^{-1} \mathbf A_{\mathbf x})^{-1}}\right]^{ij}_{kl} \left({\mathbf A_{\mathbf v}^{-1} \mathbf D {\mathbf A_{\mathbf v}^{-1}}^{\mathsf T}}\right)^{kl} + \mathcal O(\mu^{-2}) \label{eq:underdamped-xx} \\
\langle \widetilde v^i \widetilde v^j \rangle &= -2 \left[{(\mathbf A_{\mathbf v} \otimes \mathds 1 + \mathds 1 \otimes \mathbf A_{\mathbf v})^{-1}}\right]^{ij}_{kl} D^{kl} + \mathcal O(\mu^{-2}) \label{eq:underdamped-vv} \\
\langle \widetilde{\mathbf x} \widetilde{\mathbf v}^{\mathsf T} \rangle &= 2\mu^{-1} \mathbf A_{\mathbf v}^{-1} \mathbf D {\mathbf A_{\mathbf v}^{-1}}^{\mathsf T} + \mathcal O(\mu^{-3}).
\end{align}
The condition for detailed balance reads:
\begin{align}
\mathbf A_{\mathbf v}^{-1} \mathbf A_{\mathbf x} \langle \widetilde{\mathbf x} \widetilde{\mathbf x}^{\mathsf T} \rangle &= \langle \widetilde{\mathbf x} \widetilde{\mathbf x}^{\mathsf T} \rangle (\mathbf A_{\mathbf v}^{-1} \mathbf A_{\mathbf x})^{\mathsf T} \label{eq:detailed-balance-mu-x} \\
\mathbf A_{\mathbf v} \langle \widetilde{\mathbf v} \widetilde{\mathbf v}^{\mathsf T} \rangle &= \langle \widetilde{\mathbf v} \widetilde{\mathbf v}^{\mathsf T} \rangle \mathbf A_{\mathbf v}^{\mathsf T}. \label{eq:detailed-balance-mu-v}
\end{align}
The actual covariances are:
\begin{align}
\langle \mathbf x \mathbf x^{\mathsf T} \rangle &= \langle \widetilde{\mathbf x} \widetilde{\mathbf x}^{\mathsf T} \rangle + \mathcal O(\mu^{-2}) \\
\langle \mathbf v \mathbf v^{\mathsf T} \rangle &= \langle \widetilde{\mathbf v} \widetilde{\mathbf v}^{\mathsf T} \rangle + \mathcal O(\mu^{-2}) \\
\langle \mathbf x \mathbf v^{\mathsf T} \rangle &= \langle \widetilde{\mathbf x} \widetilde{\mathbf v}^{\mathsf T} \rangle - \mu^{-1} \langle \widetilde{\mathbf x} \widetilde{\mathbf x}^{\mathsf T} \rangle (\mathbf A_{\mathbf v}^{-1} \mathbf A_{\mathbf x})^{\mathsf T} + \mu^{-1} \mathbf A_{\mathbf v}^{-1} \langle \widetilde{\mathbf v} \widetilde{\mathbf v}^{\mathsf T} \rangle + \mathcal O(\mu^{-3}).
\end{align}
We see that if the detailed balance conditions Eq.\ \eqref{eq:detailed-balance-mu-x}\textendash \eqref{eq:detailed-balance-mu-v} are satisfied, then because $\langle \widetilde{\mathbf x} \widetilde{\mathbf v}^{\mathsf T} \rangle = \langle \widetilde{\mathbf v} \widetilde{\mathbf x}^{\mathsf T} \rangle$ it follows that $\langle \mathbf x \mathbf v^{\mathsf T} \rangle$ is symmetric (with error $\mathcal O(\mu^{-3})$), and therefore zero since it is also antisymmetric. We also note that $\langle \mathbf x \mathbf v^{\mathsf T} \rangle = \mathcal O(\mu^{-1})$, which makes sense because $\mathbf x$ changes on a time-scale $\mathcal O(\mu)$. It also means that its magnitude cannot be judged by the products of $\langle \mathbf x \mathbf x^{\mathsf T} \rangle$ and $\langle \mathbf v \mathbf v^{\mathsf T} \rangle$. Rather, we can write:
\begin{equation}
\begin{aligned}
\langle \mathbf x \mathbf v^{\mathsf T} \rangle &= \frac{\langle \mathbf x \mathbf v^{\mathsf T} \rangle - \langle \mathbf v \mathbf x^{\mathsf T} \rangle}{2} \\
&= \mu^{-1} \frac{\mathbf A_{\mathbf v}^{-1} \mathbf A_{\mathbf x} \langle \widetilde{\mathbf x} \widetilde{\mathbf x}^{\mathsf T} \rangle - \langle \widetilde{\mathbf x} \widetilde{\mathbf x}^{\mathsf T} \rangle (\mathbf A_{\mathbf v}^{-1} \mathbf A_{\mathbf x})^{\mathsf T}}{2} + \mu^{-1} \frac{\mathbf A_{\mathbf v}^{-1} \langle \widetilde{\mathbf v} \widetilde{\mathbf v}^{\mathsf T} \rangle - \langle \widetilde{\mathbf v} \widetilde{\mathbf v}^{\mathsf T} \rangle {\mathbf A_{\mathbf v}^{-1}}^{\mathsf T}}{2} + \mathcal O(\mu^{-3}) \\
&= \frac 1 2 \mathbf{L}(\widetilde{\mathbf x}, \widetilde{\mathbf x}^{\mathsf T}) + \frac 1 2 \mu^{-2} \mathbf A_{\mathbf v}^{-1} \mathbf{L}(\widetilde{\mathbf v}, \widetilde{\mathbf v}^{\mathsf T}) {\mathbf A_{\mathbf v}^{-1}}^{\mathsf T} + \mathcal O(\mu^{-3}).
\end{aligned}
\end{equation}
Following the discussion in the first section (``Linear Gaussian stationary systems''), we may judge the significance of this quantity by comparing to $2 \mu^{-1} \mathbf A_{\mathbf v}^{-1} \mathbf D {\mathbf A_{\mathbf v}^{-1}}^{\mathsf T}$, i.e., for the ``ensemble covariance'':
\begin{equation}
	\langle x^i v^j \rangle \langle x^k v^l \rangle \sim 2 \mu^{-2} (\mathbf A_{\mathbf v}^{-1})^i_{i'} (\mathbf A_{\mathbf v}^{-1})^j_{j'} (\mathbf A_{\mathbf v}^{-1})^k_{k'} (\mathbf A_{\mathbf v}^{-1})^l_{l'} (D^{i'k'} D^{j'l'} - D^{i'l'} D^{j'k'}),
\end{equation}
where we used, in accordance with Eq.\ \eqref{eq:L-mag}:
\begin{equation}
	L(\widetilde x^i, \widetilde x^j) L(\widetilde v^k, \widetilde v^l) \sim 2 (\mathbf A_{\mathbf v}^{-1})^i_{i'} (\mathbf A_{\mathbf v}^{-1})^j_{j'} (D^{i'k} D^{j'l} - D^{i'l} D^{j'k}).
\end{equation}
The last covariance of interest is:
\begin{equation}
\langle \dot{\mathbf v} \mathbf v^{\mathsf T} \rangle = \mu \mathbf A_{\mathbf v} \langle \mathbf v \mathbf v^{\mathsf T} \rangle + \mathcal O(\mu^{-1}).
\end{equation}
We may evaluate quantitative significance of the velocity-velocity angular momenta as:
\begin{equation}
	L(v^i, v^j) L(v^k, v^l) \sim 2 (D^{ik} D^{jl} - D^{il} D^{jk}).
\end{equation}

Now we turn to the actual covariance functions. Similarly to the covariances, we have:
\begin{equation}
\langle \mathbf x(\tau) \mathbf x(0)^{\mathsf T} \rangle = \langle \widetilde{\mathbf x}(\tau) \widetilde{\mathbf x}(0)^{\mathsf T} \rangle + \mathcal O(\mu^{-2}) = \exp(-\mu^{-1} \mathbf A_{\mathbf v}^{-1} \mathbf A_{\mathbf x} \tau) \langle \mathbf x \mathbf x^{\mathsf T} \rangle + \mathcal O(\mu^{-2}).
\end{equation}
However, the expression on the r.h.s.\ results in a covariance function whose derivative at $\tau = 0$ is not antisymmetric. The antisymmetry of the derivative of the covariance function at $\tau = 0$ is expressed as:
\begin{equation}
	\left.{\frac{\mathrm d}{\mathrm d \tau} \langle \mathbf x(\tau) \mathbf x(0)^{\mathsf T} \rangle}\right|_{\tau=0} + \left.{\frac{\mathrm d}{\mathrm d \tau} \langle \mathbf x(-\tau) \mathbf x(0)^{\mathsf T} \rangle}\right|_{\tau=0} = \mathbf 0.
\end{equation}
Thus the $\mathcal O(\mu^{-2})$ correction has derivatives of order $\mathcal O(\mu^{-1})$. Similarly, the velocity-velocity covariance function is:
\begin{equation}
\langle \mathbf v(\tau) \mathbf v(0)^{\mathsf T} \rangle = \exp(\mu \mathbf A_{\mathbf v} \tau) \langle \mathbf v \mathbf v^{\mathsf T} \rangle + \mathcal O(\mu^{-2}).
\end{equation}
However, we also have:
\begin{equation}
\int_0^{\infty} \mathrm d \tau \, \langle v^i(\tau) v^j(0) \rangle + \int_0^{\infty} \mathrm d \tau \, \langle v^i(-\tau) v^j(0) \rangle = 0,
\end{equation}
if either $x^i$ or $x^j$ is stationary, meaning that the $\mathcal O(\mu^{-2})$ correction to the velocity-velocity covariance function has an integral of order $\mathcal O(\mu^{-1})$.

Finally, we address position-velocity covariance functions with a time-lag. For $\tau \gg \mathcal O(\mu^{-1})$, we have:
\begin{equation}
	\langle v^i(\tau) x^j(0) \rangle = \langle \dot{\widetilde x}^i(\tau) \widetilde x^j(0) \rangle + \mathcal O(\mu^{-3}),
\end{equation}
which suggests the criterion:
\begin{equation}
	\Delta_{\textrm{exp\textendash theo}} \langle v^i(\tau) x^j(0) \rangle \Delta_{\textrm{exp\textendash theo}} \langle v^k(\tau) x^l(0) \rangle \sim \mu^{-2} \widetilde D^{ik} \widetilde D^{jl}
\end{equation}
where $\widetilde{\mathbf D} := \mathbf A_{\mathbf v}^{-1} \mathbf D {\mathbf A_{\mathbf v}^{-1}}^{\mathsf T}$ and, as before, $\Delta_{\textrm{exp\textendash theo}}$ denotes the deviation of experimental to theoretical values. However, extrapolating the above to $\tau = 0$ and antisymmetrizing results in one-fourth the prescription for position-velocity covariances at zero time-lag. This suggests an interpolation of the form:
\begin{equation}
	\Delta_{\textrm{exp\textendash theo}} \langle v^i(\tau) x^j(0) \rangle \Delta_{\textrm{exp\textendash theo}} \langle v^k(\tau) x^l(0) \rangle \sim \mu^{-2} (\delta^i_{i'} + \exp(\mu \mathbf A_{\mathbf v} \tau)^i_{i'}) (\delta^k_{k'} + \exp(\mu \mathbf A_{\mathbf v} \tau)^k_{k'}) \widetilde D^{i'k'} \widetilde D^{jl}.
\end{equation}
As before, if $\langle \mathbf x \mathbf v^{\mathsf T} \rangle$ or $\mathbf L(\mathbf v, \mathbf v^{\mathsf T})$ exceed the ``reference'' values, this needs to be accounted for in the above comparison. However, this prescription may be somewhat problematic if $\mathbf A_{\mathbf v}$ has complex eigenvalues. In such a case, the r.h.s.\ of the above may be oscillatory and could be close to zero for $\tau \sim \mathcal O(\mu^{-1})$\footnote{We might consider ways to eliminate the complex eigenvalues. One idea is to use the time-symmetric quantity $\mathbf D \langle \mathbf v \mathbf v^{\mathsf T} \rangle^{-1}$ in place of $\mathbf A_{\mathbf v}$. However, this will in general change the real parts of the eigenvalues (see remarks at the end of section 2.1). Another idea is to use the eigendecomposition of $\mathbf A_{\mathbf v}$ and simply change all the eigenvalues to their real parts. However, this mapping cannot be continuously extended to non-diagonalizable matrices. To see this, consider a matrix with eigenvalues $\lambda$ and $\lambda + \delta \lambda$ with corresponding eigenvectors $\mathbf w$ and $\mathbf w + \delta \mathbf w$, where $\delta \lambda \ne 0$, $\delta \mathbf w$ are infinitesimal. Then $\delta \mathbf w/\delta \lambda$ is a generalized eigenvector corresponding to $\mathbf w$. Thus the matrix in question depends only on $\delta \mathbf w/\delta \lambda$, and not on the complex argument of $\delta \lambda$. However, the proposed procedure will have different results depending on whether $\delta \lambda$ is real or imaginary, and thus it cannot be continuously extended to this case.}.

\subsection{Third-order properties}
We now consider dynamics involving inhomogeneous diffusion and nonlinear drift, as follows:
\begin{align}
\langle \dot v^i \mid \mathbf x, \mathbf v \rangle &= (A_x)^i_j x^j + \mu (A_v)^i_j v^j + (a_{xx})^i_{jk} (x^j x^k - \langle x^j x^k \rangle) \nonumber \\
&\quad {} + 2 \mu (a_{xv})^i_{jk} (x^j v^k - \langle x^j v^k \rangle) + \mu (a_{vv})^i_{jk} (v^j v^k - \langle v^j v^k \rangle), \\
\left \langle \frac{\mathrm d [v^i, v^j](t)}{\mathrm d t} \mid \mathbf x(t), \mathbf v(t) \right \rangle &= 2 \mu D^{ij} + 2 \mu (b_x)^{ij}_k x^k + 2 \mu (b_v)^{ij}_k v^k,
\end{align}
where all coefficients are constant.

\subsubsection{Inference from data} We first note that the estimation of quantities $\langle f(\mathbf x, \mathbf v) \dot{\mathbf v} \rangle$ from measured time-series data $\mathbf x(n \Delta t)$, $n = 0, 1, 2, \ldots$ is not what might be na\"{\i}vely expected from time-discretization of $\mathbf v$ and $\dot{\mathbf v}$, but has a correction due to fluctuations \cite{Bruckner}. Aside from that, there is the issue of measuring quantities that are robust against measurement noise, for which the authors of Ref.\ \cite{Bruckner} consider quantities of the form:
\begin{equation}
\left \langle f\left({\mathbf x(t), \frac{\mathbf x(t+\Delta t) - \mathbf x(t-\Delta t)}{2 \Delta t}}\right) \frac{\mathbf x(t+\Delta t) - 2 \mathbf x(t) + \mathbf x(t-\Delta t)}{(\Delta t)^2} \right \rangle
\end{equation}
where this is computed for a number of ``basis functions'' $f(\mathbf x, \mathbf v)$ to obtain information about $\langle \dot{\mathbf v} \mid \mathbf x, \mathbf v \rangle$. For our purposes, we consider polynomial basis functions of degree at most 2. If we choose monomials, the resulting quantities are either even or odd under time reversal. To characterize third-order properties of the dynamics using such quantities, we can use $\langle x^i x^j x^k \rangle$, $\langle x^i x^j v^k \rangle$ which satisfy:
\begin{equation}
0 = \frac{\mathrm d}{\mathrm d t} \langle x^i(t) x^j(t) x^k(t) \rangle = \langle x^i x^j v^k \rangle + \langle x^i v^j x^k \rangle + \langle v^i x^j x^k \rangle,
\end{equation}
$\langle x^i v^j v^k \rangle$, $\widetilde{L}(x^i, v^j, v^k)$, $\langle v^i v^j v^k \rangle$, $L(v^i v^j, v^k)$ (satisfying Eq.\ \eqref{eq:LLL0}), $\langle x^i \, \mathrm d [v^j, v^k]/\mathrm d t \rangle$, and $\langle v^i \, \mathrm d [v^j, v^k]/\mathrm d t \rangle$. With the exception of the last two, these quantities can be straightforwardly estimated from time-lapse data without corrections due to fluctuations. The estimation of $\bm{\mathcal B}$ is derived in the Supplementary Material in \cite{Bruckner} and corrected in \cite{LowCommentBruckner}.

\subsubsection{Solution of third-order quantities} Next, we analyze how the aforementioned quantities depend on the $a$ and $b$ coefficients. We have, to order $\mathcal O(a,b)$:
\begin{equation}
\begin{aligned}
\langle \widetilde x^i \widetilde x^j \widetilde x^k \rangle &= 2 \left[{(\mathbf A_{\mathbf v}^{-1} \mathbf A_{\mathbf x} \otimes \mathds 1 \otimes \mathds 1 + \mathds 1 \otimes \mathbf A_{\mathbf v}^{-1} \mathbf A_{\mathbf x} \otimes \mathds 1 + \mathds 1 \otimes \mathds 1 \otimes \mathbf A_{\mathbf v}^{-1} \mathbf A_{\mathbf x})^{-1}}\right]^{ijk}_{i'j'k'} \\
&\quad {} \times \Big[{-(\mathbf A_{\mathbf v}^{-1})^{i'}_{i''} \left({(\widetilde a_{xx})^{i''}_{lm} \langle \widetilde x^{j'} \widetilde x^l \rangle \langle \widetilde x^{k'} \widetilde x^m \rangle + \mu (a_{xv})^{i''}_{lm} \langle \widetilde x^{j'} \widetilde x^l \rangle \langle \widetilde x^{k'} \widetilde v^m \rangle + \mu (a_{xv})^{i''}_{lm} \langle \widetilde x^{j'} \widetilde v^m \rangle \langle \widetilde x^{k'} \widetilde x^l \rangle}\right)}\Big. \\
&\qquad {} - (\mathbf A_{\mathbf v}^{-1})^{j'}_{j''} \left({(\widetilde a_{xx})^{j''}_{lm} \langle \widetilde x^{i'} \widetilde x^l \rangle \langle \widetilde x^{k'} \widetilde x^m \rangle + \mu (a_{xv})^{j''}_{lm} \langle \widetilde x^{i'} \widetilde x^l \rangle \langle \widetilde x^{k'} \widetilde v^m \rangle + \mu (a_{xv})^{j''}_{lm} \langle \widetilde x^{i'} \widetilde v^m \rangle \langle \widetilde x^{k'} \widetilde x^l \rangle}\right) \\
&\qquad {} - (\mathbf A_{\mathbf v}^{-1})^{k'}_{k''} \left({(\widetilde a_{xx})^{k''}_{lm} \langle \widetilde x^{i'} \widetilde x^l \rangle \langle \widetilde x^{j'} \widetilde x^m \rangle + \mu (a_{xv})^{k''}_{lm} \langle \widetilde x^{i'} \widetilde x^l \rangle \langle \widetilde x^{j'} \widetilde v^m \rangle + \mu (a_{xv})^{k''}_{lm} \langle \widetilde x^{i'} \widetilde v^m \rangle \langle \widetilde x^{j'} \widetilde x^l \rangle}\right) \\
&\qquad {} + (\mathbf A_{\mathbf v}^{-1})^{i'}_{i''} (\mathbf A_{\mathbf v}^{-1})^{j'}_{j''} (b_x)^{i''j''}_l \langle \widetilde x^{k'} \widetilde x^l \rangle + (\mathbf A_{\mathbf v}^{-1})^{i'}_{i''} (\mathbf A_{\mathbf v}^{-1})^{k'}_{k''} (b_x)^{i''k''}_l \langle \widetilde x^{j'} \widetilde x^l \rangle \\
&\qquad \Big.{{} + (\mathbf A_{\mathbf v}^{-1})^{j'}_{j''} (\mathbf A_{\mathbf v}^{-1})^{k'}_{k''} (b_x)^{j''k''}_l \langle \widetilde x^{i'} \widetilde x^l \rangle}\Big] + \mathcal O(\mu^{-1}) (a_{vv}, b_v) + \mathcal O(\mu^{-2}) (a_{xx}, a_{xv}, b_x)
\end{aligned}
\end{equation}
where
\begin{equation}
(\widetilde a_{xx})^i_{jk} := (a_{xx})^i_{jk} - (\mathbf A_{\mathbf v}^{-1} \mathbf A_{\mathbf x})^{l}_k (a_{xv})^{i}_{jl} - (\mathbf A_{\mathbf v}^{-1} \mathbf A_{\mathbf x})^{l}_j (a_{xv})^{i}_{kl}.
\end{equation}
We also have:
\begin{equation}
\begin{aligned}
\langle \widetilde x^i \widetilde x^j \widetilde v^k \rangle &= -2\mu^{-1}(\mathbf A_{\mathbf v}^{-1})^{k}_{k'} \Big({-(\mathbf A_{\mathbf v}^{-1})^{i}_{i'} (a_{xv})^{i'}_{lm} \langle \widetilde x^j \widetilde x^l \rangle \langle \widetilde v^k \widetilde v^m \rangle - (\mathbf A_{\mathbf v}^{-1})^{j}_{j'} (a_{xv})^{j'}_{lm} \langle \widetilde x^i \widetilde x^l \rangle \langle \widetilde v^k \widetilde v^m \rangle}\Big. \\
&\qquad {} + (\widetilde a_{xx})^{k'}_{lm} \langle \widetilde x^i \widetilde x^l \rangle \langle \widetilde x^j \widetilde x^m \rangle + \mu (a_{xv})^{k'}_{lm} \langle \widetilde x^i \widetilde x^l \rangle \langle \widetilde x^j \widetilde v^m \rangle + \mu (a_{xv})^{k'}_{lm} \langle \widetilde x^i \widetilde v^m \rangle \langle \widetilde x^j \widetilde x^l \rangle \\
&\qquad \Big.{{} - (\mathbf A_{\mathbf v}^{-1})^{i}_{i'} (b_x)^{i'k'}_l \langle \widetilde x^j \widetilde x^l \rangle - (\mathbf A_{\mathbf v}^{-1})^{j}_{j'} (b_x)^{j'k'}_l \langle \widetilde x^i \widetilde x^l \rangle}\Big) \\
&\quad {} + \mathcal O(\mu^{-2}) (a_{vv}, b_v) + \mathcal O(\mu^{-3}) (a_{xx}, a_{xv}, b_x),
\end{aligned} 
\end{equation}
\begin{equation}
\begin{aligned}
\langle \widetilde x^i \widetilde v^j \widetilde v^k \rangle &= -2 \left[{(\mathbf A_{\mathbf v} \otimes \mathds 1 + \mathds 1 \otimes \mathbf A_{\mathbf v})^{-1}}\right]^{jk}_{j'k'} \left({(a_{xv})^{j'}_{lm} \langle \widetilde x^i \widetilde x^l \rangle \langle \widetilde v^{k'} \widetilde v^m \rangle + (a_{xv})^{k'}_{lm} \langle \widetilde x^i \widetilde x^l \rangle \langle \widetilde v^{j'} \widetilde v^m \rangle + (b_x)^{j'k'}_{l} \langle \widetilde x^i \widetilde x^l \rangle}\right) \\
&\quad{} + \mathcal O(\mu^{-1}) (a_{vv}, b_v) + \mathcal O(\mu^{-2}) (a_{xx}, a_{xv}, b_x),
\end{aligned}
\end{equation}
\begin{equation}
\begin{aligned}
\langle \widetilde v^i \widetilde v^j \widetilde v^k \rangle &= -2 \left[{(\mathbf A_{\mathbf v} \otimes \mathds 1 \otimes \mathds 1 + \mathds 1 \otimes \mathbf A_{\mathbf v} \otimes \mathds 1 + \mathds 1 \otimes \mathds 1 \otimes \mathbf A_{\mathbf v})^{-1}}\right]^{ijk}_{i'j'k'} \\
&\quad{} \times \Big({(a_{vv})^{i'}_{lm} \langle \widetilde v^{j'} \widetilde v^l \rangle \langle \widetilde v^{k'} \widetilde v^m \rangle + (a_{vv})^{j'}_{lm} \langle \widetilde v^{i'} \widetilde v^l \rangle \langle \widetilde v^{k'} \widetilde v^m \rangle + (a_{vv})^{k'}_{lm} \langle \widetilde v^{i'} \widetilde v^l \rangle \langle \widetilde v^{j'} \widetilde v^m \rangle}\Big. \\
&\qquad \Big.{{} + (b_v)^{i'j'}_l \langle \widetilde v^{k'} \widetilde v^l \rangle + (b_v)^{i'k'}_l \langle \widetilde v^{j'} \widetilde v^l \rangle + (b_v)^{j'k'}_l \langle \widetilde v^{i'} \widetilde v^l \rangle}\Big) \\
&\quad{} + \mathcal O(\mu^{-1}) (a_{xv}, b_x) + \mathcal O(\mu^{-2}) (a_{vv}, b_v) + \mathcal O(\mu^{-3}) (a_{xx}, a_{xv}, b_x),
\end{aligned}
\end{equation}
\begin{align}
\langle x^i x^j x^k \rangle &= \langle \widetilde x^i \widetilde x^j \widetilde x^k \rangle + \mathcal O(\mu^{-2}), \label{eq:under-x-x-x} \\
\langle x^i x^j v^k \rangle &= \langle \widetilde x^i \widetilde x^j \widetilde v^k \rangle + \mu^{-1}(\mathbf A_{\mathbf v}^{-1})^i_{i'} \langle \widetilde v^i \widetilde x^j \widetilde v^k \rangle + \mu^{-1}(\mathbf A_{\mathbf v}^{-1})^j_{j'} \langle \widetilde x^i \widetilde v^j \widetilde v^k \rangle \\
&\quad{} - \mu^{-1}(\mathbf A_{\mathbf v}^{-1} \mathbf A_{\mathbf x})^k_{k'} \langle \widetilde x^i \widetilde x^j \widetilde x^{k'} \rangle + \mu^{-2} (\mathbf A_{\mathbf v}^{-1})^i_{i'} (\mathbf A_{\mathbf v}^{-1})^j_{j'} \langle \widetilde v^{i'} \widetilde v^{j'} \widetilde v^k \rangle + \mathcal O(\mu^{-3}), \nonumber \\
\langle x^i v^j v^k \rangle &= \langle \widetilde x^i \widetilde v^j \widetilde v^k \rangle + \mu^{-1} (\mathbf A_{\mathbf v}^{-1})^i_{i'} \langle \widetilde v^{i'} \widetilde v^j \widetilde v^k \rangle + \mathcal O(\mu^{-2}), \\
\langle v^i v^j v^k \rangle &= \langle \widetilde v^i \widetilde v^j \widetilde v^k \rangle - \mu^{-1} (\mathbf A_{\mathbf v}^{-1} \mathbf A_{\mathbf x})^i_{i'} \langle \widetilde x^{i'} \widetilde v^j \widetilde v^k \rangle - \mu^{-1} (\mathbf A_{\mathbf v}^{-1} \mathbf A_{\mathbf x})^j_{j'} \langle \widetilde v^i \widetilde x^{j'} \widetilde v^k \rangle \label{eq:under-v-v-v} \\
&\quad{} - \mu^{-1} (\mathbf A_{\mathbf v}^{-1} \mathbf A_{\mathbf x})^k_{k'} \langle \widetilde v^i \widetilde v^j \widetilde x^{k'} \rangle + \mathcal O(\mu^{-2}), \nonumber
\end{align}
where we have omitted factors of $a,b$ in Eqs.\ \eqref{eq:under-x-x-x}\textendash \eqref{eq:under-v-v-v}. From Eq.\ \eqref{eq:L-tilde} we can compute $\widetilde{L}(x^i, v^j, v^k)$:
\begin{equation}
\begin{aligned}
\widetilde{L}(x^i, v^j, v^k) &= \mu (\mathbf A_{\mathbf v})^k_{k'} \langle x^i v^j v^{k'} \rangle + \mu (a_{xv})^k_{lm} \langle x^i x^l \rangle \langle v^j v^m \rangle - \mu (\mathbf A_{\mathbf v})^j_{j'} \langle x^i v^{j'} v^k \rangle \\
&\quad{} - \mu (a_{xv})^j_{lm} \langle x^i x^l \rangle \langle v^k v^m \rangle + \mathcal O(\mu^0) (a_{vv}, b_v) + \mathcal O(\mu^{-1}) (a_{xx}, a_{xv}, b_x).
\end{aligned}
\end{equation}
Also:
\begin{equation}
\begin{aligned}
L(v^i v^j, v^k) &= \mu (\mathbf A_{\mathbf v})^k_{k'} \langle v^i v^j v^{k'} \rangle + 2 \mu (a_{vv})^k_{lm} \langle v^i v^l \rangle \langle v^j v^m \rangle - \mu (\mathbf A_{\mathbf v})^i_{i'} \langle v^{i'} v^j v^k \rangle \\
&\quad{} - 2 \mu (a_{vv})^i_{lm} \langle v^j v^l \rangle \langle v^k v^m \rangle - \mu (\mathbf A_{\mathbf v})^j_{j'} \langle v^i v^{j'} v^k \rangle - 2 \mu (a_{vv})^j_{lm} \langle v^i v^l \rangle \langle v^k v^m \rangle \\
&\quad{} - 2 \mu (b_v)^{ij}_l \langle v^k v^l \rangle  + \mathcal O(\mu^0) (a_{xv}, b_x) + \mathcal O(\mu^{-1}) (a_{vv}, b_v) + \mathcal O(\mu^{-2}) (a_{xx}, a_{xv}, b_x).
\end{aligned}
\end{equation}
We see that the considered quantities are partitioned into two groups: one which, to leading order in $\mu$, depends only on $a_{xx}$, $a_{xv}$, and $b_x$, and the other which, to leading order in $\mu$, depends only on $a_{vv}$ and $b_v$.

\subsubsection{Quantitative significance} To judge quantitative significance, we first consider the quantities $\langle x^i x^j x^k \rangle$ and $\langle v^i v^j v^k \rangle$. These are judged according to the procedure for third moments. For $\langle x^i v^j v^k \rangle$, we may use:
\begin{equation}
	\langle x^i v^j v^k \rangle \langle x^{i'} v^{j'} v^{k'} \rangle \sim \langle x^i x^{i'} \rangle \frac{\langle v^j v^{j'} \rangle \langle v^k v^{k'} \rangle + \langle v^j v^{k'} \rangle \langle v^k v^{j'} \rangle}{2}.
\end{equation}
The remaining quantities are angular momenta which exist only in multiple dimensions, including $\langle x^i x^j v^k \rangle = L(x^i x^j, x^k)/2$. For this quantity, we write $\mathbf x$ in terms of $\widetilde{\mathbf x}$ and $\widetilde{\mathbf v}$ and expand. We first note the magnitudes of the angular momenta:
\begin{align}
	L(\widetilde x^i \widetilde x^j, \widetilde x^k) &= \mathcal O(\mu^{-1}), \\
	L(\widetilde x^i \widetilde v^j, \widetilde x^k) &= \mathcal O(\mu^0), \\
	\widetilde{L}(\widetilde x^i, \widetilde v^j, \widetilde v^k) &= \mathcal O(\mu), \\
	L(\widetilde v^i \widetilde v^j, \widetilde x^k) &= \mathcal O(\mu^0), \\
	L(\widetilde v^i \widetilde v^j, \widetilde v^k) &= \mathcal O(\mu),
\end{align}
where we have omitted factors of $a,b$, and we can write:
\begin{align}
	L(\widetilde x^i \widetilde x^j, \widetilde v^k) &= -L(\widetilde x^i \widetilde v^k, \widetilde x^j) - L(\widetilde x^j \widetilde v^k, \widetilde x^i), \\
	L(\widetilde x^i \widetilde v^j, \widetilde v^k) &= \widetilde{L}(\widetilde x^i, \widetilde v^j, \widetilde v^k) - \frac 1 2 L(\widetilde v^j \widetilde v^k, \widetilde x^i).
\end{align}
At this point, if we use the ensemble covariance of angular momenta, we will get zero since we have merely rewritten $L(x^i x^j, x^k)$. Therefore, we need to introduce some equalities, similarly to the case of linear Gaussian dynamics where we have:
\begin{equation}
	L(\widetilde x^i, (\mathbf A_{\mathbf v}^{-1} \widetilde{\mathbf v})^j) = L(\widetilde x^j, (\mathbf A_{\mathbf v}^{-1} \widetilde{\mathbf v})^i) + \mathcal O(\mu^{-2}).
\end{equation}
In the case of third-order dynamics, we have, to order $\mathcal O(a,b)$:
\begin{align}
	\widetilde{L}(\widetilde x^i, \widetilde v^j, \widetilde v^k) &= \mu \langle \widetilde x^i \widetilde v^j (\mathbf A_{\mathbf v} \widetilde{\mathbf v})^k \rangle - \mu \langle \widetilde x^i (\mathbf A_{\mathbf v} \widetilde{\mathbf v})^j \widetilde v^k \rangle - \frac \mu 2 \left[{L(\widetilde x^i \widetilde v^j, (\mathbf A_{\mathbf v} \widetilde{\mathbf x})^k) - L(\widetilde x^i \widetilde v^k, (\mathbf A_{\mathbf v} \widetilde{\mathbf x})^j)}\right] + \mathcal O(\mu^{-1}), \label{eq:Ltilde-xvv} \\
	\langle \widetilde x^i \widetilde v^j \widetilde v^k \rangle &= \frac 1 2 \left[{(\mathbf A_{\mathbf v} \otimes \mathds 1 + \mathds 1 \otimes \mathbf A_{\mathbf v})^{-1}}\right]^{jk}_{j'k'} \nonumber \\
	&\quad{} \times \left[{L(\widetilde x^i \widetilde v^{j'}, (\mathbf A_{\mathbf v} \widetilde{\mathbf x})^{k'}) + L(\widetilde x^i \widetilde v^{k'}, (\mathbf A_{\mathbf v} \widetilde{\mathbf x})^{j'}) - \mu^{-1} L(\widetilde v^{j'} \widetilde v^{k'}, \widetilde x^i)}\right] + \mathcal O(\mu^{-2}). \label{eq:xvv}
\end{align}
where Eq.\ \eqref{eq:xvv} is substituted into Eq.\ \eqref{eq:Ltilde-xvv}. The resulting equality can be substituted into the expression for $L(x^i x^j, x^k)$ and the ensemble covariance of angular momenta applied. It should be mentioned that the $\mu^{-2} L(\widetilde v^i \widetilde v^j, \widetilde x^k)$ and $\mu^{-3} L(\widetilde v^i \widetilde v^j, \widetilde v^k)$ terms contribute only $\mathcal O(\mu^{-4})$ to the ensemble covariance of $L(x^i x^j, x^k)$. For $\widetilde{L}(x^i, v^j, v^k)$ and $L(v^i v^j, v^k)$, we can use the usual prescriptions without issue. It is worth noting that using the above equality and using the prescription for the angular momenta $L(\widetilde x^i \widetilde v^j, \widetilde x^k)$ gives a different expression for the ``ensemble covariance'':
\begin{equation}
	\begin{aligned}
		\widetilde{L}(x^i, v^j, v^k) \widetilde{L}(x^{i'}, v^{j'}, v^{k'}) &\sim 2 \mu^2 \left[{(\mathbf A_{\mathbf v}^{-1} \otimes \mathds 1 + \mathds 1 \otimes \mathbf A_{\mathbf v}^{-1})^{-1}}\right]^{jk}_{lm} \left[{(\mathbf A_{\mathbf v}^{-1} \otimes \mathds 1 + \mathds 1 \otimes \mathbf A_{\mathbf v}^{-1})^{-1}}\right]^{j'k'}_{l'm'} \\
		&\quad{} \times \langle x^i x^{i'} \rangle \Big[{D^{ll'} \left({\mathbf A_{\mathbf v}^{-1} \mathbf D {\mathbf A_{\mathbf v}^{-1}}^{\mathsf T}}\right)^{mm'} - \left({\mathbf D {\mathbf A_{\mathbf v}^{-1}}^{\mathsf T}}\right)^{lm'} (\mathbf A_{\mathbf v}^{-1} \mathbf D)^{ml'}}\Big. \\
		&\qquad{} - D^{lm'} \left({\mathbf A_{\mathbf v}^{-1} \mathbf D {\mathbf A_{\mathbf v}^{-1}}^{\mathsf T}}\right)^{ml'} + \left({\mathbf D {\mathbf A_{\mathbf v}^{-1}}^{\mathsf T}}\right)^{ll'} (\mathbf A_{\mathbf v}^{-1} \mathbf D)^{mm'} \\
		&\qquad{} - D^{ml'} \left({\mathbf A_{\mathbf v}^{-1} \mathbf D {\mathbf A_{\mathbf v}^{-1}}^{\mathsf T}}\right)^{lm'} + \left({\mathbf D {\mathbf A_{\mathbf v}^{-1}}^{\mathsf T}}\right)^{mm'} (\mathbf A_{\mathbf v}^{-1} \mathbf D)^{ll'} \\
		&\qquad{} + \Big.{D^{mm'} \left({\mathbf A_{\mathbf v}^{-1} \mathbf D {\mathbf A_{\mathbf v}^{-1}}^{\mathsf T}}\right)^{ll'} - \left({\mathbf D {\mathbf A_{\mathbf v}^{-1}}^{\mathsf T}}\right)^{ml'} (\mathbf A_{\mathbf v}^{-1} \mathbf D)^{lm'}}\Big] + \mathcal O(\mu^0),
	\end{aligned}
\end{equation}
which is also a reasonable choice. Lastly, it should be noted that $\langle x^i v^j v^k \rangle = L(x^i v^j, x^k)/2$ is also an angular momentum, but the calculated ensemble covariance using the angular momenta is $\mathcal O(\mu^{-2})$ when $\mathbf x$ and $\mathbf v$ are expressed in terms of $\widetilde {\mathbf x}$ and $\widetilde {\mathbf v}$ and when Eqs.\ \eqref{eq:Ltilde-xvv}\textendash \eqref{eq:xvv} are applied.

For the inhomogeneous diffusion, we have for the ``ensemble covariance'':
\begin{align}
	\left \langle x^i \frac{\mathrm d [v^j, v^k]}{\mathrm d t} \right \rangle \left \langle x^{i'} \frac{\mathrm d [v^{j'}, v^{k'}]}{\mathrm d t} \right \rangle &\sim 2 \mu^2 \langle x^i x^{i'} \rangle (D^{jj'} D^{kk'} + D^{jk'} D^{kj'}), \\
	\left \langle v^i \frac{\mathrm d [v^j, v^k]}{\mathrm d t} \right \rangle \left \langle v^{i'} \frac{\mathrm d [v^{j'}, v^{k'}]}{\mathrm d t} \right \rangle &\sim 2 \mu^2 \langle v^i v^{i'} \rangle (D^{jj'} D^{kk'} + D^{jk'} D^{kj'}).
\end{align}
Finally, we note that from the formula for the inverse of a $2 \times 2$ block matrix \cite{BernsteinMatrixMathematics}, the quantities $b_x$ and $b_v$ are even and odd under time reversal, respectively.

\subsection{Dynamics on long time-scales} We now show that under the assumption $\left|(a,b)\right| \lesssim \mathcal O(\mu^{-1})$, the dynamics of $\mathbf x$ is Markovian on $\mathcal O(\mu)$ time-scales with $\mathcal O(\mu^{-2})$ error. We do this by solving for the generalized Koopman eigenfunctions as in subsection 6.4. First, we have $\mathbf x = (\widetilde{\mathbf x} + \mu^{-1} \mathbf A_{\mathbf v}^{-1} \widetilde{\mathbf v})(1 + \mathcal O(\mu^{-2}))$. We will do computations with the transformed variables $\widetilde{\mathbf x}$, $\widetilde{\mathbf v}$. The generalized Koopman eigenfunctions with leading terms $\widetilde{\mathbf x}$ and $\widetilde{\mathbf v}$ are, respectively:
\begin{align}
	\mathbf f_{1 \widetilde{\mathbf x}} (\widetilde{\mathbf x}, \widetilde{\mathbf v}) &= \widetilde{\mathbf x} + \mathcal O(\mu^{-1}) a \mathcal O(a,b) \widetilde{\mathbf v} + \mathcal O(a) (\widetilde x^i \widetilde x^j - \langle \widetilde x^i \widetilde x^j \rangle) \nonumber \\
	&\quad{} + \mathcal O(\mu^{-1}) \mathcal O(a) (\widetilde x^i \widetilde v^j - \langle \widetilde x^i \widetilde v^j \rangle) + \mathcal O(\mu^{-1}) \mathcal O(a) (\widetilde v^i \widetilde v^j - \langle \widetilde v^i \widetilde v^j \rangle) + \cdots, \\
	\mathbf f_{1 \widetilde{\mathbf v}} (\widetilde{\mathbf x}, \widetilde{\mathbf v}) &= \widetilde{\mathbf v} + a \mathcal O(a,b) \widetilde{\mathbf x} + \mathcal O(\mu^{-1}) \mathcal O(a) (\widetilde x^i \widetilde x^j - \langle \widetilde x^i \widetilde x^j \rangle) \nonumber \\
	&\quad{} + \mathcal O(a) (\widetilde x^i \widetilde v^j - \langle \widetilde x^i \widetilde v^j \rangle) + \mathcal O(a) (\widetilde v^i \widetilde v^j - \langle \widetilde v^i \widetilde v^j \rangle) + \cdots,
\end{align}
with dynamics:
\begin{align}
	\left \langle \frac{\mathrm d}{\mathrm d t} \mathbf f_{1 \widetilde{\mathbf x}} (\widetilde{\mathbf x}, \widetilde{\mathbf v}) \mid \widetilde{\mathbf x}, \widetilde{\mathbf v} \right \rangle &= -\mu^{-1} \mathbf A_{\mathbf v}^{-1} \mathbf A_{\mathbf x} [1 + \mathcal O(\mu^{-2}) + \mathcal O(\mu) a \mathcal O(a,b)] \mathbf f_{1 \widetilde{\mathbf x}} (\widetilde{\mathbf x}, \widetilde{\mathbf v}), \\
	\left \langle \frac{\mathrm d}{\mathrm d t} \mathbf f_{1 \widetilde{\mathbf v}} (\widetilde{\mathbf x}, \widetilde{\mathbf v}) \mid \widetilde{\mathbf x}, \widetilde{\mathbf v} \right \rangle &= \mu \mathbf A_{\mathbf v} [1 + \mathcal O(\mu^{-2}) + a \mathcal O(a,b)] \mathbf f_{1 \widetilde{\mathbf v}} (\widetilde{\mathbf x}, \widetilde{\mathbf v}).
\end{align}
With these expressions in hand, we can proceed similarly to Eq.\ \eqref{eq:expec-x} and obtain, for $\tau > 0$:
\begin{align}
	\langle \widetilde{\mathbf x}(\tau) \mid \widetilde{\mathbf x}(0), \widetilde{\mathbf v}(0) \rangle &= \mathcal O(1) \widetilde{\mathbf x}(0) + \mathcal O(\mu^{-1}) a \mathcal O(a,b) \widetilde{\mathbf v}(0) + \mathcal O(a) (\widetilde x^i(0) \widetilde x^j(0) - \langle \widetilde x^i \widetilde x^j \rangle) \nonumber \\
	&\quad{} + \mathcal O(\mu^{-1}) \mathcal O(a) (\widetilde x^i(0) \widetilde v^j(0) - \langle \widetilde x^i \widetilde v^j \rangle) + \mathcal O(\mu^{-1}) \mathcal O(a) (\widetilde v^i(0) \widetilde v^j(0) - \langle \widetilde v^i \widetilde v^j \rangle) + \cdots, \label{eq:expec-x-tilde} \\
	\langle \widetilde{\mathbf v}(\tau) \mid \widetilde{\mathbf x}(0), \widetilde{\mathbf v}(0) \rangle &= \mathcal O(1) \widetilde{\mathbf v}(0) + a \mathcal O(a,b) \widetilde{\mathbf x}(0) + \mathcal O(\mu^{-1}) \mathcal O(a) (\widetilde x^i(0) \widetilde x^j(0) - \langle \widetilde x^i \widetilde x^j \rangle) \nonumber \\
	&\quad{} + \mathcal O(a) (\widetilde x^i(0) \widetilde v^j(0) - \langle \widetilde x^i \widetilde v^j \rangle) + \mathcal O(a) (\widetilde v^i \widetilde v^j - \langle \widetilde v^i(0) \widetilde v^j(0) \rangle) + \cdots. \label{eq:expec-v-tilde}
\end{align}
We now introduce another time-lag $\tau' > 0$ and consider the quantities:
\begin{equation}
	\langle x^i(\tau+\tau') x^j (\tau') \rangle, \quad \langle x^i (\tau+\tau') (x^j(\tau')-x^j(0)) \rangle.
\end{equation}
For $\tau, \tau' \sim \mathcal O(\mu)$, the above quantities vary by $\mathcal O(1)$. We translate Eqs.\ \eqref{eq:expec-x-tilde}\textendash \eqref{eq:expec-v-tilde} in time by $\tau'$ and apply the law of iterated expectations. We see that all terms in $\mathbf x(\tau')$ except for $\widetilde{\mathbf x}(\tau')$ contribute only $\mathcal O(\mu^{-2})$. We now consider the quantities:
\begin{equation}
	\langle x^i(\tau+\tau') x^j(\tau') x^k(\tau') \rangle, \quad \langle x^i(\tau+\tau') x^j(\tau') (x^k(\tau') - x^k(0)) \rangle, \quad \langle x^i(\tau+\tau') (x^j(\tau') - x^j(0)) (x^k(\tau') - x^k(0)) \rangle.
\end{equation}
We seek to show that all terms in $\mathbf x(\tau')$ involving $\widetilde{\mathbf v}(\tau')$ contribute $\mathcal O(\mu^{-2}) \mathcal O(a,b)$ to the above quantities. For this, we need to use:
\begin{align}
	\langle \widetilde v^i \widetilde x^j \widetilde x^k \widetilde x^l \rangle &= \mathcal O(\mu^{-1}), \\
	\langle \widetilde v^i(\tau') \widetilde x^j(0) \widetilde x^k(0) \widetilde x^l(0) \rangle &= \mathcal O(\mu^{-1}),
\end{align}
and similarly whenever any subset of the factors $\widetilde{\mathbf x}(0)$ is replaced by $\widetilde{\mathbf x}(\tau')$. Also,
\begin{align}
	\langle (\widetilde v^i \widetilde v^j - \langle \widetilde v^i \widetilde v^j \rangle) \widetilde x^k \widetilde x^l \rangle &= \mathcal O(\mu^{-2}) + \mathcal O((a,b)^2), \\
	\langle (\widetilde v^i(\tau') \widetilde v^j(\tau') - \langle \widetilde v^i \widetilde v^j \rangle) \widetilde x^k(0) \widetilde x^l(0) \rangle &= O(\mu^{-2}) + \mathcal O((a,b)^2),
\end{align}
and similarly when either factor $\widetilde{\mathbf x}(0)$ is replaced by $\widetilde{\mathbf x}(\tau')$. We have thus established that on $\mathcal O(\mu)$ time-scales, the conditional expectation of $\mathbf x$ is Markovian with $\mathcal O(\mu^{-2})$ error. Higher-order Hermite polynomials (e.g.\ Eq.\ \eqref{eq:expec-x-x}) can be handled similarly.

\section{Detection of non-Markovianity}

Stochastic processes exhibiting temporal heterogeneity have been detected in animal locomotion \cite{StephensLongTimescalesCElegans2011} and cell migration \cite{MetznerSuperstatisticalRandomWalks2015}. These may take the form of temporal trends \cite{StephensLongTimescalesCElegans2011} or stochastic variability in the parameters characterizing the stochastic process \cite{MetznerSuperstatisticalRandomWalks2015}. The latter case is known as ``superstatistics'' and usually incorporates variability in the process variance leading to leptokurtic distributions \cite{TimeSeriesToSuperstatistics, StatMechSuperstatistics}. However, the concept of superstatistics need not involve non-Gaussianity. For example, consider a linear Gaussian process with a fluctuating unobserved mean which itself varies according to a linear Gaussian process. This type of superstatistical process has been used to model cell migration in \cite{MitterwallnerNonMarkovianMotility2020}. In this study, the additional term due to the unobserved process was interpreted as a contribution to the effective noise, leading to a colored noise term.

The most general zero-mean linear Gaussian process in one variable ($x$) corresponding to a linear Markov Gaussian process in two variables, one of which is unobserved ($y$), is given by \cite{FerrettiRG2022, LowCommentFerretti2023}:
\begin{align}
	\begin{pmatrix}
		\dot x \\
		\dot y
	\end{pmatrix} &= \begin{pmatrix}
	-\lambda & 1 \\
	-\kappa & 0
	\end{pmatrix} \begin{pmatrix}
	x \\
	y
	\end{pmatrix} + \boldsymbol{\xi}, \\
	\langle \boldsymbol{\xi}(t) \boldsymbol{\xi}^{\mathsf T}(t') \rangle &= 2 \begin{pmatrix}
		D & 0 \\
		0 & D'
	\end{pmatrix} \delta (t-t'),
\end{align}
where $\boldsymbol \xi$ is zero-mean Gaussian white noise and all coefficients are constant. For stationarity, we require $\lambda>0$, $\kappa>0$. (We discuss the non-stationary case later.) This formulation allows for feedback $x \to y$, which means that the state of the system could potentially influence the unobserved process. If no such feedback is allowed, then the dynamics instead satisfies:
\begin{equation}
	\begin{pmatrix}
		\dot x \\
		\dot y
	\end{pmatrix} = \begin{pmatrix}
	-\lambda & 1 \\
	0 & -\eta
	\end{pmatrix} \begin{pmatrix}
	x \\
	y
	\end{pmatrix} + \boldsymbol{\xi},
\end{equation}
for constants $\lambda$, $\eta$, which means that the eigenvalues of the matrix of coefficients multiplying the state vector (called $\mathbf A$ in previous sections) are constrained to be real. We will hereafter work with the first formulation as the second is a special case.

The Gaussian process $x(t)$ is completely characterized by its covariance function $C(\tau) := \langle x(0) x(\tau) \rangle$:
\begin{equation}
	C(\tau) = \frac{1}{\mu_{+}-\mu_{-}} \left[{ \left({D + \frac{D'}{\kappa}}\right) \left({ \frac{\mu_{+}}{\lambda} e^{-\mu_{+}\tau} - \frac{\mu_{-}}{\lambda} e^{-\mu_{-}\tau} }\right) - \frac{D'}{\kappa} (e^{-\mu_{+}\tau} - e^{-\mu_{-}\tau}) }\right]
\end{equation}
($\tau \ge 0$), where $-\mu_{\pm}$ are the eigenvalues of $\mathbf A$:
\begin{equation}
	\mu_{\pm} = \frac{\lambda \pm \sqrt{\lambda^2 - 4 \kappa}}{2}.
\end{equation}
We define the correlation function $R(\tau) := C(\tau)/C(0)$. We now compare $R(\tau)$ with the correlation function predicted from a linear Markov process, i.e., $e^{\dot R(0) \tau}$. If these two are equal, then the process $x(t)$ is Markovian.

We see that $C(\tau)$ reduces to a single exponential decay when:
\begin{equation}
	\frac{D}{D'} = \frac{1}{\kappa} \left({\frac{\lambda}{\mu_{\pm}} - 1}\right).
\end{equation}
This case only occurs for real $\mu_{\pm}$. Writing $\delta_{\mp}$ for the r.h.s.\ of the above, it can be easily verified that when $\delta_{-} < D/D' < \delta_{+}$, we have $\ddot R(0) > (\dot R(0))^2$. Otherwise, if $0 \le D/D' < \delta_{-}$ or $\delta_{+} < D/D' \le \infty$, or if $\mu_{\pm}$ are complex conjugates, then we have $\ddot R(0) < (\dot R(0))^2$. Thus, we see that $\ddot R(0)$ provides a proxy for non-Markovianity. Although $C(\tau)$ is characterized by four parameters while a single exponential decay is characterized by two parameters, we only need a single equality to test for non-Markovianity. This occurs because when the coefficient of one of the exponential decays is zero, the value of the corresponding decay rate is irrelevant.

Alternatively, instead of $\ddot R(0)$, we may consider the integral $\int_0^T \mathrm d \tau \, R(\tau)$ for some $T>0$. As discussed in section 3.3, for the estimation of such integrals from experimental data, $T$ must be significantly less than the trajectory duration. We write:
\begin{equation}
	R(\tau) = c_{+} e^{-\mu_{+} \tau} + c_{-} e^{-\mu_{-} \tau}
\end{equation}
where $c_{+} + c_{-} = 1$. First, we consider the case of $\mu_{\pm}$ real and distinct. If $\delta_{-} < D/D' < \delta_{+}$, then $c_{+}>0$ and $c_{-}>0$. Hence, we can use Jensen's inequality to conclude that $R(\tau) > e^{\dot R(0) \tau}$ for $\tau>0$. If $0 \le D/D' < \delta_{-}$, then $c_{+}<0$ and $c_{-}>1$, and the inequality $R(\tau) < e^{\dot R(0) \tau}$ is equivalent to:
\begin{equation}
	c_{-} (e^{\Delta \mu \tau} - 1) < e^{c_{-} \Delta \mu \tau} - 1
\end{equation}
where $\Delta \mu := \mu_{+} - \mu_{-}$. The above inequality follows immediately from taking derivatives. Similarly, if $\delta_{+} < D/D' \le \infty$, then $c_{+}>1$ and $c_{-}<0$, and the same inequality holds. For the case $\mu_{+} = \mu_{-} = \lambda/2$, we have:
\begin{equation}
	R(\tau) = (1 + c' \tau) e^{-\lambda \tau/2},
\end{equation}
where $c'$ is a constant with $c' \ne 0$ unless $D/D' = 1/\kappa$. The inequality $R(\tau) < e^{\dot R(0) \tau}$ is then equivalent to:
\begin{equation}
	1 + c' \tau < e^{c' \tau},
\end{equation}
which again follows immediately from taking derivatives.

For the case of $\mu_{\pm}$ complex, we no longer have $R(\tau) < e^{\dot R(0) \tau}$ for all $\tau>0$. However, we have:
\begin{equation}
	\int_0^T \mathrm d \tau \, R(\tau) < \int_0^T \mathrm d \tau \, e^{\dot R(0) \tau}
\end{equation}
for all $0<T \le \infty$. To show this, we note that $R(\tau)$ is the product of a sinusoid and an exponential decay, which implies that:
\begin{equation}
	\int_0^T \mathrm d \tau \, R(\tau) < \int_0^{\tau_0} \mathrm d \tau \, R(\tau)
\end{equation}
for $\tau_0 < T \le \infty$, where $\tau_0$ is the first zero of $R(\tau)$. Thus it suffices to show that $R(\tau) < e^{\dot R(0) \tau}$ for $0<\tau<\tau_0$. Writing $\mu_{\pm} = \lambda/2 \pm i \omega$ and $c_{\pm} = (1 \pm ic)/2$, this inequality is equivalent to:
\begin{equation}
	\cos (\omega \tau) + c \sin (\omega \tau) < e^{c \omega \tau},
\end{equation}
which is true because the second derivative of the l.h.s.\ is negative for $0<\tau<\tau_0$, whereas the second derivative of the r.h.s.\ is non-negative.

Now, we discuss the case where $x(t)$ is integrated of order 1, i.e., $x(0)$ does not possess a stationary distribution but its differences $x(\tau)-x(0)$ do. We also assume that $x(t)$ has been detrended so that $\langle \dot x(t) \rangle = 0$. We treat this case by setting one of the eigenvalues of $\mathbf A$ to zero. We do not set both eigenvalues to zero simultaneously, since this would result in $x(t)$ being integrated of order 2, i.e., its differences $x(\tau)-x(0)$ would not possess a stationary distribution while its second differences $x(2\tau)-2x(\tau)+x(0)$ would. Thus we take $\kappa \to 0$ while $\lambda$ remains finite. The process $x(t)$ is now characterized by the mean squared displacement function\footnote{Alternatively $\langle \dot x(0) \dot x(\tau) \rangle$, which is one-half the second derivative of the mean squared displacement (Eq.\ \eqref{eq:long-time-D}).}:
\begin{equation}
	V(\tau) := \langle (x(\tau)-x(0))^2 \rangle = 2 [C(0) - C(\tau)],
\end{equation}
where the last equality holds for a stationary process. We may evaluate $V(\tau)$ using this equality and then taking the limit as $\kappa \to 0$. In this limit, $\mu_{+} = \lambda - \kappa/\lambda$ and $\mu_{-} = \kappa/\lambda$, and hence:
\begin{equation}
	V(\tau) = \frac{2}{\lambda} \left[{ \left({D - \frac{D'}{\lambda^2}}\right) (1 - e^{-\lambda \tau}) + \frac{D'}{\lambda} \tau }\right].
\end{equation}
For a linear Markov process, $V(\tau) = 2D\tau$. We see that $\ddot V(0) = 0$ if and only if $V(\tau) = 2D\tau$. Furthermore, $\ddot V(\tau)$ has the same sign as $\ddot V(0)$ and thus we may instead test using an integral of the velocity autocovariance function.

Of course, the aforementioned tests can only detect the particular form of non-Markovianity occurring for the simple model investigated, and not more complicated models. However, the results presented indicate that these are reasonable tests for detecting non-Markovianity. This approach can be used not only for the ``raw'' state variables, but also for the quantities characterizing a Markov process, i.e., quadratic functions, angular momenta, and diffusivities.

\section{Conclusions}
We have analyzed Langevin equations obeying or deviating from linear Gaussian dynamics. One of our main results is that, from a purely theoretical standpoint, lower-order covariance functions are insensitive to higher-order effects, whereas higher-order covariance functions are affected by lower-order effects. In particular, corrections to second-order covariance functions are quadratic in third-order coefficients (as required by symmetry, although this is also true for fourth-order coefficients), whereas second-order time-irreversibility affects third-order covariance functions to leading order.

We have used Koopman eigenfunctions to facilitate calculations in the case of integrated variables (written in terms of martingales) and nonlinear drift. In the latter case, we might consider these to be new coordinates under a nonlinear coordinate transformation, as these quantities obey a law of linear drift. However, the time-reversed dynamics of Koopman eigenfunctions do not necessarily obey a law of linear drift\footnote{The time-reversed dynamics has the same eigenvalues of the Perron\textendash Frobenius and Koopman operators. To see this, start from the Kolmogorov backward equation \cite{RiskenFokkerPlanck}: $\partial p(\mathbf x, t \mid \mathbf x', t')/\partial t' = -\mathcal K p(\mathbf x, t \mid \mathbf x', t')$ for $t'<t$, where $\mathcal K$ acts on $\mathbf x'$. Now use Bayes's rule to write $p(\mathbf x', t' \mid \mathbf x, t) = p(\mathbf x, t \mid \mathbf x', t') p(\mathbf x')/p(\mathbf x)$ and therefore $\partial p(\mathbf x', t' \mid \mathbf x, t)/\partial t' = -\mathcal L p(\mathbf x', t' \mid \mathbf x, t)$ ($\mathcal L$ acting on $\mathbf x'$) where $(\mathcal Lf)(\mathbf x) := p(\mathbf x) \mathcal K [f(\mathbf x)/p(\mathbf x)]$. It is easily seen that $\mathcal L$ is the adjoint of $\mathcal P$ with respect to the inner product defined by $(f, g) := \int \mathrm d \mathbf x \, f(\mathbf x)^* g(\mathbf x)/p(\mathbf x)$, and therefore $\mathcal L$ has the same eigenvalues as $\mathcal P$. (The time-reversed process is also Markov, as the backward time-evolution of the probability density depends only on its current values.)}. (For example, in the two-dimensional stationary dynamics presented in section 5, if $x$ and $y$ are Koopman eigenfunctions, then third-order covariance functions would be time-symmetric, which is not necessarily the case.) Hence, our use of Koopman eigenfunctions is restricted to being a calculational tool.

We have introduced criteria for judging quantitative significance, both for interesting effects such as non-Gaussianity or time-irreversibility, and for comparing experiment with theory. Our analysis takes linear Gaussian models as a reference point and utilizes second- and third-order covariance functions. Due to moment closure problem, our analysis is limited to asymptotic expansion in the drift nonlinearity. For strongly nonlinear drift, other methods may be more suitable \cite{StephensLongTimescalesCElegans2011, BrucknerLearning2024}.

We have used a limited form of ``stochastic force inference'' \cite{Frishman,Bruckner} where drift functions are fitted by quadratic polynomials and diffusion functions by linear polynomials. We have identified dimensionality as being a limiting factor in inference of dynamics. We note that in this framework generally, entropy production and information content are not well estimated because it involves the inverse of the diffusion matrix. In contrast to a Bayesian analysis \cite{Ferretti2020}, in our analysis, effects are evaluated on quantitative rather than statistical grounds. Instead of trying to estimate entropy production, we use moments of probability current density to quantify time-irreversible dynamics. Additionally, we have identified a characterization of stochastic dynamics based on time-symmetric and time-antisymmetric quantities, as an alternative to coefficients in the Langevin equation.

We expect that the framework presented in this work will be useful in analyses of high-dimensional stochastic systems. We note that extensions to our analysis are possible only up to fourth order (cubic polynomial for drift and quadratic polynomial for diffusion). Afterwards, the analysis procedure may fail (see Appendix B). Besides, analysis of moments beyond fourth order may be infeasible in biological applications.

\section{Appendix A}
For a complex variable $w$, we now have the possibility for angular momentum of a single variable:
\begin{equation}
	L(w, w^*) = -L(w^*, w) = -(L(w, w^*))^*
\end{equation}
($w^*$ being the complex conjugate of $w$), which is purely imaginary, but not necessarily zero. If $x$ is real, we also have that $\widetilde{L}(x, w, w^*)$ is purely imaginary. For judging quantitative significance, we need to treat $w$ and $w^*$ as separate variables. Otherwise, we use the same procedure. For example, suppose we have a complex variable $w$ obeying the symmetry $w \to e^{i \phi} w$ for any real $\phi$. We then have, for the ``ensemble variance'':
\begin{equation}
	\left({\Delta_{\textrm{exp\textendash theo}} \langle ww^* \rangle}\right)^2 \sim \frac{\langle w^2 \rangle \langle {w^*}^2 \rangle + \langle ww^* \rangle^2}{2} = \frac{\langle ww^* \rangle^2}{2},
\end{equation}
where $\Delta_{\textrm{exp\textendash theo}}$ denotes the deviation of experiment to theory, and the r.h.s.\ refers to theoretical values. The result is half of what might be expected from the case of real variables. To further justify this choice, we separate the real and imaginary parts: $w =: x + iy$. By symmetry, we have $\langle xy \rangle = 0$ and $\langle x^2 \rangle = \langle y^2 \rangle = \langle ww^* \rangle/2$. We have:
\begin{equation}
	\begin{aligned}
		\left({\Delta_{\textrm{exp\textendash theo}} \langle ww^* \rangle}\right)^2 &= \left({\Delta_{\textrm{exp\textendash theo}} \langle x^2 \rangle + \Delta_{\textrm{exp\textendash theo}} \langle y^2 \rangle}\right)^2 \\
		&= \left({\Delta_{\textrm{exp\textendash theo}} \langle x^2 \rangle}\right)^2 + \left({\Delta_{\textrm{exp\textendash theo}} \langle y^2 \rangle}\right)^2 + 2 \left({\Delta_{\textrm{exp\textendash theo}} \langle x^2 \rangle}\right) \left({\Delta_{\textrm{exp\textendash theo}} \langle y^2 \rangle}\right) \\
		&\sim \langle x^2 \rangle^2 + \langle y^2 \rangle^2 + 2 \langle xy \rangle^2 \\
		&= \frac{\langle ww^* \rangle^2}{2},
	\end{aligned}
\end{equation}
which matches our previous result. The objection may be raised, however, that under assumption of symmetry $w \to e^{i \phi} w$, we cannot consider the experimental measurements of $\langle x^2 \rangle$ and $\langle y^2 \rangle$ as uncorrelated. On the other hand, to maintain continuity with the case where symmetry is not obeyed, we have to accept the above result.

If we are interested in the real parts alone (or any combination of real and imaginary parts) of complex quantities, we may use the formula:
\begin{equation}
	\Re z_1 \Re z_2 = \frac{\Re [z_1 z_2^*] + \Re [z_1 z_2]}{2}.
\end{equation}
For example, if we have complex variables $w_1,w_2$ obeying the symmetry $w_i \to e^{i \phi} w_i$ for any real $\phi$ (where $w_1,w_2$ are simultaneously transformed), then we have:
\begin{equation}
	\left({\Re \Delta_{\textrm{exp\textendash theo}} \langle w_1 w_2^* \rangle}\right)^2 \sim \frac{\langle w_1 w_1^* \rangle \langle w_2 w_2^* \rangle + |\langle w_1 w_2^* \rangle|^2}{4}.
\end{equation}

Also, for complex Gaussian variables $z_1,z_2,z_3,z_4$, we have the complex version of Isserlis's theorem:
\begin{equation}
	\langle z_1 z_2 z_3 z_4 \rangle = \langle z_1 z_2 \rangle \langle z_3 z_4 \rangle + \langle z_1 z_3 \rangle \langle z_2 z_4 \rangle + \langle z_1 z_4 \rangle \langle z_2 z_3 \rangle,
\end{equation}
which follows from the real case using that the above is (with some abuse of terminology) a multilinear map (over $\mathbb C$). Explicitly, we can prove the case where $z_m$, $m \ge n$ are real by performing mathematical induction on $n$, where the induction step is done by separating real and imaginary parts of $z_n$.

\section{Appendix B}
\begin{figure}
	\begin{center}
		\includegraphics[scale=0.5]{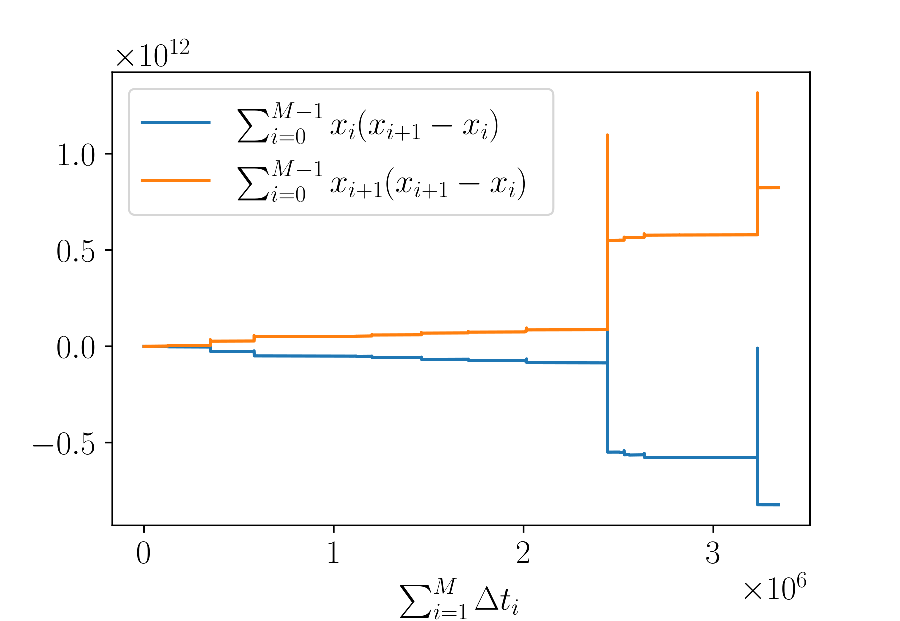}
	\end{center}
	\caption{Divergence of $\langle x \dot x \rangle$ for Eq.\ \eqref{eq:Langevin-weird}.}
	\label{fig:infinite}
\end{figure}

\begin{figure}
	\begin{center}
		\includegraphics[scale=0.5]{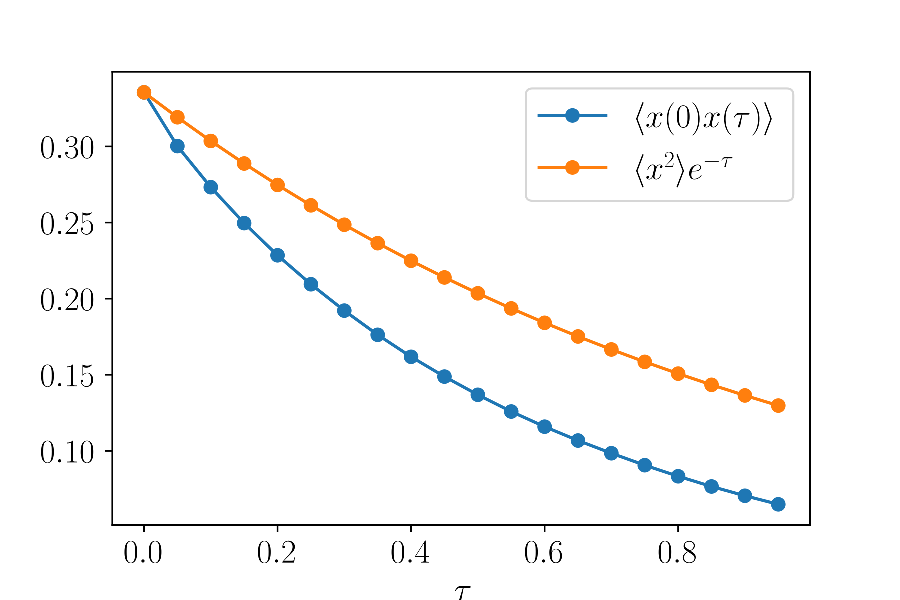}
	\end{center}
	\caption{Covariance function for Eq.\ \eqref{eq:Langevin-weird}.}
	\label{fig:cov-weird}
\end{figure}

\begin{figure}
	\begin{center}
		\includegraphics[scale=0.5]{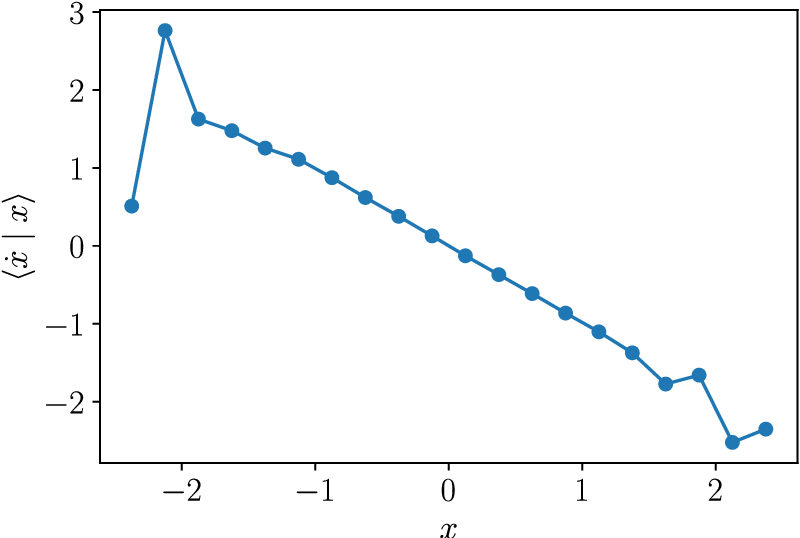}
	\end{center}
	\caption{Binned drift function for Eq.\ \eqref{eq:Langevin-weird}.}
	\label{fig:binned}
\end{figure}

Consider the Langevin equation:
\begin{equation} \label{eq:Langevin-weird}
	\langle \dot x \mid x \rangle = -x, \quad \left \langle \frac{\mathrm d[x,x]}{\mathrm d t} \mid x \right \rangle = 1+x^6.
\end{equation}
The stationary probability distribution is given by \cite{GardinerStochasticMethods} (calculated by WolframAlpha):
\begin{equation}
	p(x) \propto \frac{1}{1+x^6} \exp \left({-\frac{1}{3} \ln (x^2+1) + \frac{1}{6} \ln (x^4 - x^2 + 1) + \frac{\arctan(2x+\sqrt{3}) - \arctan(2x - \sqrt{3})}{\sqrt{3}}}\right).
\end{equation}
Applying It\^o's lemma and taking expectations apparently gives $2 \langle x^2 \rangle = -2 \langle x \dot x \rangle = 1 + \langle x^6 \rangle$. However, the l.h.s.\ is finite whereas the r.h.s.\ is infinite\footnote{The same problem is encountered if $x^6$ is replaced by $x^4$ in Eq.\ \eqref{eq:Langevin-weird}.}. The problem lies in the middle expression. Upon Euler\textendash Maruyama discretization with a fixed time-step, i.e., $x_{i+1} = (1 - \Delta t) x_i + \sqrt{(1+x_i^6) \Delta t} \cdot \zeta_i$ with $\zeta_i \sim \mathcal{N}(0,1)$ i.i.d., it would seem that $\langle x_i (x_{i+1}-x_i)/\Delta t \rangle$ should be $-\langle x^2 \rangle$ because $\langle x^4 \rangle < \infty$. However, with a fixed time-step $\Delta t$, the sequence $x_i$ diverges. To properly simulate this system, an adaptive time-step must be used, i.e., $\Delta t_i$ depending on $x_i$. We simulated this system using $\Delta t_i = 0.05 (1+x_i^4)^{-1}$ for $10^8$ time-steps. Numerically, it appears that $\langle x \dot x \rangle = -\infty$ (Fig.\ \ref{fig:infinite}), in accordance with Eq.\ \eqref{eq:Lyapunov}. Moreover, although the covariance function $\langle x(0) x(\tau) \rangle$ exists for all $\tau$, it is apparently not equal to $\langle x^2 \rangle e^{-\tau}$ (Fig.\ \ref{fig:cov-weird}, calculated using $3 \times 10^5$ time-steps), as might be expected from the drift function. Thus stochastic force inference fails when using a linear basis function. However, binning reproduces the correct drift (Fig.\ \ref{fig:binned}, calculated using $10^7$ time-steps). We thus conclude that there is a non-commutation of limits:
\begin{equation}
	\lim_{E \to \infty} \lim_{\tau \to 0^+} \int_{-E}^E \mathrm d x \, p(x) x \frac{\langle x(\tau) - x(0) \mid x(0) = x \rangle}{\tau} \ne \lim_{\tau \to 0^+} \lim_{E \to \infty} \int_{-E}^E \mathrm d x \, p(x) x \frac{\langle x(\tau) - x(0) \mid x(0) = x \rangle}{\tau}.
\end{equation}

\section{Code availability}
Code for the simulations is available at \texttt{https://github.com/yeerenlow/langevin}.

\section{Acknowledgments}
Y.I.L.\ acknowledges support from the McGill University Department of Physics. The idea of ignoring quantitatively small deviations between experiment and theory is due to Paul Fran\c{c}ois (Universit\'{e} de Montr\'{e}al).

\bibliographystyle{unsrt}
\bibliography{document2}

\begin{thebibliography}{10}

\bibitem{StephensDimensionalityDynamicsCElegans2008}
G.~J. Stephens, B.~Johnson-Kerner, W.~Bialek, and W.~S. Ryu.
\newblock Dimensionality and dynamics in the behavior of \textit{C. elegans}.
\newblock {\em PLoS Comput. Biol.}, 4:e1000028, 2008.

\bibitem{BrucknerLearning2024}
D.~B. {Br\"{u}ckner} and C.~P. Broedersz.
\newblock Learning dynamical models of single and collective cell migration: a
  review.
\newblock {\em Rep. Prog. Phys.}, 87:056601, 2024.

\bibitem{GardinerStochasticMethods}
C.~Gardiner.
\newblock {\em Stochastic Methods}.
\newblock Springer Berlin, Heidelberg, 4 edition, 2009.

\bibitem{Weiss}
J.~B. Weiss.
\newblock Coordinate invariance in stochastic dynamical systems.
\newblock {\em Tellus A}, 55:208--218, 2003.

\bibitem{Seifert2012}
U.~Seifert.
\newblock Stochastic thermodynamics, fluctuation theorems and molecular
  machines.
\newblock {\em Rep. Prog. Phys.}, 75:126001, 2012.

\bibitem{Dieball2022}
C.~Dieball and A.~Godec.
\newblock Mathematical, thermodynamical, and experimental necessity for corase
  graining empirical densities and currents in continuous space.
\newblock {\em Phys. Rev. Lett.}, 129:140601, 2022.

\bibitem{Battle2016}
C.~Battle, C.~P. Broedersz, N.~Fakhri, V.~F. Geyer, J.~Howard, C.~F. Schmidt,
  and F.~C. MacKintosh.
\newblock Broken detailed balance at mesoscopic scales in active biological
  systems.
\newblock {\em Science}, 352:604--607, 2016.

\bibitem{Gonzalez2019}
J.~P. Gonzalez, J.~C. Neu, and S.~W. Teitsworth.
\newblock Experimental metrics for detection of detailed balance violation.
\newblock {\em Phys. Rev. E}, 99:022143, 2019.

\bibitem{Frishman}
A.~Frishman and P.~Ronceray.
\newblock Learning force fields from stochastic trajectories.
\newblock {\em Phys. Rev. X}, 10:021009, 2020.

\bibitem{Bruckner}
D.~B. {Br\"{u}ckner}, P.~Ronceray, and C.~P. Broedersz.
\newblock Inferring the dynamics of underdamped stochastic systems.
\newblock {\em Phys. Rev. Lett.}, 125:058103, 2020.

\bibitem{SelmecziCellMotility2008}
D.~Selmeczi, L.~Li, L.~I.~I. Pedersen, S.~F. N\o{}rrelykke, P.~H. Hagedorn,
  S.~Mosler, N.~B. Larsen, E.~C. Cox, and H.~Flyvbjerg.
\newblock Cell motility as random motion: a review.
\newblock {\em Eur. Phys. J.: Spec. Top.}, 157:1--15, 2008.

\bibitem{LiDictyDynamics2011}
L.~Li, E.~C. Cox, and H.~Flyvbjerg.
\newblock ``{D}icty dynamics'': \textit{Dictyostelium} motility as persistent
  random motion.
\newblock {\em Phys. Biol.}, 8:046006, 2011.

\bibitem{Gnesotto}
F.~S. Gnesotto, F.~Mura, J.~Gladrow, and C.~P. Broedersz.
\newblock Broken detailed balance and non-equilibrium dynamics in living
  systems: a review.
\newblock {\em Rep. Prog. Phys.}, 81:066601, 2018.

\bibitem{52588}
D.~Serre (https://mathoverflow.net/users/8799/denis serre).
\newblock Eigenvalues of sum of a non-symmetric matrix and its transpose
  $({A}+{A}^{T})$.
\newblock MathOverflow.
\newblock URL: https://mathoverflow.net/q/52588 (version: 2011-01-20).

\bibitem{Gladrow}
J.~Gladrow, N.~Fakhri, F.~C. MacKintosh, C.~F. Schmidt, and C.~P. Broedersz.
\newblock Broken detailed balance of filament dynamics in active networks.
\newblock {\em Phys. Rev. Lett.}, 116:248301, 2016.

\bibitem{Goldenfeld}
N.~Goldenfeld.
\newblock {\em Lectures on Phase Transitions and the Renormalization Group}.
\newblock Taylor \& Francis Group LLC, 1992.

\bibitem{PriestleyTimeSeries}
M.~B. Priestley.
\newblock {\em Spectral Analysis and Time Series}, volume~1.
\newblock Academic Press Inc., 1981.

\bibitem{VanKampenStochasticProcesses}
N.~G. {van Kampen}.
\newblock {\em Stochastic Processes in Physics and Chemistry}.
\newblock Elsevier, 3 edition, 2007.

\bibitem{MezicKoopman}
I.~Mezi\'{c}.
\newblock Spectral properties of dynamical systems, model reduction and
  decompositions.
\newblock {\em Nonlinear Dyn.}, 41:309--325, 2005.

\bibitem{WilliamsKoopman}
M.~O. Williams, I.~G. Kevrekidis, and C.~W. Rowley.
\newblock A data-driven approximation of the {K}oopman operator: extending
  dynamic mode decomposition.
\newblock {\em J. Nonlinear Sci.}, 25:1304--1346, 2015.

\bibitem{LearningDynamicalSystemsKoopman}
V.~R. Kostic, P.~Novelli, A.~Maurer, C.~Ciliberto, L.~Rosasco, and M.~Pontil.
\newblock Learning dynamical systems via {K}oopman operator regression in
  reproducing kernel {H}ilbert spaces.
\newblock {\em Advances in Neural Information Processing Systems},
  35:4017--4031, 2022.

\bibitem{SharpSpectralRatesKoopman}
V.~R. Kostic, K.~Lounici, P.~Novelli, and M.~Pontil.
\newblock Sharp spectral rates for {K}oopman operator learning.
\newblock In {\em Thirty-seventh Conference on Neural Information Processing
  Systems}, 2023.

\bibitem{RiskenFokkerPlanck}
H.~Risken.
\newblock {\em The Fokker\textendash Planck Equation}.
\newblock Springer Berlin, Heidelberg, 2 edition, 1989.

\bibitem{LowCommentBruckner}
Y.~I. Low.
\newblock Comment on {"Inferring the dynamics of underdamped stochastic
  systems"}, 2026.

\bibitem{BernsteinMatrixMathematics}
D.~S. Bernstein.
\newblock {\em Matrix Mathematics}.
\newblock Princeton University Press, 2 edition, 2009.

\bibitem{StephensLongTimescalesCElegans2011}
G.~J. Stephens, M.~{Bueno de Mesquita}, W.~S. Ryu, and W.~Bialek.
\newblock Emergence of long timescales and stereotyped behaviors in
  \textit{Caenorhabditis elegans}.
\newblock {\em Proc. Natl. Acad. Sci. U.S.A.}, 108:7286--7289, 2011.

\bibitem{MetznerSuperstatisticalRandomWalks2015}
C.~Metzner, C.~Mark, J.~Steinwachs, L.~Lautscham, F.~Stadler, and B.~Fabry.
\newblock Superstatistical analysis and modelling of heterogeneous random
  walks.
\newblock {\em Nat. Commun.}, 6:7516, 2015.

\bibitem{TimeSeriesToSuperstatistics}
C.~Beck, E.~G.~D. Cohen, and H.~L. Swinney.
\newblock From time series to superstatistics.
\newblock {\em Phys. Rev. E}, 72:056133, 2005.

\bibitem{StatMechSuperstatistics}
C.~Beck.
\newblock Generalized statistical mechanics for superstatistical systems.
\newblock {\em Phil. Trans. R. Soc. A}, 369:453--465, 2011.

\bibitem{MitterwallnerNonMarkovianMotility2020}
B.~G. Mitterwallner, C.~Schreiber, J.~O. Daldrop, J.~O. {R\"{a}dler}, and R.~R.
  Netz.
\newblock Non-{M}arkovian data-driven modeling of single-cell motility.
\newblock {\em Phys. Rev. E}, 101:032408, 2020.

\bibitem{FerrettiRG2022}
F.~Ferretti, V.~Chard\`{e}s, T.~Mora, A.~M. Walczak, and I.~Giardina.
\newblock Renormalization group approach to connect discrete- and
  continuous-time descriptions of {G}aussian processes.
\newblock {\em Phys. Rev. E}, 105:044133, 2022.

\bibitem{LowCommentFerretti2023}
Y.~I. Low.
\newblock Comment on ``{R}enormalization group approach to connect discrete-
  and continuous-time descriptions of {G}aussian processes''.
\newblock {\em Phys. Rev. E}, 107:046102, 2023.

\bibitem{Ferretti2020}
F.~Ferretti, V.~Chard\`{e}s, T.~Mora, A.~M. Walczak, and I.~Giardina.
\newblock Building general {L}angevin models from discrete datasets.
\newblock {\em Phys. Rev. X}, 10:031018, 2020.

\end{thebibliography}
\end{document}